\documentclass[aps,pre,twocolumn,showpacs,superscriptaddress]{revtex4-1}  %
\usepackage{graphicx}  %
\usepackage{dcolumn}   %
\usepackage{bm}        %
\usepackage{amssymb}   %
\usepackage{color}
\usepackage {tikz}
\usepackage{epstopdf}
\usetikzlibrary {positioning}
\usepackage{amsmath}
\usepackage{caption}
\usepackage{subcaption}
\usepackage{lipsum}
\usepackage{kantlipsum} %
\makeatletter
\renewcommand\section{\@startsection{section}{1}{\z@}%
  {-3.5ex \@plus -1ex \@minus -.2ex}{2.3ex \@plus.2ex}%
  {\normalfont\large\bfseries\raggedright}}
\renewcommand\subsection{\@startsection{subsection}{2}{\z@}%
  {-3.25ex\@plus -1ex \@minus -.2ex}{1.5ex \@plus .2ex}%
  {\normalfont\normalsize\bfseries\raggedright}}
\renewcommand\subsubsection{\@startsection{subsubsection}{3}{\z@}%
  {-2.5ex\@plus -1ex \@minus -.2ex}{1ex \@plus .2ex}%
  {\normalfont\normalsize\itshape\raggedright}}
\makeatother
\begin{document}

\widetext
\title{Periodically modulated traveling waves in integrate-and-fire networks: recursive speed law and propagation failure}
\author{Jie Nissel}
\thanks{These authors contributed equally to this work.}
\affiliation{Department of Mathematics and Statistics, Georgia State University, Atlanta, Georgia, USA}
\author{Ricardo Erazo-Toscano}
\thanks{These authors contributed equally to this work.}
\affiliation{Norcliffe Foundation Center for Integrative Brain Research, Seattle Children's Research Institute, Seattle, Washington, USA}
\author{Rosahn Bhattarai}  %
\email{rbhattarai1@gsu.edu}  %
\affiliation{Department of Mathematics, Perimeter College, Georgia State University, Atlanta, Georgia, USA}
\author{Marius Osan}
\email{mariusosan11@gmail.com}  %
\affiliation{Universal Alloy Corporation, Baia Mare, Romania}
\author{Mandeep Chauhan}  %
\email{mandeep@pyrsquare.com}
\affiliation{Pyrsquare Analytics, Mountain View, California, USA}
\author{Remus Osan}
\email{remus.osan@tins.ro}
\affiliation{Transylvanian Institute of Neuroscience, Cluj-Napoca, Romania}
\date{\today}

\begin{abstract}

Traveling waves of activity in neural tissue can be halted by spatial
inhomogeneity in synaptic coupling. We study an integrate-and-fire network in which
each neuron fires once and the coupling decays exponentially. For this model the
leading-edge firing map reduces exactly to a scalar equation for the wave speed in
space, with a slow unstable and a fast stable homogeneous speed, $c_1$ and $c_2$, and
a periodic modulation of the coupling enters this equation pointwise. Positive
periodic waves terminate in folds. For slowly varying modulation the fold amplitude
approaches a plateau set by the trough of the modulation, where the local bottleneck
speed is $\sqrt{c_1c_2}$; the approach to the plateau is set by the local geometry
of the trough. For rapidly varying modulation the fold amplitude grows linearly with
frequency, with a slope given by a full-amplitude average over the integrated
modulation profile; a weak-ripple truncation of that average overestimates the slope
about $2.5$-fold at the default coupling. A perturbative recursion in the amplitude
gives the speed profile explicitly. We distinguish loss of the periodic wave from
failure of a particular launch, and the excitatory regime $\epsilon\le1$ from its
sign-changing extension. The results are checked against direct integration of the
reduced equation and first-spike firing-map simulations of the network.

Keywords: Traveling waves; Inhomogeneity; Perturbative recursion; Propagation failure; Speed approximations
\end{abstract}

\pacs{87.19.lj, 87.19.ll, 87.19.lq, 02.60.-x}
\maketitle

\section{Introduction}

The analysis of synaptically generated traveling waves in integrate-and-fire networks begins with Ermentrout~\cite{Ermentrout1998}, who derived a self-consistency condition for the velocity of a solitary pulse and showed that it admits two branches, one fast and one slow. Bressloff~\cite{Bressloff2000} then proved, for a one-dimensional network of excitable integrate-and-fire neurons in which each neuron fires once, that the fast branch is stable and the slow branch unstable, and identified a critical coupling below which no solitary pulse exists. An evolution equation describing the approach to these constant-speed solutions was obtained by Osan and Ermentrout~\cite{OsanErmentrout2002}, and related one-dimensional networks, with finite-support connectivity or with several spikes per neuron, were analyzed in Refs.~\cite{OsanRubinCurtuErmentrout2003, OsanCurtuRubinErmentrout2004}.

In Ref.~\cite{PRE2016} that evolution equation was sharpened, for exponentially decaying coupling, into an exact and purely local law. The result is worth stating plainly, since it is what the present work modulates. The firing map is a global object: the neuron now approaching threshold integrates the synaptic input generated by every neuron that has already fired, so a priori the motion of the leading edge depends on the entire history of the wave. For exponential coupling, however, differentiating the firing map twice with respect to position annihilates that integral, and what survives is a relation between the acceleration of the leading edge and its \emph{instantaneous} speed alone: the acceleration is quadratic in the speed, vanishing at a slow unstable speed $c_1$ and at a fast stable speed $c_2$. A system with unbounded memory thus reduces exactly to a scalar autonomous ordinary differential equation in space. The statement is structural rather than numerical: simulation produces trajectories, and no finite collection of them establishes that the evolution depends on the speed alone. In the homogeneous network the consequence is an all-or-none criterion, propagation failing when the speed falls below $c_1$ and otherwise relaxing to $c_2$. One of the results below is that this criterion does not survive spatial modulation.

Locality is a property of the kernel rather than a modeling convenience. For finite-support coupling the leading edge depends on a window of past firing times and the acceleration is correspondingly window-delayed~\cite{OsanRubinCurtuErmentrout2003, ErazoToscano2023}, so no scalar law of this kind is available; the two-branch $c_1$/$c_2$ structure and the all-or-none failure picture nevertheless persist~\cite{ErazoToscano2023}. It is the memorylessness of exponential coupling that makes the explicit analysis below possible.

This framework has produced insights on the mechanisms of stable constant-speed traveling wave solutions in homogeneous media, relevant to areas of the brain with a similar architecture throughout, such as neocortical layer IV~\cite{Staiger2004, Vitalis2018}. However, other brain architectures show microstructural inhomogeneities, such as the columnar organization of the visual cortex and the barrel organization of the somatosensory cortex~\cite{Rockland2010, AdamsHorton2009, Ermentrout2009, Swindale1996, Paul2001}; these approximately periodic inhomogeneities could modify the dynamics of traveling wave propagation, which motivated us to extend the framework to an inhomogeneous periodic modulation of the coupling. 

So far, inhomogeneity in the synaptic connections likely to exist in brain tissue has received much less attention, since it substantially increases the complexity of the mathematical models. Existing analyses have relied on spatial averaging and homogenization theory to obtain the average wave speed and the transition between propagation success and failure~\cite{kilpatrick2008, Paul2001, Keener2000-1}, under the assumption that the modulation amplitude $\epsilon$ is small and varies on a length scale shorter than the coupling range. As we show below, at the failure transition the speed ripple is of order unity at the default coupling, so a weak-ripple truncation of the averaged equation misestimates the failure conditions in the rapidly varying regime, and averaging over the modulation does not apply at all in the slowly varying regime, where failure is governed by the modulation's \emph{extreme} (the trough, through a local saddle-node) rather than by its mean. That naive homogenization fails for bistable propagation failure, and that a more careful averaging repairs it, has long been recognized, by Keener for the bistable equation~\cite{Keener1987, Keener2000} and by Bressloff and Kilpatrick--Folias--Bressloff for inhomogeneous neural media~\cite{Paul2001, kilpatrick2008}, with Coombes and Laing analyzing pulsating fronts under periodic modulation with an interface method~\cite{CoombesLaing2011}; a recent large-period analysis of bistable fronts in reaction--diffusion media gives an explicit limiting-speed formula in that setting~\cite{DingHamelLiang2025}. Our contribution is to ask what periodic modulation does to the exact local law described above, and to answer at two levels. Perturbatively, the modulated speed obeys a recursion whose structure is independent of the modulation shape: the shape enters once, at first order, while every higher order is forced only by products of lower orders passed through the same linear operator, so that each order follows from an explicit Fourier recursion. For a cosine modulation we obtain the harmonic content of every order explicitly. At its boundary, the periodic wave is lost at a fold, and for the exponential kernel the slow-limit bottleneck speed $c_\ast=\sqrt{c_1c_2}$, the fast-limit slope as a Stieltjes transform of the sojourn density of the modulation's centered primitive, and the finite-period corrections set by the local geometry of the trough are all explicit. The two descriptions meet in the quasi-static limit, where, for the cosine, the radius of convergence of the expansion of the quasi-static speed is the slow-limit failure amplitude.

The closest prior analysis of the same question is that of Coombes and Laing~\cite{CoombesLaing2011}, who study a coarse-grained neural field with a periodically modulated, translationally invariant kernel and develop an interface description that improves on homogenization by removing the requirement that the modulation be rapid. For modulation of the connectivity their interface calculation already yields a wavelength-dependent threshold at small amplitude, $|\epsilon|<|2h-1|\sqrt{1+(2\pi/\sigma)^2}$ in their notation [their Eqs.~(39) and (42)], which has a plateau at long period and grows linearly at short period; the qualitative plateau-to-linear crossover is therefore not new here. (Their wavelength-independent thresholds concern other forms of modulation.) The models and quantities differ. Theirs is a rate description, in which the front connects two steady states of a bistable medium and has a single homogeneous speed; here the network is spiking, each neuron fires once, and the homogeneous problem has two speed branches, $c_1$ and $c_2$. What this paper adds for that model is an exact finite-amplitude reduction to a scalar speed equation, explicit constants for its slow and fast limits, the dependence of the fast limit on the waveform, and the distinction between loss of the periodic wave and failure of a particular launch.  A preliminary account of this program, without the derivations presented here, appeared as a conference abstract~\cite{ZhangOsan2012}. A different sense of modulation is studied by Kerr, Ashwin and Wedgwood~\cite{KerrAshwinWedgwood2025}, whose periodicity is intrinsic and temporal---subthreshold oscillations from a linearized voltage-gated current---rather than imposed on the medium, and whose coupling is a balanced Mexican hat, a difference of Gaussians with $\int w=0$. Being sign-indefinite and Gaussian-tailed, that kernel admits no memoryless reduction of the firing map, and they locate the travelling waves and their folds by numerical continuation; the exponential decay assumed here is what makes the scalar law, and with it explicit asymptotics of the failure boundary, available. (The two asymptotic regimes below, slowly- and rapidly-varying, are the large- and small-period limits familiar from pulsating-front theory in periodic media~\cite{BerestyckiHamel2002}.)

Here we extend the analytically tractable integrate-and-fire framework discussed
above~\cite{OsanErmentrout2002, OsanRubinCurtuErmentrout2003, Osan2003neurocomputing2, OsanCurtuRubinErmentrout2004, PRE2016, ErazoToscano2023}
to periodically inhomogeneous coupling. Starting from the exact firing-map
equation, we derive a reduced equation in which the wave acceleration depends
quadratically on the speed plus a term linear in the local inhomogeneity, and we
treat two modulations: a piecewise-constant alternating inhomogeneity, for which
the periodic speed profile and the failure bifurcation are obtained in closed
form; and a cosine inhomogeneity, for which we construct a small-amplitude
perturbation series for the speed (its harmonic support fixed order by order)
and determine the failure boundary $\epsilon_f(\omega)$ analytically in the
large- and small-wavelength limits. We distinguish two failure observables that
are often conflated: the \emph{fold} amplitude at which the positive periodic
speed profile ceases to exist, and the \emph{launch} threshold at which a wave
started at the homogeneous speed collapses; the two differ by a few percent and
we report both. The predictions derived from the reduced equation are confirmed throughout by direct first-spike firing-map simulations of the network. For modulation amplitudes $\epsilon\le1$
the coupling stays excitatory everywhere; we treat $\epsilon>1$, where the
modulated coupling changes sign, as an extension of the model whose biological
counterpart is hypothesized rather than established (Sec.~\ref{sec:domain}).

\section{\label{sec:model} The model and its reduction}

The network is the integrate-and-fire model of Ref.~\cite{PRE2016}, modified in one
respect: the coupling carries a periodic modulation. Neurons are arranged on a line,
each fires once, and the membrane potential at position $x$ is the accumulated
synaptic input from all neurons that have already fired,
$V(x,t)=g_{syn}\int_{-\infty}^{x}J(x,y)A(t-t^*_y)\,dy$, with $t^*_y$ the firing time
at $y$ and $g_{syn}$ the global excitability. The spatial and temporal factors are
those of Ref.~\cite{PRE2016} up to the modulation,
\begin{equation}\label{eq:JA}
J(x,y)= \frac{e^{-\frac{|x-y|}{\sigma}}}{2\sigma}(1+K(y)),\; A(t) = \frac{e^{-\frac{t}{\tau_2}} - e^{-\frac{t}{\tau_1}}}{1-\tau_1/\tau_2},
\end{equation}
where $\sigma$ is the connectivity scale, $\tau_1$ the membrane integration time,
$\tau_2>\tau_1$ the synaptic decay time, and $K$ the modulation, absent in
Ref.~\cite{PRE2016}. Firing times are monotonic in position, so the leading edge is
fixed by the consistency condition
\begin{equation}\label{eq:main}
V_T=g_{syn}\int_{-\infty}^{x} J(x, y)\,A(t_x^* - t_y^*)\,dy .
\end{equation}

\paragraph*{Domain of the modulation.}\label{sec:domain}
Throughout, $K$ is a bounded, zero-mean periodic modulation normalized so that
$\min_y K=-1$, times an amplitude $\epsilon\ge0$ (so $K(y)=\epsilon k(\omega y)$ with
$\min k=-1$). Two regimes must be kept apart. For $\epsilon\le1$ the factor $1+K(y)$
stays nonnegative: the modulation weakens the excitatory coupling in its troughs,
down to a silent gap at $\epsilon=1$ (the case treated in Ref.~\cite{PRE2016}), and
this is the regime we regard as the neural model, since any nonnegative periodic
coupling $g(y)$ can be written in this form by absorbing its offset into $g_{syn}$.
For $\epsilon>1$ the factor changes sign in the troughs, and the presynaptic
neurons there deliver a negative current with the excitatory time course $A(t)$.
This is a well-defined signed-weight extension of the model, and the reduced
equation below holds unchanged, but its biological reading is hypothesized rather
than established: it would correspond to patches whose net effective drive within
the leading-edge window is inhibitory (fast feedforward inhibition, or
compartments such as barrel septa with a different interneuron composition), it
respects Dale's law only at the population level, it uses the excitatory time
course for the negative part, and it cannot describe the disinhibited slice
preparations in which propagating activity is classically recorded. We mark
$\epsilon=1$ on every phase diagram and state for each result which regime it
concerns; the same restriction to nonnegative weights was imposed in
Ref.~\cite{kilpatrick2008}.

The reduction of that firing map to a closed acceleration--speed relation follows
Ref.~\cite{PRE2016} step for step; the inhomogeneity enters only through the factor
$1+K(y)$, and the algebra is otherwise unchanged. Writing the speed as
$c=dx/dt^*_x$, the acceleration of the leading edge is
$a(x)=d^2x/d{t^*_x}^2=-c^3\,d^2t^*_x/dx^2$. Because the coupling is exponential,
differentiating the firing map does not generate additional spatial terms: the map is
effectively memoryless, unlike the finite-support case, where the leading edge depends
on a window of past firing times and the acceleration is correspondingly
window-delayed~\cite{OsanRubinCurtuErmentrout2003, ErazoToscano2023}. Differentiating
Eq.~\ref{eq:main} once and twice with respect to $x$ and eliminating the auxiliary
accumulators then yields (Sec.~\ref{S-app:accel-deriv} of the Supplemental Material~\cite{SM})
\begin{equation} \label{eq:simple}
a(x) = -\frac{(c(x)-c_1)(c(x)-c_2)}{\sigma}+  Bc(x)K(x)
\end{equation}
where $c_1$ and $c_2$, the speed for the slow-unstable and the fast-stable constant speed traveling wave solutions respectively, depend only on network parameters $\sigma$, $\tau_1$, $\tau_2$, $V_T$, $g_{syn}$ as shown here explicitly. 
\begin{align}
c_{1, 2} = \sigma/2 \Big((B-\beta) \mp 
\sqrt{(B-\beta)^2-\frac{4}{\tau_1\tau_2}} \ \Big),
\end{align}
where $B = g_{syn}/(2V_T\tau_1)$ (below we write $g$ for $g_{syn}$) and $\beta = (\tau_1+\tau_2)/(\tau_1\tau_2)$. The
analysis below holds for any admissible parameters, and we give the explicit
parameter dependence wherever a closed form exists; for concreteness, every numerical value quoted in the text and figures uses,
unless stated otherwise, the \emph{default parameter set}\label{eq:defaults}
$\tau_1=1$, $\tau_2=2$, $\sigma=1$, $V_T=1$ and $g_{syn}=10$, for which $B=5$,
$c_1=0.1492$ and $c_2=3.3508$. All quantities are dimensionless: space is measured in units of the coupling length
$\sigma$ and time in units of $\tau_1$, so speeds are in units of $\sigma/\tau_1$; since $\sigma=\tau_1=1$ in the
default set, the numbers quoted in the text and on the figure axes are directly in these units.

\section{Analysis of propagation failure under constant inhomogeneity}\label{sec:const}
In this section we first consider a simple form of inhomogeneity: a piecewise-constant modulation that alternates between $+\epsilon$ and $-\epsilon$ in blocks of length $\lambda$. We call $\lambda$ the \emph{phase length}; the full spatial period of the modulation is $2\lambda$, so the equivalent frequency is $\omega=\pi/\lambda$,
 
\begin{equation} 
K(x) = (-1)^{[x/\lambda]} \epsilon
\end{equation} 

where $[\cdot]$ denotes the floor function. (For $\epsilon>1$ the negative blocks carry sign-changing coupling, Sec.~\ref{sec:domain}; the closed forms below hold for all $\epsilon$.) The differential equation for the speed then becomes
  \begin{equation} \label{eq:alt}
\sigma\frac{dc}{dx} = -\frac{(c(x)-c_1)(c(x)-c_2)}{c(x)}+ \sigma B (-1)^{[x/\lambda]}\epsilon
\end{equation}
The solution of Eq.~(\ref{eq:alt}) is most compactly written in terms of the
per-phase quadratic $Q(c)=c^2-sc+c_1c_2$ and its discriminant, where
\begin{equation}\label{eq:sdelta}
s = c_1+c_2+m, \qquad \Delta=\sqrt{4c_1c_2-s^2},
\end{equation}
with $m = \sigma B \epsilon$ in the positive phase and $m= -\sigma B \epsilon$ in the
negative phase. Relating space $x$, speed $c$, $\epsilon$ and $\sigma$, it reads
\begin{equation}\label{eq:x(c)}
-\frac{x}{\sigma} = f(m,c) = \frac{s}{\Delta}\arctan\frac{2c-s}{\Delta}
+ k_1 + \tfrac12\ln\big|Q(c)\big| ,
\end{equation}
the integration constant being fixed by $c(x=0)=c_0$,
\begin{equation}\label{eq:k1}
k_1 = -\frac{s}{\Delta}\arctan\frac{2c_0-s}{\Delta} - \tfrac12\ln\big|Q(c_0)\big| .
\end{equation}
The discriminant $\Delta$ is real only when $Q$ has complex roots; when its roots are real (always the
case in the positive phase, and in the negative phase outside the window
$(\sqrt{c_2}-\sqrt{c_1})^2<\sigma B\epsilon<(\sqrt{c_2}+\sqrt{c_1})^2$) the arctangent
is to be read as the equivalent real logarithm of the two roots, the form written
out explicitly in Eqs.~(\ref{eq:lambda-p})--(\ref{eq:lambda-m}) below.

Equation~(\ref{eq:x(c)}) gives $x$ as a function of the speed $c$, that is, the
\emph{inverse} of the speed profile $c(x)$. Because it is transcendental (an
$\arctan$ plus a logarithm), it cannot be inverted in closed form; the profile
$c(x)$ is instead reconstructed parametrically, by sampling $c\in[c_0,c_f]$,
evaluating $x(c)$, and plotting the pairs $(x(c),c)$ piecewise, with $m =
+\sigma B\epsilon$ on the positive phase and $m = -\sigma B\epsilon$ on the
negative phase, the segments matched in $c$ at each phase boundary (the constant
$k_1$ only sets the horizontal placement of each segment). Whether the per-phase
roots are real or complex changes only the form of the primitive, not the
existence of the periodic wave: that is decided by the period-closure conditions
derived next (cf.\ Fig.~\ref{S-fig:3}).

\begin{figure*}[htb]
 \centering
  \includegraphics[width=0.96\textwidth]{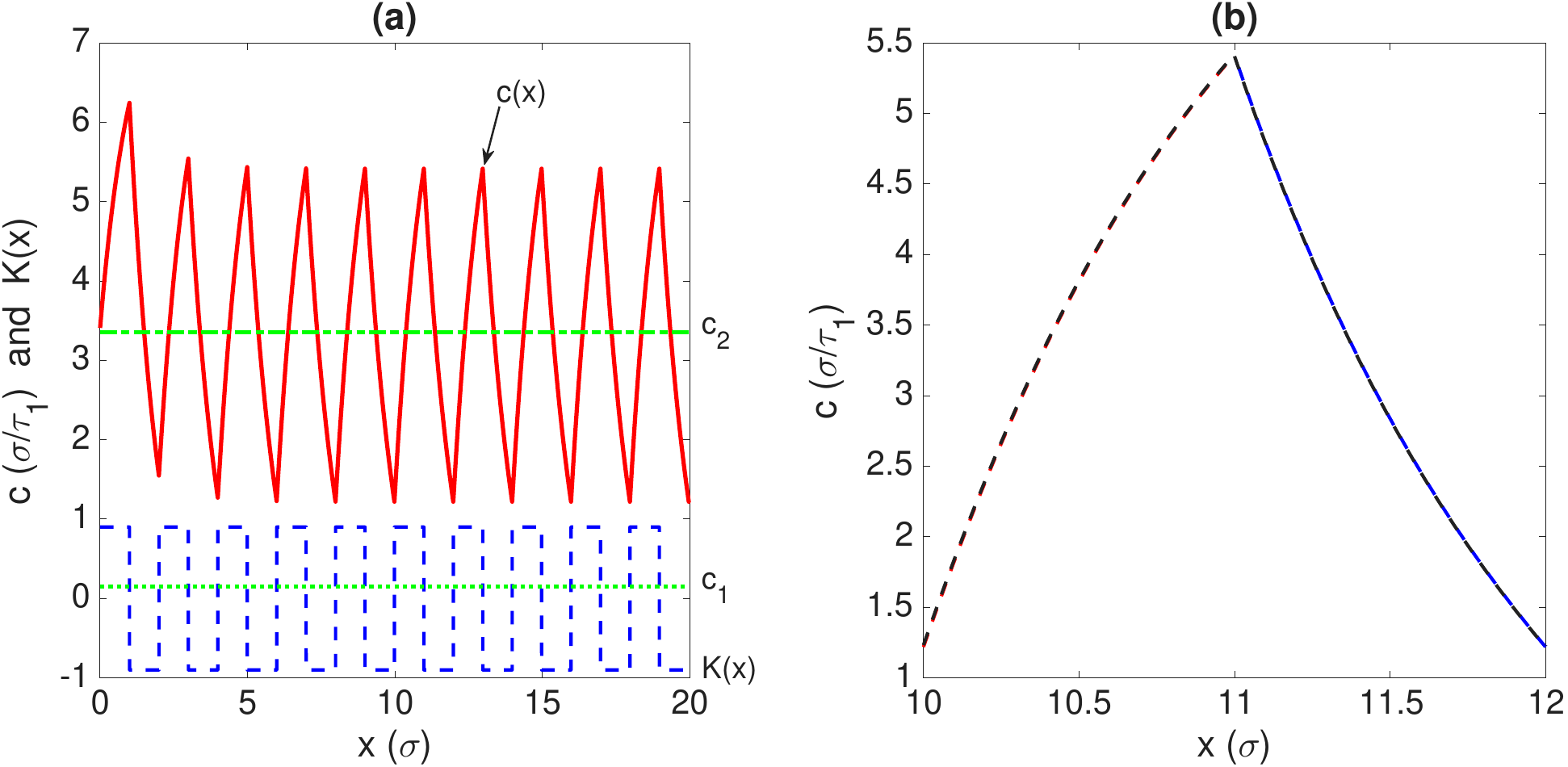}
  \caption{\label{fig:1} {\bf Traveling-wave speed with piecewise-constant (alternating $\pm\epsilon$) inhomogeneity.}  {\bf (a) Speed versus space under the alternating modulation.} The solid red line denotes the propagating periodic speed versus space ($x$) under the influence of the piecewise-constant inhomogeneity $K(x)$, drawn as the dashed blue line; the dotted and dash-dot green lines mark the homogeneous speeds $c_1$ and $c_2$.
  {\bf (b) Firing-map simulation vs.\ closed-form solution $x(c)$ (Eq.~\ref{eq:x(c)}).} The solid black line denotes
  the first-spike firing-map simulation of the network; the dashed red and blue lines are the closed-form solution (Eq.~\ref{eq:x(c)})
  for the positive ($+\epsilon$) and negative ($-\epsilon$) phases, respectively.
  Parameters: $\lambda = 1$, $\epsilon = 0.9$, default network otherwise. }
\end{figure*}

Figure~\ref{fig:1}(a) shows how the traveling-wave speed (solid red curve) varies in
space under the alternating modulation $K(x)$ (dashed blue line). The speed eventually
settles into a stable periodic profile whose minimum and maximum we denote $c_0$
and $c_f$. For $\lambda = 1$, $\epsilon = 0.9$, $\tau_1 = 1$,
  $\tau_2 = 2$, $\sigma = 1$, $V_T = 1$, $g_{syn} = 10$ this gives $c_1 =
  0.1492$, $c_2 = 3.3508$, $c_0 = 1.22$ and $c_f = 5.41$.
The minimum and maximum speeds $c_0$ and $c_f$ are not free parameters:
periodicity requires that the positive phase carry the speed from $c_0$ to $c_f$
and the negative phase return it from $c_f$ to $c_0$, each over a length
$\lambda$. Setting both phase integrals (Eqs.~\ref{eq:cp12} and \ref{eq:cm12},
below) equal to $\lambda$ yields two transcendental equations for $(c_0,c_f)$,
solved numerically. The speed profile is then the closed form
Eq.~(\ref{eq:x(c)}) reconstructed parametrically as described above; no spatial
integration of the firing map is required to draw it.
In Fig.~\ref{fig:1}(b) the analytical solution (Eq.~\ref{eq:x(c)}) of
Eq.~\ref{eq:alt} is compared with numerical simulation over one excitatory
period ($+\epsilon$) and one inhibitory period ($-\epsilon$); the two are in
excellent agreement. The simulations both confirm the analytical solution and,
by exploiting it, reduce the computational cost.
In the numerical simulation we used a ``shocked'' (initially firing) region of
length one unit. Neurons occupy a one-dimensional slice with discretization
$\delta = 10^{-3}$ over a total domain of length $20$, coupled through the
exponentially decaying kernel. Under the alternating inhomogeneity the
propagating speed varies periodically with period $2\lambda$: the positive phase
increases the wave speed and the negative phase decreases it. Whether the
traveling wave is sustained depends on $\epsilon$ and $\lambda$; intuitively, a
large amplitude and a long decreasing phase promote propagation failure. The
failure is abrupt rather than gradual, however: the wave holds a nearly periodic
profile (essentially fixed $c_0$ and $c_f$) right up until the down-swing first
crosses the critical threshold, after which there is no longer a fixed point to
arrest the descent and the speed collapses to zero within a single, violently
short final phase, rather than coasting gently down over many shrinking periods.
This is the signature of the underlying saddle-node: a slow, healthy-looking
approach followed by a sudden death.
 
To locate the periodic speeds $c_0$ and $c_f$ we read the same separable ODE,
$dx = \sigma\,c\,dc/N(c)$, as a transit-length condition rather than a
trajectory: integrated across a full phase, its left side is the phase length,
fixed at $\lambda$. Within each constant phase $N(c)$ has its own equilibrium
speeds, the roots $c_{p1,p2}$ ($+\epsilon$, raised) and $c_{m1,m2}$
($-\epsilon$, lowered), toward which the speed is drawn but which it reaches only
in the limit of infinite length; over a finite $\lambda$ it slides part-way, and
the integral measures how far. A steady periodic wave requires the up-swing
($c_0\!\to\!c_f$ in $+\epsilon$) and the down-swing ($c_f\!\to\!c_0$ in
$-\epsilon$) to close into a repeating cycle, so both phase integrals must equal
$\lambda$. This is a period-closure (return-map) condition: two equations in the
two unknowns $(c_0,c_f)$. Integrating one phase of length $\lambda$ when $K(x) = \epsilon$,
where the speed increases from $c_0$ to $c_f$,
\begin{equation}\label{eq:cp12}
\lambda = \int_{0}^{\lambda}\!dx
= \int_{c_0}^{c_f}\!\frac{\sigma c\,dc}{-(c-c_{p1})(c-c_{p2})},
\end{equation}
where $c_{p1},c_{p2}$ are the roots of the per-phase quadratic
$c^2-s_+c+c_1c_2$ obtained by multiplying the integrand through by $c$,
\begin{equation}\label{eq:cpdef}
c_{p1,p2} = \tfrac12\big(s_+ \mp \sqrt{s_+^2-4c_1c_2}\,\big),
\end{equation}
with $s_\pm = c_1+c_2\pm\sigma B\epsilon$; index 1 denotes the smaller root throughout.
Then continued with the next phase when $K(x) = -\epsilon$, where speed decreases from $c_f$ to $c_0$, 
\begin{equation}\label{eq:cm12}
\lambda = \int_{\lambda}^{2\lambda}\!dx
= \int_{c_f}^{c_0}\!\frac{\sigma c\,dc}{-(c-c_{m1})(c-c_{m2})},
\end{equation} 
where 
 \begin{equation}\label{eq:cmdef}
c_{m1,m2} = \tfrac12\big(s_- \mp \sqrt{s_-^2-4c_1c_2}\,\big).
\end{equation}

Both integrals are elementary: partial fractions evaluate them in closed form,
\begin{align}
\lambda &= \frac{\sigma}{c_{p1}-c_{p2}}\left[-c_{p1}\ln\tfrac{c_f-c_{p1}}{c_0-c_{p1}}
          + c_{p2}\ln\tfrac{c_f-c_{p2}}{c_0-c_{p2}}\right], \label{eq:lambda-p}\\
\lambda &= \frac{\sigma}{c_{m1}-c_{m2}}\left[-c_{m1}\ln\tfrac{c_0-c_{m1}}{c_f-c_{m1}}
          + c_{m2}\ln\tfrac{c_0-c_{m2}}{c_f-c_{m2}}\right], \label{eq:lambda-m}
\end{align}
for the positive and negative phases, respectively. These are nothing but
Eq.~(\ref{eq:x(c)}) evaluated across one phase, $\lambda = |x(c_f)-x(c_0)|$, with
the constant $k_1$ cancelling, so the period-closure conditions reduce to two
explicit transcendental equations in $(c_0,c_f)$, with no integral remaining.
When $c_{m1},c_{m2}$ are complex conjugates (the shaded window of
Fig.~\ref{S-fig:3}), the logarithms in Eq.~(\ref{eq:lambda-m}) combine into a real
arctangent, the same form as Eq.~(\ref{eq:x(c)}), so no separate formula is
needed.

\begin{figure}[htb]
 \centering
  \includegraphics[width=\columnwidth]{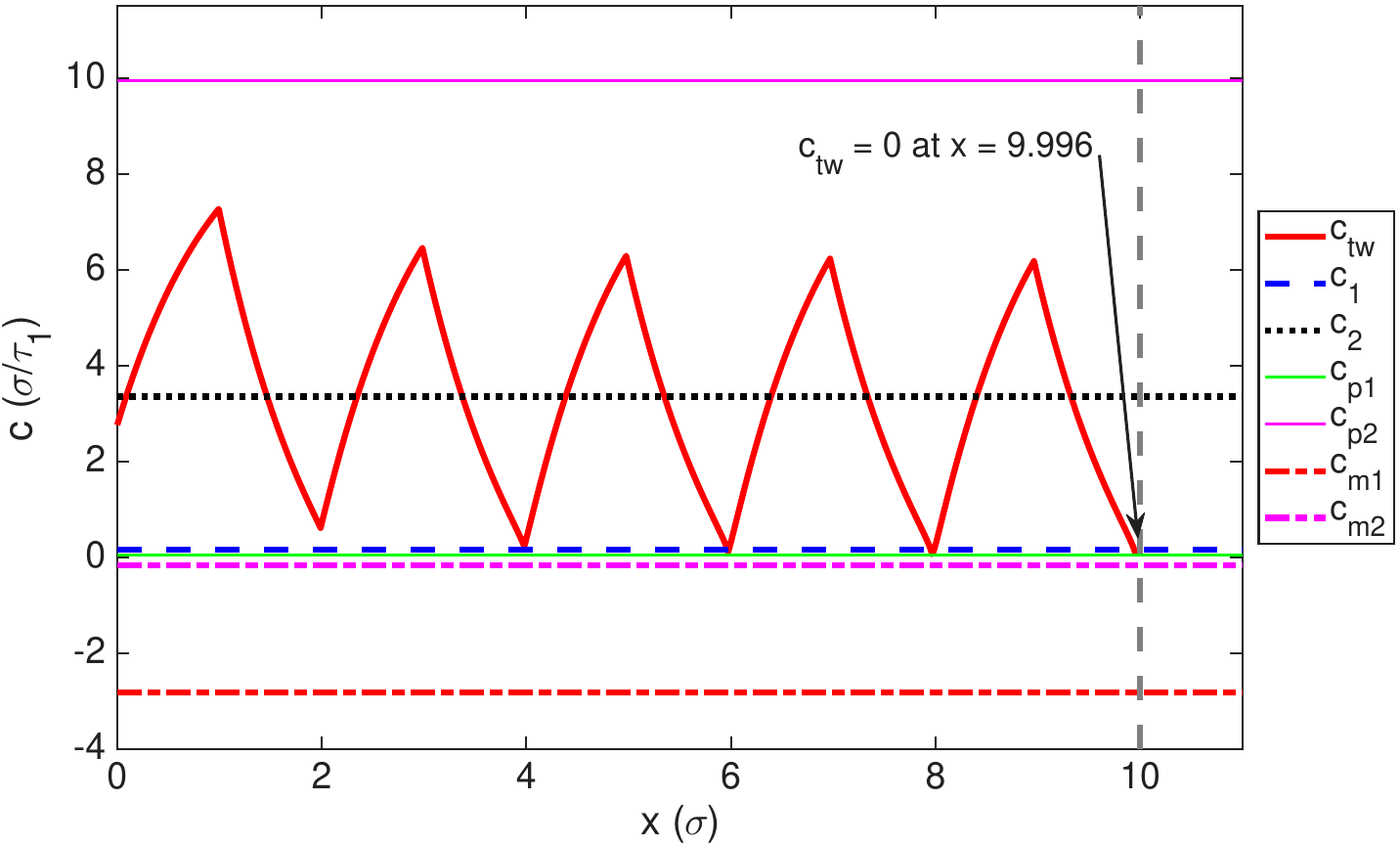}
  \caption{\label{fig:2} {\bf Traveling-wave speed and the per-phase critical speeds
  under alternating constant inhomogeneity.} The thick solid red curve is the traveling-wave speed
  $c_{tw}(x)$ for the piecewise-constant $\pm\epsilon$ modulation ($\lambda=1$,
  $\epsilon=1.3$, default network). During each excitatory ($+\epsilon$) half-period the
  speed climbs toward the stable positive-phase equilibrium $c_{p2}\approx10$ (upper
  magenta, thin solid) and away from the unstable $c_{p1}\approx0.05$ (green, thin solid); the homogeneous speeds
  $c_1\approx0.15$ (blue dashed) and $c_2\approx3.35$ (black dotted) are drawn for reference. At this
  amplitude the inhibitory-phase roots are \emph{both negative}, $c_{m1}\approx-2.8$ (red
  dash-dot) and $c_{m2}\approx-0.2$ (magenta dash-dot), so there is no positive equilibrium in the $-\epsilon$
  phase and the speed there can only fall. A speed below $c_{p1}$ cannot be recovered
  (it falls in both phases), so staying above $c_{p1}$ is necessary for survival; it is
  not sufficient, since the basin boundary is the unstable periodic profile, which lies
  above $c_{p1}$. Here the speed is recovered for several periods until the down-swing
  finally drops it below $c_{p1}$ and the wave collapses to zero at $x\approx9.996$ (gray vertical line, $c_{tw}=0$). Thus for $\lambda=1$
  the wave fails only at $\epsilon\approx1.3$, well above the slow-limit floor
  $\epsilon_f=0.42$ (Fig.~\ref{S-fig:3}): a finite period postpones failure.}
\end{figure}

The construction thus proceeds along a short dependency chain, only the last link
of which is numerical:
\begin{equation*}
g,\tau_1,\tau_2,V_T\ \rightarrow\ c_1,c_2\
\xrightarrow{+\epsilon}\ c_{p1,p2},c_{m1,m2}\
\xrightarrow{(\ref{eq:cp12}),(\ref{eq:cm12})}\ c_0,c_f .
\end{equation*}
The homogeneous fixed points $c_1,c_2$ are closed form in the network
parameters; adding the amplitude $\epsilon$ fixes the per-phase equilibria
$c_{p1,p2},c_{m1,m2}$ algebraically; and the period-closure conditions then pin
down the periodic extrema $c_0,c_f$ through a single $2\times2$ numerical solve.
Propagation failure, in this exactly solvable case, is the loss of any admissible
pair $(c_0,c_f)$ solving Eqs.~(\ref{eq:lambda-p})--(\ref{eq:lambda-m}): the stable
and unstable periodic profiles merge in a fold and disappear (Fig.~\ref{fig:4}).
Complex negative-phase roots do not by themselves imply failure (the wave of
Fig.~\ref{fig:2} survives with $c_{m1},c_{m2}$ complex for several phase lengths);
they only mean that the negative phase has no equilibrium to arrest the fall, so
that survival depends on the next positive phase catching the speed in time.

We now examine the critical speeds $c_1, c_2, c_{p1}, c_{p2}, c_{m1}, c_{m2}$ as
functions of $x$ and $\epsilon$ and explain how they shape the traveling-wave
speed (Fig.~\ref{fig:2}; Fig.~\ref{S-fig:3} of the Supplemental Material~\cite{SM}). In Fig.~\ref{fig:2}, for $\epsilon =
1.3$ the traveling wave $c_{tw}$ fails at $x = 9.996$. In the positive phase
$c_{p1}$ and $c_{p2}$ are the unstable and stable speed solutions, and if the
speed drops below $c_{p1}$ propagation fails. In the negative phase the unstable
and stable solutions are $c_{m1}$ and $c_{m2}$. When the speed falls below
$c_{m1}$, or when $c_{m1}$ and $c_{m2}$ are complex (no real solution), failure
is not determined within that phase alone.

Because the positive phase of the following period raises the speed again, the
negative-phase roots do not by themselves decide failure. Staying above $c_{p1}$ is
necessary but not sufficient: the survival boundary is the unstable periodic
profile, not a frozen-phase root. (At $\epsilon=1.29$, $\lambda=1$, a launch at
$c=0.06>c_{p1}=0.0505$ at the start of the positive phase still collapses, at
$x\approx2.01$.) For $\lambda = 1$ the periodic wave is lost at
$\epsilon \approx 1.29$, rather than where the negative-phase equilibria disappear.
These equilibria, $c_{m1},c_{m2}$, merge at $\epsilon_0=0.42$ and have
no positive value beyond it (Fig.~\ref{S-fig:3} of the Supplemental
Material~\cite{SM}); their disappearance does \emph{not}, by itself, cause failure.
For $\epsilon>\epsilon_0$ the speed falls throughout the negative phase, so an
infinitely long phase would fail at $\epsilon_0$, but a finite phase ends before
the collapse completes and the next positive phase restores the speed. Failure
therefore depends on both amplitude and phase length.

\begin{figure}[htb]
 \centering
  \includegraphics[width=\columnwidth]{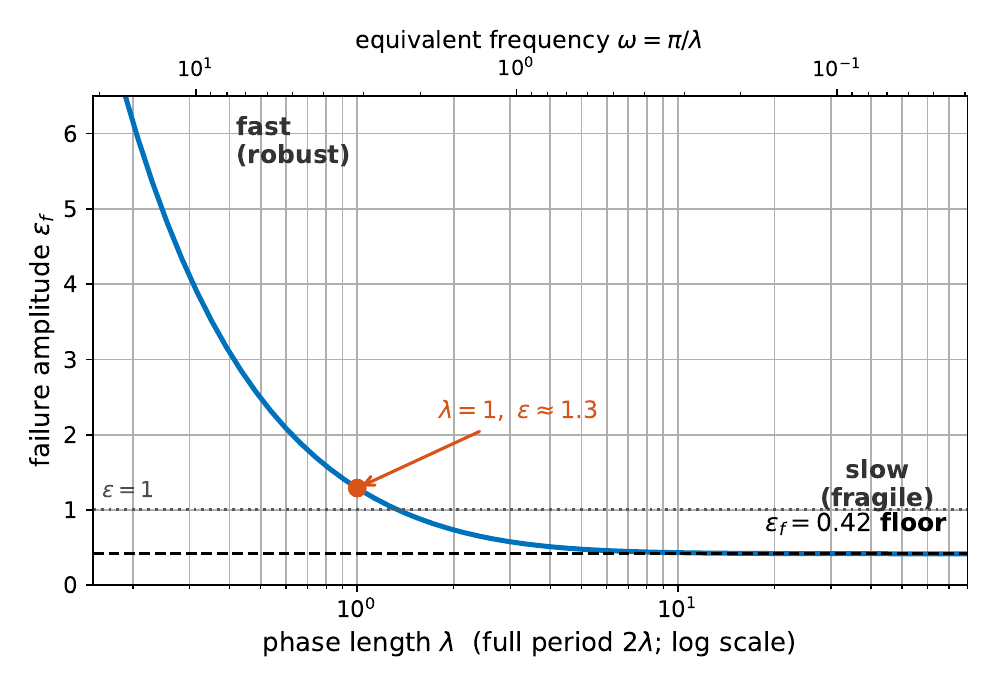}
  \caption{\label{fig:epsf-lambda} {\bf Failure amplitude $\epsilon_f$ vs.\
  phase length $\lambda$ (full period $2\lambda$, equivalent frequency $\omega=\pi/\lambda$ on the top axis; log scale), alternating $\pm\epsilon$
  inhomogeneity.} $\epsilon_f(\lambda)$ is the largest amplitude admitting a
  periodic wave, from the period-closure conditions
  (Eqs.~\ref{eq:lambda-p}--\ref{eq:lambda-m}). It decreases monotonically from a
  $1/\lambda$-like divergence at small $\lambda$ (fast modulation, robust) to the
  slow-limit floor $\epsilon_f=(\sqrt{c_2}-\sqrt{c_1})^2/(\sigma B)=0.42$ (dashed)
  as $\lambda\to\infty$. The orange marker at $\lambda=1$ ($\epsilon\approx1.3$)
  is the case of Fig.~\ref{fig:2}. Above the dotted line $\epsilon=1$ the
  modulated coupling changes sign in the negative blocks (Sec.~\ref{sec:domain}). The slow limit is the least
  robust, with failure set by the trough (extreme), not the average. Default network parameters.}
\end{figure}
Figure~\ref{fig:epsf-lambda} makes this period dependence explicit, showing the
failure amplitude $\epsilon_f(\lambda)$ obtained by solving the period-closure
conditions (Eqs.~\ref{eq:lambda-p}--\ref{eq:lambda-m}) at the fold where the
periodic orbit ceases to exist. We plot it against $\log\lambda$, since $\lambda$
ranges over orders of magnitude and the interesting behaviour is the approach to
the floor. Two limits bracket the curve. As $\lambda\to\infty$ (slow, adiabatic
modulation) $\epsilon_f$ falls to the geometric-mean floor
$\epsilon_f=(\sqrt{c_2}-\sqrt{c_1})^2/(\sigma B)=0.42$, the trough saddle-node:
a long negative phase gives the speed all the length it needs to fall off the
vanishing equilibrium, so failure is governed entirely by the extreme of the
modulation. As $\lambda\to0$ (fast modulation) $\epsilon_f$ diverges: the
negative phase is over before the speed can fall appreciably, and arbitrarily
large amplitudes are tolerated. The whole curve lies \emph{above} the floor, so
the slow limit is the most dangerous case: the wave is least robust when the
inhomogeneity varies slowly, exactly the regime in which homogenization or
averaging would be invoked, and exactly where it fails.

The way $\epsilon_f$ relaxes onto this floor follows from a short coasting
argument that also quantifies how far past threshold a finite period reaches
(the derivation is given in full in Sec.~\ref{S-app:coasting} of the Supplemental Material~\cite{SM}). For
$\epsilon$ just above $\epsilon_0$ the negative-phase equilibria $c_{m1},c_{m2}$
(which merge at the speed $\sqrt{c_1c_2}$ as $\epsilon\to\epsilon_0$) have
annihilated, so within that phase the speed has no fixed point and can only fall,
but it falls at a \emph{finite} rate, and a short enough negative phase ends
before the collapse completes. Near the merge the down-phase dynamics
($\sigma\,dc/dx=-(c-c_1)(c-c_2)/c-\sigma B\epsilon$) reduce to a saddle-node
normal form: writing $\varphi=c-\sqrt{c_1c_2}$ (the right-hand side and its
$c$-derivative both vanish at the merge, with second derivative
$-2/\sqrt{c_1c_2}$), to leading order
\begin{equation}\label{eq:coast}
\sigma\,\varphi'=-\frac{\varphi^2}{\sqrt{c_1c_2}}-D,\qquad D=\sigma B(\epsilon-\epsilon_0)>0,
\end{equation}
where $D$ is the depth below threshold (the exact nonlinear term is
$-\varphi^2/(\sqrt{c_1c_2}+\varphi)$). This Riccati flow carries the speed down to
failure ($c\to0$) over a finite length $L_{\rm coast}\propto(c_1c_2)^{1/4}/\sqrt{D}$,
almost all of it spent in the slow passage through the bottleneck near
$c=\sqrt{c_1c_2}$ (the final plunge to $c=0$ adds only an $O(\sigma)$ correction; the
$O(1)$ prefactor depends on the entry speed). The wave therefore survives as long as
the negative phase is shorter, $\lambda\lesssim L_{\rm coast}$. Equating the two gives the asymptotic boundary
\begin{equation}\label{eq:floor-approach}
\epsilon_f-\epsilon_0\ \simeq\ \kappa\,\frac{\sigma\sqrt{c_1c_2}}{B\,\lambda^2},
\qquad \kappa\approx\pi^2,
\end{equation}
where the order-unity prefactor $\kappa$ is set by the coasting integral: in the slow
limit both ends of the down-swing lie deep in the tails of the arctangent in
$L_{\rm coast}$ (each contributing $\pi/2$), giving $\kappa\to\pi^2$
(Sec.~\ref{S-app:coasting} of the Supplemental Material~\cite{SM}), which agrees to within ${\sim}10\%$ with the directly
computed boundary (Fig.~\ref{fig:epsf-lambda}). Thus $\epsilon_f$
relaxes to the floor quadratically in $1/\lambda$, always from
above: the strict extreme criterion $\epsilon_f=\epsilon_0$ is the
$\lambda\to\infty$ envelope, and a finite gradient lets the wave coast a little
past threshold by an amount that vanishes as $\lambda^{-2}$. This $\lambda^{-2}$
approach is specific to the \emph{flat} trough of the alternating profile, where
$K$ holds its extreme value throughout the down-phase, so the drift $D$ in
Eq.~(\ref{eq:coast}) is strictly constant.
A \emph{smoothly} modulated coupling behaves differently: its nonzero trough
curvature $k''$ curtails the dangerous window, and the floor is then
approached only \emph{linearly}, $\epsilon_f-\epsilon_0\propto\lambda^{-1}$,
through a Weber (parabolic-cylinder) reduction of the same normal form
(Sec.~\ref{S-app:coasting} of the Supplemental Material~\cite{SM}). The exponent of the
approach is thus set by whether the trough is flat or curved, another quantity
fixed by the local shape of the extreme, not by any spatial average.

\paragraph*{\label{sec:cosh}The geometric mean.} Written as
\begin{equation}\label{eq:isolated}
\sigma c' + c + \frac{c_1c_2}{c} = (c_1+c_2) + \sigma B\epsilon\,k(\omega x) \equiv s(x),
\end{equation}
the speed law has local equilibria where $c+c_1c_2/c=s$. The left side is
minimized, with value $2\sqrt{c_1c_2}$, at $c_\ast=\sqrt{c_1c_2}$, so under slow
modulation the two local equilibria merge at $c_\ast$ where the trough of $s$
reaches $2\sqrt{c_1c_2}$, which gives $\epsilon_0=(\sqrt{c_2}-\sqrt{c_1})^2/(\sigma
B)$ directly. This is a property of the algebraic equilibrium condition; the
differential equation itself is not invariant under $c\mapsto c_1c_2/c$.

\subsection{Speed band of the sustained wave}
\begin{figure}[htb]
 \centering
  \includegraphics[width=\columnwidth]{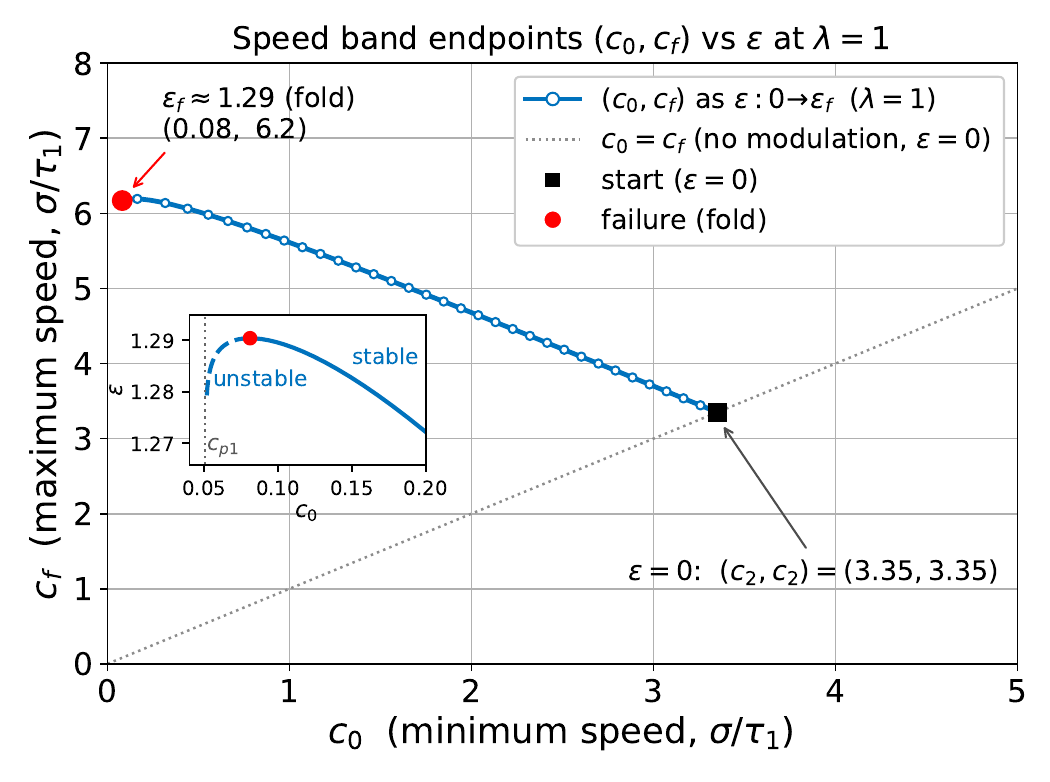}
  \caption{\label{fig:4} {\bf Speed-band endpoints $(c_0,c_f)$ traced as
  $\epsilon$ increases, at fixed $\lambda=1$
  (Eqs.~\ref{eq:lambda-p}--\ref{eq:lambda-m}).} The blue curve is the locus of the
  minimum ($c_0$) and maximum ($c_f$) wave speeds as $\epsilon$ grows from $0$ to
  $\epsilon_f$. At $\epsilon=0$ both equal the homogeneous speed $c_2$ (black
  square, on the $c_0=c_f$ diagonal); as $\epsilon$ grows the band widens ($c_0$
  falls, $c_f$ rises) until propagation fails at $\epsilon_f\approx1.29$, where the stable
  periodic wave meets an unstable periodic orbit in a fold ($(c_0,c_f)\approx(0.08,6.2)$, red).
  Inset: $\epsilon$ along the branch near the fold; the stable wave (solid) and the unstable orbit
  (dashed) meet at $\epsilon_f$, where $c_0$ is still above the threshold $c_{p1}\approx0.05$ (dotted).}
\end{figure}

When propagation is sustained, we are interested in the range of speed the wave
can reach, namely the periodic minimum $c_0$ and maximum $c_f$. These two extrema are not
free parameters: for a given amplitude $\epsilon$ and phase length $\lambda$ they are
pinned by the two closed-form period-closure conditions derived above
(Eqs.~\ref{eq:lambda-p}--\ref{eq:lambda-m}), a $2\times2$ system. When $c_0$ and $c_f$
coincide the wave travels at constant speed ($\epsilon=0$); once the $(c_0,c_f)$ solution
ceases to exist, propagation fails. It disappears in a fold, where the stable periodic wave meets an
unstable periodic orbit. At $\lambda=1$ this occurs at $\epsilon_f\approx1.29$, with
$c_0\approx0.08$ and $c_f\approx6.2$; $c_0$ is still above the threshold $c_{p1}\approx0.05$
at the fold, as illustrated in Fig.~\ref{fig:4}.

The speed band across the whole $(\epsilon,\lambda)$ plane, and for a family of
phase lengths, is shown in Figs.~\ref{S-fig:5} and~\ref{S-fig:c0cf-fam} of the
Supplemental Material~\cite{SM}.

In summary, for this alternating constant inhomogeneity the traveling-wave speed
oscillates about $c_2$ with period $2\lambda$ and amplitude set by $\epsilon$;
the periodic wave is lost at a fold of the period-closure conditions, where it
meets the unstable periodic profile that bounds its basin; $c_{p1}$
(Eq.~\ref{eq:cpdef}) is only a lower bound for that boundary. The larger $\epsilon$
and $\lambda$, the wider the speed oscillation.

\section{A general theory of propagation failure under periodic perturbation}
Having solved the constant (square-wave) inhomogeneity exactly
(Sec.~\ref{sec:const}), we now ask what survives for an \emph{arbitrary} periodic
perturbation. Let $K(x)=\epsilon\,k(\omega x)$ be any bounded, zero-mean periodic
modulation of amplitude $\epsilon$ and spatial frequency $\omega$ (wavelength
$2\pi/\omega$), normalized so that $\min k=-1$; the zero-mean condition is
essential for the fast limit below, since a nonzero mean simply shifts the
coupling and survives averaging. Through
$a=c\,dc/dx$ the leading-edge speed obeys the reduced equation
\begin{equation}\label{eq:cx-gen}
\sigma\frac{dc}{dx}=-\frac{(c-c_1)(c-c_2)}{c}+\sigma B\epsilon\,k(\omega x),
\end{equation}
the acceleration form of Eq.~\ref{eq:simple}. The square wave of
Sec.~\ref{sec:const} is the one shape for which this is autonomous between
switches and hence exactly solvable; for a general $K$ no closed form exists.
Nonetheless the qualitative behaviour is fixed by a simple competition of
scales, which we now describe before extracting the boundary.

Read as a dynamical system in the spatial variable $x$, Eq.~\ref{eq:cx-gen} is a
one-dimensional flow for the speed $c$ that is continually pushed around by the
modulation $\epsilon k(\omega x)$. Its unforced part ($\epsilon=0$) carries the two fixed points
already met above: the stable $c_2$, toward which the speed relaxes over a
characteristic length $\ell=\sigma c_2/(c_2-c_1)$, and the unstable $c_1$, the
floor of the basin below which the wave collapses. The modulation repeatedly
drives the speed away from $c_2$; whether a sustained (periodic) speed profile
survives, or the speed is forced across $c_1$ and the wave dies, is decided by
the ratio of two lengths, the relaxation length $\ell$ and the modulation
wavelength $\lambda=2\pi/\omega$. This single ratio splits the problem into two
opposite regimes, and in each the failure boundary $\epsilon_f(\omega)$ turns out
to be fixed by just one functional of the modulation shape, while the
small-amplitude speed series has a harmonic support that does not depend on the
shape's amplitude at all:
\begin{itemize}
\item[(i)] a \emph{trough-controlled} slow plateau (large $\lambda$, $\ell\ll\lambda$):
when the modulation varies on a scale $\lambda$ much longer than the speed's
relaxation length $\ell=\sigma c_2/(c_2-c_1)$, the speed has time, at every $x$, to
settle onto the instantaneous stable fixed point of Eq.~\ref{eq:cx-gen}: it tracks
the slowly moving equilibrium \emph{adiabatically}. We use the term in its
dynamical-systems sense: the equilibrium drifts slowly compared with the rate
$1/\ell$ at which $c$ relaxes onto it, so away from the trough the speed follows it
with a lag of order $\ell/\lambda$ (no thermodynamic ``heat'' is implied; the word is
borrowed only for this slow-parameter following). The tracking is not uniform: as
the two local equilibria approach each other the local relaxation rate goes to zero,
and near the trough the local saddle-node normal form (Sec.~\ref{S-app:coasting} of the Supplemental
Material~\cite{SM}) replaces adiabatic following. As $K$ descends into its trough the
instantaneous stable and unstable speeds approach one another, and at the deepest
point, where $k=\min k=-1$, they collide and annihilate in a saddle-node;
once the stable branch is gone nothing holds the speed up and the wave dies. In
this limit the wave fully samples every trough, so failure is decided by the
\emph{single deepest point} of the modulation (the wave is only as robust as the
weakest patch it must cross), giving a threshold that depends on the trough alone,
\begin{equation}\label{eq:gen-slow}
\epsilon_0=\frac{(\sqrt{c_2}-\sqrt{c_1})^2}{\sigma B}
\end{equation}
(the same for every shape under the normalization $\min k=-1$, with $\epsilon_0$ a
fixed function of the coupling and time constants, $\epsilon_0=1-g_{\min}/g$,
Eq.~\ref{eq:eps0g}, equal to $0.42$ at the default coupling, while an asymmetric
modulation normalized instead by its crest fails in proportion to its trough depth);
\item[(ii)] a \emph{profile-controlled} fast slope (small $\lambda$, $\ell\gg\lambda$):
when the modulation switches faster than the wave can respond, the speed stops
following each wiggle. It splits into a slowly varying mean $C(x)$ plus a small fast
ripple $\xi(x)$ that simply mirrors the forcing, $c=C+\xi$ with
$\sigma\xi'=\sigma B\epsilon k$ (so $\xi=(B\epsilon/\omega)\!\int\!k\,d\theta$). Averaging
the speed equation over one fast period leaves a homogenized equation for the mean,
$\sigma C'=-C+(c_1+c_2)-c_1c_2\,\langle1/c\rangle$, and everything hinges on the one
average $\langle1/c\rangle=\langle1/(C+\xi)\rangle$. The ripple averages out of the
speed itself, $\langle c\rangle=\langle C+\xi\rangle=C$ exactly, so one might guess
the same holds for the restoring term, $\langle1/c\rangle=1/C$. It does not: the
average of a reciprocal is not the reciprocal of the average,
$\langle1/c\rangle\neq1/\langle c\rangle$, and that gap is the whole story. (It also
means that the spatial mean $C$ is not the speed measured from transit time,
$v_{\rm travel}=1/\langle1/c\rangle$; for the cosine $v_{\rm travel}=\sqrt{C^2-a^2}$,
which at the averaged fold is $1.15$ against $C=3.07$. Comparisons with measured or
literature speeds must use the latter.) Because
$1/c$ curves upward, the ripple's dips, where $c$ is small, inflate
$1/c$ more than its peaks deflate it; averaged over a period the restoring term
$c_1c_2/c$ is therefore \emph{larger} than $c_1c_2/C$, so the fast ripple makes the
effective medium look \emph{weaker} than its mean would suggest. How much weaker
depends on where the ripple spends its time: it lingers near its turning points
(where it momentarily stops) and races through the middle, so the extreme values
count most. For the cosine, $\xi=a\sin(\omega x)$ dwells near $\pm a$ (the arcsine
law), and the average closes in elementary form,
\begin{equation}\label{eq:gen-fast}
\mathcal{S}(C)\equiv\Big\langle\frac{1}{C+\xi}\Big\rangle=\int\frac{\rho(\xi)}{C+\xi}\,d\xi
=\frac{1}{\sqrt{C^2-a^2}},
\end{equation}
where $\rho(\xi)$ is the fraction of each spatial period (not of physical time)
over which the ripple lies near the level $\xi$ (its occupation, or \emph{sojourn},
density), and the integral $\mathcal{S}(C)$
is its Stieltjes transform. The key point is that this average reflects the \emph{whole} waveform
through $\rho$, not just its deepest point, so the high-frequency failure slope is
shape-specific, fixed by the saddle-node of the averaged drift (a critical ripple
amplitude $a_c$ and slope $a_c/B$). And because at the boundary the ripple is of
order unity rather than small, the full average must be kept: linearizing it is
exactly where homogenization breaks down (Sec.~\ref{sec:homog});
\item[(iii)] a \emph{shape-independent} recursion: expanding the speed in
powers of $\epsilon$ about $c_2$, only the first term $h_1$ is forced directly by
$K$ (a linear, low-pass response to its Fourier content), while every higher
order is forced solely by products of lower-order terms passed through the
\emph{same} linear operator. The modulation shape thus enters once, at first
order, and the harmonic content of each order follows from that of the orders
below it. The series is analytic in $\epsilon$ while the periodic solution stays
hyperbolic, with a radius of convergence that depends on $\omega$ and on the
network parameters; we do not claim a universal value for it.
\end{itemize}
The contrast between (i) and (ii) is the conceptual core of what follows: slow
failure is governed by the modulation's \emph{extreme} (its trough) and fast
failure by its \emph{full profile} (through $\rho$), but \emph{neither} by its
spatial mean. An averaging that keeps only the mean of $K$, or that truncates the
average $\langle1/c\rangle$ at its weak-ripple quadratic term, therefore cannot
locate the boundary at either end; the fast limit itself is an average, but of
the full nonlinear restoring term (made quantitative in Sec.~\ref{sec:homog}).
We establish these three statements in turn and realize them on three standard
shapes: the smooth cosine (our principal worked example, validated against the
discrete first-spike firing map), the piecewise-linear triangle, and the piecewise-constant
square already solved in Sec.~\ref{sec:const}. Together these span smooth, kinked, and
discontinuous modulation (Fig.~\ref{fig:shapes}, Table~\ref{tab:shapeslope}).
We carry the analysis through in full for the smooth cosine $k(\omega x)=\cos(\omega x)$,
returning to the triangle and square shapes when we compare profiles below; for the
cosine the acceleration--speed relation reads
\begin{equation} \label{eq:cosine}
a(x) = -\frac{(c(x)-c_1)(c(x)-c_2)}{\sigma}+  \frac{g\,c(x)}{2V_T\tau_1}\,\epsilon \cos(\omega x).
\end{equation}

Numerically, one can shoot an initial speed and observe whether the wave
stabilizes or fails. Under forcing, the constant $c_1$ is no longer an invariant
separatrix: at $c=c_1$ the drift is $B\epsilon k(\omega x)$, which is positive on the
crests, so a trajectory can dip below $c_1$ and recover (at $\epsilon=1.1865$,
$\omega=2$ the launch from $c_2$ reaches $0.143<c_1=0.149$ and then settles onto a
periodic profile with minimum $0.483$). The basins are separated instead by the
\emph{unstable} periodic profile that coexists with the stable one; a launch
survives if it lands above it. Two failure observables must therefore be kept
apart: the \emph{fold} $\epsilon_{\rm fold}(\omega)$ at which the stable and
unstable periodic profiles merge and disappear, and the \emph{launch threshold}
$\epsilon_{\rm launch}(\omega)$ at which the particular launch $c(0)=c_2$, started
at the forcing maximum, collapses within a prescribed distance. At $\epsilon=1.2$,
$\omega=2$, for instance, the $c_2$ launch fails at $x\approx2.2$ while a launch
from $c=6$ converges to a periodic wave with minimum $0.431$: the medium supports
a wave that this launch does not reach. We compute both boundaries below
(Fig.~\ref{fig:epsf}); they differ by at most $\sim4\%$, the launch threshold
lying below the fold at intermediate frequencies (a basin effect) and slightly
above it at high frequencies (a finite-horizon effect). The two inhomogeneity
parameters $\epsilon$ and $\omega$ jointly govern propagation, which is more
easily halted for larger $\epsilon$ or smaller $\omega$ (larger wavelength).

Because the speed cannot be solved for in closed form, we attack it with two
complementary analytical tools, each suited to a different regime. When $\epsilon$
is small the speed stays near the stable value $c_2$, and a \emph{perturbation
(Taylor) series} in $\epsilon$ converges rapidly (Sec.~\ref{sec:approx}): it gives
the speed profile and its harmonic content explicitly. The periodic solution is
analytic in $\epsilon$ near $\epsilon=0$, and its radius of convergence depends on
$\omega$. A generic square-root fold on the real axis bounds the radius from
\emph{above}; a complex singularity may lie closer, so the radius is not in general
the fold amplitude (in the quasi-static limit of the cosine the two coincide at
$\epsilon_0$; at $\omega=\pi$ a contraction argument gives at least $1.25$, below the
fold at $1.85$), and a truncation, being a polynomial in $\epsilon$, cannot by
itself display the fold, at which the solution ceases to be analytic. To follow the branch
through failure we use a nonperturbative \emph{harmonic-balance} representation
(Sec.~\ref{sec:hb}), a
trigonometric expansion with a \emph{free} mean speed in which the fold appears
directly as the saddle-node of the projected equations, and, as the reference,
direct continuation of the periodic profile with a Floquet (neutral-multiplier)
condition. Harmonic balance reduces to the Taylor series in the small-$\epsilon$
limit (with the mean frozen at $c_2$), so the two descriptions agree where both
apply. The failure boundary itself we also pin analytically in the
two limiting regimes, slowly- and rapidly-varying inhomogeneity
(Secs.~\ref{sec:r1}--\ref{sec:r2}), and stitch with a compact interpolation
(Sec.~\ref{sec:compact}). We begin with the small-$\epsilon$ series.
\subsection{\label{sec:approx} Speed approximations of the traveling wave}
The transient speed is an important indicator of propagation status, but the
cosine inhomogeneity makes an analytical solution considerably harder to obtain.
Indeed, Eq.~\ref{eq:cx}, obtained from Eq.~\ref{eq:cosine} through the relation
$a(x) = c(x)\,dc/dx$, has no exact closed-form solution,
\begin{equation} \label{eq:cx}
\sigma\frac{dc}{dx} = -\frac{(c-c_1)(c-c_2)}{c}+ \sigma B\epsilon \cos(\omega x).
\end{equation}

When $\epsilon$ is small the speed stays near the stable state $c_2$, and we
expand
\begin{eqnarray} \label{eq:c_approx}
c = c_2+\epsilon h_1(x)+\epsilon^2 h_2(x) + ...+\epsilon^n h_n(x)+...
\end{eqnarray}
Substituting into Eq.~\ref{eq:cx} and collecting powers of $\epsilon$ gives, at
each order, a linear equation $\sigma h_n' = \lambda_0 h_n + (\text{forcing from
lower orders})$ with the same homogeneous rate
$\lambda_0=-(c_2-c_1)/c_2\equiv-\gamma$ (derivation in Sec.~\ref{S-app:pert} of the Supplemental Material~\cite{SM}).
At first order the forcing is the drive itself,
\begin{equation}\label{eq:h1ode}
\sigma h_1'(x) = \lambda_0 h_1(x) + \sigma B\cos(\omega x),
\end{equation}
so $h_1$ is a damped, phase-lagged linear response to the cosine,
\begin{equation}\label{eq:h1}
h_1(x)=A_1\cos(\omega x+\phi_1),\;
A_1=\frac{\sigma B}{\sqrt{\gamma^2+(\sigma\omega)^2}},\; \phi_1=-\delta_1,
\end{equation}
with phase lag $\delta_k\equiv\arctan(k\sigma\omega/\gamma)$. The interpretation is
transparent: the wave responds to the inhomogeneity through a first-order
low-pass filter: the $k$th harmonic is attenuated by
$1/\sqrt{\gamma^2+(k\sigma\omega)^2}$ and delayed by $\delta_k$.

\begin{figure}[htb]
\centering
\includegraphics[width=\columnwidth]{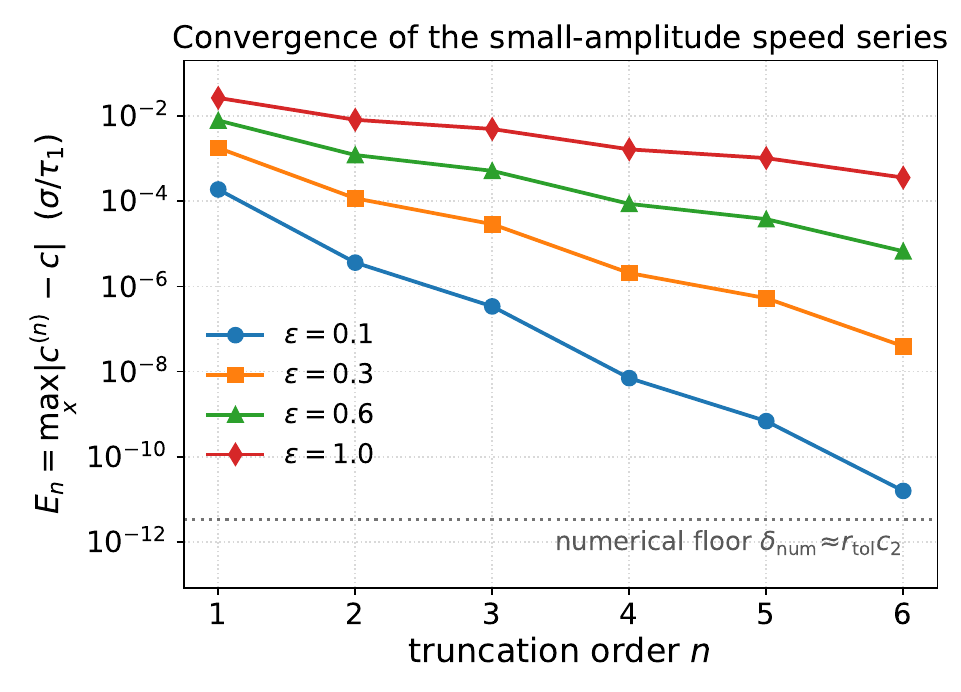}
\caption{\label{fig:conv} {\bf Convergence of the speed series.} Maximum error
$E_n=\max_x|c^{(n)}-c|$ over one period of the order-$n$ truncation versus the
numerically integrated speed ($\omega=\pi$, default parameters), for several
amplitudes $\epsilon$. The error decreases geometrically with order, rapidly for
small $\epsilon$ (reaching ${\sim}10^{-11}$ by sixth order at $\epsilon=0.1$) and
progressively more slowly as $\epsilon$ approaches $1$; six orders do not determine
the radius of convergence (see text). The dotted line is the numerical floor
$\delta_{\rm num}\approx r_{\rm tol}c_2$ set by the reference-integration tolerance;
errors at or below it reflect integration/interpolation noise rather than
truncation.}
\end{figure}

Higher orders follow the same pattern, the nonlinearity feeding products of
lower-order harmonics back through that same response: $h_2$ adds a $2\omega$
harmonic and a constant mean shift, and $h_3$ adds $\omega$ and $3\omega$
harmonics, each carrying the attenuation and lag of its order. Below, this is cast as an explicit recursion valid for every $n$
(Eqs.~\ref{eq:hn}--\ref{eq:hngen}) and bounds the truncation error
(Fig.~\ref{fig:conv}): over the orders computed at $\omega=\pi$ the error falls
roughly geometrically; at $\epsilon=1$ the first three orders are already within
$5\times10^{-3}$ of the exact speed (Fig.~\ref{S-fig:7} of the Supplemental
Material~\cite{SM}). The worked
low-order coefficients ($A_2,\phi_2,C_1$ and $A_3,A_4,\phi_3,\phi_4$) are
collected in Sec.~\ref{S-app:pert} of the Supplemental Material~\cite{SM}.

Collecting all orders (the inductive harmonic structure is verified in Sec.~\ref{S-app:pert} of the Supplemental Material~\cite{SM}), $h_n$ obeys the linear equation
\begin{align}
\sigma h_n' - \lambda_0 h_n &= -c_1 c_2\,\tilde P_n, \nonumber\\[2pt]
\tilde P_n &= \frac1{c_2}\sum_{j\ge2}\frac{(-1)^j}{c_2^{\,j}}
\!\!\sum_{\substack{m_1+\cdots+m_j=n\\ m_i\ge1}}\!\! h_{m_1}\cdots h_{m_j},
\label{eq:hn}
\end{align}
i.e.\ it is forced only by products of lower-order terms. Because each such
product is a trigonometric polynomial, $h_n$ follows harmonic-by-harmonic from
the same first-order response as $h_1$: if $f_{n,k}$ is the complex amplitude of
the $k$-th harmonic of $-c_1c_2\tilde P_n$, then
\begin{equation}\label{eq:hngen}
h_n(x)=\sum_{k\in S_n}
\frac{|f_{n,k}|\,\cos\!\big(k\omega x+\arg f_{n,k}-\delta_k\big)}
     {\sqrt{\gamma^2+(k\sigma\omega)^2}},
\end{equation}
with $S_n=\{0,2,\dots,n\}$ for even $n$ and $\{1,3,\dots,n\}$ for odd $n$.
Equations~\ref{eq:hn}--\ref{eq:hngen} (for $n\ge2$; $h_1$ comes from the forcing equation) reproduce $h_2,h_3$ above and generate
every higher order recursively.

Figure~\ref{fig:conv} shows
the maximum error $E_n=\max_x|c^{(n)}-c|$ over one period of the order-$n$
truncation $c^{(n)}=c_2+\sum_{m=1}^n\epsilon^m h_m$ against the numerically
integrated speed, at $\omega=\pi$. For small $\epsilon$ a few terms suffice (at
$\epsilon=0.1$ the third order already reaches $3\times10^{-7}$), and over the
six orders shown the error falls roughly geometrically, more slowly as $\epsilon$
approaches $1$. These are measured errors at one frequency and one parameter set;
six decreasing errors do not fix a radius of convergence, and we make no general
claim for it. A fold on the real $\epsilon$ axis is generically a square-root branch
point and so bounds the radius from above, but a nearer complex singularity is not
excluded in general (an asymmetric shape such as $k\propto4\cos\theta+\cos2\theta$
already has one, on the negative-$\epsilon$ side, in the quasi-static limit). Two
statements are available for the cosine. In the quasi-static limit $\omega\to0$ the speed tracks
the instantaneous stable fixed point
$c_+(\epsilon)=\tfrac12\big[s+\sqrt{s^2-4c_1c_2}\,\big]$, $s=c_1+c_2+\sigma
B\epsilon\cos\omega x$, whose nearest singularity in the complex-$\epsilon$ plane
is the trough saddle-node, where the square root branches, at exactly
$\epsilon=\epsilon_0=(\sqrt{c_2}-\sqrt{c_1})^2/(\sigma B)$ (the slow-limit failure
amplitude, Eq.~\ref{eq:r1}); there the radius is $\epsilon_0$ and the failure
point is the nearest singularity. This is where the two descriptions of this
paper meet: for the quasi-static cosine problem, the series loses analyticity at
the amplitude where the stable branch ceases to exist. (This is a statement about
the algebraic quasi-static branch, not a theorem about the limit $\omega\to0$ of the
radius at finite $\omega$.) At $\omega=\pi$ a contraction-mapping argument on the periodic inverse of
$d/dx+\gamma$, $\gamma=(c_2-c_1)/c_2$, shows analyticity for $|\epsilon|<1.25$, so the radius there lies between $1.25$
and the fold amplitude $1.85$, and
the order-40 recursion still converges at $\epsilon=1.2$ (error
$9\times10^{-11}$). The reference-integration floor
$\delta_{\rm num}\approx r_{\rm tol}c_2$ (dotted line in Fig.~\ref{fig:conv})
marks where the measured error becomes interpolation noise rather than truncation.
Whatever the radius, a finite truncation is a polynomial in $\epsilon$ and
displays no fold; the branch of periodic solutions is followed through failure
by the methods of Sec.~\ref{sec:hb}.

The recursion above is written for a cosine, and the harmonic support $S_n$ of
Eq.~\ref{eq:hngen} is specific to a single-harmonic drive; its \emph{structure}
is not, since the modulation enters only as the forcing at first order while
every higher order is forced by products of lower orders through the same linear
operator. The same recursion therefore applies to an arbitrary periodic
modulation, with the first-order response computed from its Fourier series.

The perturbation series truncates by power of $\epsilon$, so a finite truncation
cannot display the fold; harmonic balance (Sec.~\ref{sec:hb}) truncates instead by
the number of harmonics, keeps the mean speed free, and follows the branch through
the fold. The two agree at first order (Sec.~\ref{S-app:hb} of the Supplemental
Material~\cite{SM}).

\subsection{Conditions inducing propagation failure}
The asymptotic speed series is established in a compact form for $\epsilon < 1$, but the
question of when propagation fails is not restricted to that range. We therefore
return to the full reduced equation (Eq.~\ref{eq:cx}) and characterize the
critical perturbation amplitude $\epsilon_f$ that separates sustained
propagation from failure, as a function of the inhomogeneity frequency
$\omega$. Two opposite limits admit closed-form analysis: a slowly varying
(large-wavelength, $\omega\to0$) regime and a rapidly varying
(small-wavelength, $\omega\to\infty$) regime. They bracket the numerically
determined boundary $\epsilon_f(\omega)$ (Fig.~\ref{fig:epsf}).

\subsubsection{\label{sec:r1} Large wavelength ($\omega\to0$): quasi-static failure}
This is the cosine realization of statement~(i): with $\min k=-1$,
Eq.~\ref{eq:gen-slow} predicts the plateau, which we now derive directly. When
the modulation varies slowly, the speed adiabatically tracks the instantaneous
fixed point of Eq.~\ref{eq:cx}, $-(c-c_1)(c-c_2)+ c\,\sigma
B\epsilon\cos(\omega x) = 0$, that is,
\begin{equation} \label{eq:r1quad}
c^2 - \big(c_1+c_2+\sigma B\epsilon\cos(\omega x)\big)c + c_1 c_2 = 0.
\end{equation}
The instantaneous stable and unstable speeds collide in a saddle-node
bifurcation at the trough $\cos(\omega x) = -1$, where the discriminant of
Eq.~\ref{eq:r1quad} vanishes, $(c_1+c_2-\sigma B\epsilon)^2 - 4c_1c_2 = 0$. This
fixes the critical amplitude and the speed at which the wave dies,
\begin{equation} \label{eq:r1}
\begin{aligned}
\epsilon_f &= \frac{c_1+c_2-2\sqrt{c_1c_2}}{\sigma B}
            = \frac{(\sqrt{c_2}-\sqrt{c_1})^2}{\sigma B}, \\[2pt]
c_\ast &= \sqrt{c_1 c_2} = \frac{\sigma}{\sqrt{\tau_1\tau_2}},
\end{aligned}
\end{equation}
using $c_1 c_2 = \sigma^2/(\tau_1\tau_2)$. Substituting the microscopic
parameters ($c_1+c_2=\sigma(B-\beta)$, $B=g/2V_T\tau_1$,
$\beta=(\tau_1+\tau_2)/\tau_1\tau_2$) collapses this to a remarkably simple,
$\sigma$-independent form,
\begin{equation}\label{eq:eps0g}
\epsilon_f = 1-\frac{g_{\min}}{g},\qquad
g_{\min}=\frac{2V_T(\sqrt{\tau_1}+\sqrt{\tau_2})^2}{\tau_2},
\end{equation}
where $g_{\min}$ is precisely the minimal coupling that supports a homogeneous
wave (the value at which $c_1$ and $c_2$ coalesce; Sec.~\ref{S-sec:phase}). The
quasi-static failure floor therefore rises continuously from zero at the marginal
coupling $g_{\min}$ toward unity as the network is made more excitable; it is
\emph{not} a universal number but a fixed function of the synaptic strength and
time constants. For the default parameters $g_{\min}=5.83$, so
$\epsilon_f = 1-5.83/10 = 0.4172$ and $c_\ast = 0.7071$. This recovers and explains the bound
already noted for the negative phase of the constant inhomogeneity
(Fig.~\ref{S-fig:3}). In this limit the local stable and unstable speeds merge at
$c_\ast=\sqrt{c_1c_2} > c_1$ (Eq.~\ref{eq:isolated}): the local bottleneck is lost
at a speed strictly above the homogeneous unstable speed $c_1$. This is a
statement about the local equilibria in the slow limit. It does not make $c_\ast$
the minimum of the critical periodic profile (at the computed cosine folds that
minimum is $0.20$--$0.23$ for $0.05\le\omega\le20$) or a final stopping speed:
once the bottleneck is lost the speed keeps falling.

\subsubsection{\label{sec:r2} Small wavelength ($\omega\to\infty$): averaged failure}
This is the cosine realization of statement~(ii): the Stieltjes transform
$\mathcal{S}(C)$ of Eq.~\ref{eq:gen-fast} is evaluated for the cosine's arcsine
sojourn density. For rapid modulation we split $c = C(x) + \xi(x)$ into a slow
mean $C$ and a fast oscillation set by the forcing, $\sigma\xi' = \sigma
B\epsilon\cos(\omega x)$, giving $\xi = a\sin(\omega x)$ with amplitude $a =
B\epsilon/\omega$. At the failure boundary $a = B\epsilon_f/\omega$ is of order
unity rather than small, so the small-amplitude expansion of
Sec.~\ref{sec:approx} does not apply and the full period average must be
retained, $\mathcal{S}(C)=\langle (C+a\sin\theta)^{-1}\rangle =
(C^2-a^2)^{-1/2}$ for $C>a$. The homogenized slow dynamics become
\begin{equation} \label{eq:slow}
\sigma\frac{dC}{dx} = -C + (c_1+c_2) - \frac{c_1 c_2}{\sqrt{C^2-a^2}}.
\end{equation}
The slow stable and unstable fixed points merge at a critical amplitude $a_c$,
obtained by setting the right-hand side of Eq.~\ref{eq:slow} and its derivative
with respect to $C$ simultaneously to zero,
\begin{equation} \label{eq:r2}
C_*(c_1+c_2-C_*)^3 = (c_1 c_2)^2, \quad
a_c^2 = C_*^2 - \Big(\frac{c_1 c_2}{c_1+c_2-C_*}\Big)^2,
\end{equation}
so that the failure boundary is \emph{linear} in $\omega$ at high frequency,
\begin{equation} \label{eq:r2line}
\epsilon_f(\omega) \simeq \frac{a_c}{B}\,\omega.
\end{equation}
For the default parameters $C_* = 3.0664$, $a_c = 2.8413$, yielding a slope
$a_c/B = 0.5683$.

\subsubsection{\label{sec:compact} A compact interpolation}
The two limits can be combined into a compact empirical interpolation. Writing
$\epsilon_0=(\sqrt{c_2}-\sqrt{c_1})^2/(\sigma B)$ for the quasi-static plateau
(Eq.~\ref{eq:r1}) and $m_\infty=a_c/B$ for the large-$\omega$ slope
(Eq.~\ref{eq:r2line}), the quadrature blend
\begin{equation}\label{eq:blend}
\epsilon_f(\omega)\ \approx\ \sqrt{\epsilon_0^2+(m_\infty\omega)^2}
\end{equation}
is exact in the two limiting \emph{values}, the floor $\epsilon_0$ and the
large-$\omega$ slope $m_\infty$. It is an empirical interpolation: at the default
parameters its largest error against the computed fold on the sampled range
$0.05\le\omega\le20$ is $6.6\%$, at $\omega\approx0.44$ (Fig.~\ref{fig:epsf}, orange);
we make no claim of a uniform bound. Being even in
$\omega$, the blend approaches the floor as $O(\omega^2)$; the true slow-limit
approach for this smooth trough is instead linear, $O(\omega)$
(Sec.~\ref{S-app:coasting} of the Supplemental Material~\cite{SM}), so the blend slightly
underestimates $\epsilon_f$ just above the floor. Retaining the slow slope
$m_1=\sqrt{\sigma\sqrt{c_1c_2}\,\epsilon_0/2B}$ (Sec.~\ref{S-app:coasting} of the Supplemental Material~\cite{SM}) in a
refined interpolation,
\begin{equation}\label{eq:blend2}
\epsilon_f(\omega)\ \approx\
\sqrt{\,\epsilon_0^2+(m_\infty\omega)^2+\frac{2\epsilon_0^3\,m_1\,\omega}{\epsilon_0^2+(m_\infty\omega)^2}\,},
\end{equation}
removes this bias: it is exact in the floor value, the slow slope, \emph{and} the
fast slope (the added term decays as $\omega\to\infty$, restoring $m_\infty\omega$),
and its largest error on the sampled range is $2.1\%$ (again an empirical
interpolation, not a bound). The simpler form
Eq.~\ref{eq:blend} is enough for the qualitative argument that follows. It is
illuminating to recast it in terms of the oscillation amplitude
$a=B\epsilon/\omega$ that the forcing injects in the fast regime
($\xi=a\sin\omega x$): within the interpolation Eq.~\ref{eq:blend},
\begin{equation}\label{eq:af}
a_f(\omega)=\frac{B\epsilon_f}{\omega}\approx\sqrt{\frac{B^2\epsilon_0^2}{\omega^2}+a_c^2},
\qquad a_c\approx2.84 .
\end{equation}
In the fast regime, where $a$ is the actual ripple amplitude, failure is thus an
$O(1)$-amplitude event, $a_f\to a_c$ as $\omega\to\infty$; outside that regime
$B\epsilon/\omega$ is not the speed ripple and Eq.~\ref{eq:af} is only a
rewriting of the interpolation. The order-unity fast-regime ripple
underlies the breakdown of homogenization discussed next. The blend (C) stitches the two
asymptotic limits and stays within $6.6\%$ of the computed fold over the sampled
range (Fig.~\ref{fig:epsf}); harmonic balance (Sec.~\ref{sec:hb}) gives a
finite-order approximation of the boundary across the whole range.

\subsubsection{\label{sec:homog} Comparison with homogenization theory}
Homogenization does not merely misestimate the boundary here; it predicts a
different shape for it. Equation~\ref{eq:homog} is linear through the origin, so
it asserts that modulation of arbitrarily small amplitude halts the wave if it is
slow enough, whereas the true boundary has a floor $\epsilon_0$ below which no
modulation stops it at any wavelength. Table~\ref{tab:homog} gives the size of
the discrepancy against the blend Eq.~\ref{eq:blend}. The rest of this subsection derives the estimate and explains why.

\begin{table}[htb]
\caption{\label{tab:homog} Failure boundary $\epsilon_f(\omega)$ at the default
coupling: homogenization (Eq.~\ref{eq:homog}) and the blend (Eq.~\ref{eq:blend})
against the fold computed by periodic collocation (neutral Floquet multiplier,
$|\log M|<10^{-6}$). Near the homogeneous onset $g_{\min}=5.83$, where
$a_c/C_\ast=0.047$, the homogenization slope is instead accurate to $0.1\%$.}
\begin{ruledtabular}
\begin{tabular}{lccccc}
$\omega$ & Boundary & Homog.\ & Error & Blend & Error \\
\hline
$1$ & $0.736$ & $1.415$ & $+92\%$  & $0.705$ & $-4.2\%$ \\
$2$ & $1.230$ & $2.829$ & $+130\%$ & $1.211$ & $-1.6\%$ \\
$3$ & $1.769$ & $4.244$ & $+140\%$ & $1.755$ & $-0.8\%$ \\
$5$ & $2.880$ & $7.074$ & $+146\%$ & $2.872$ & $-0.3\%$ \\
\hline
$\to0$ (floor $\epsilon_0$) & $0.42$ & $0$ & $-100\%$ & $0.42$ & exact \\
$\to\infty$ (slope) & $0.568$ & $1.415$ & $+149\%$ & $0.568$ & exact \\
\end{tabular}
\end{ruledtabular}
\end{table}

\begin{figure*}[htb]
 \centering
  \includegraphics[width=0.96\textwidth]{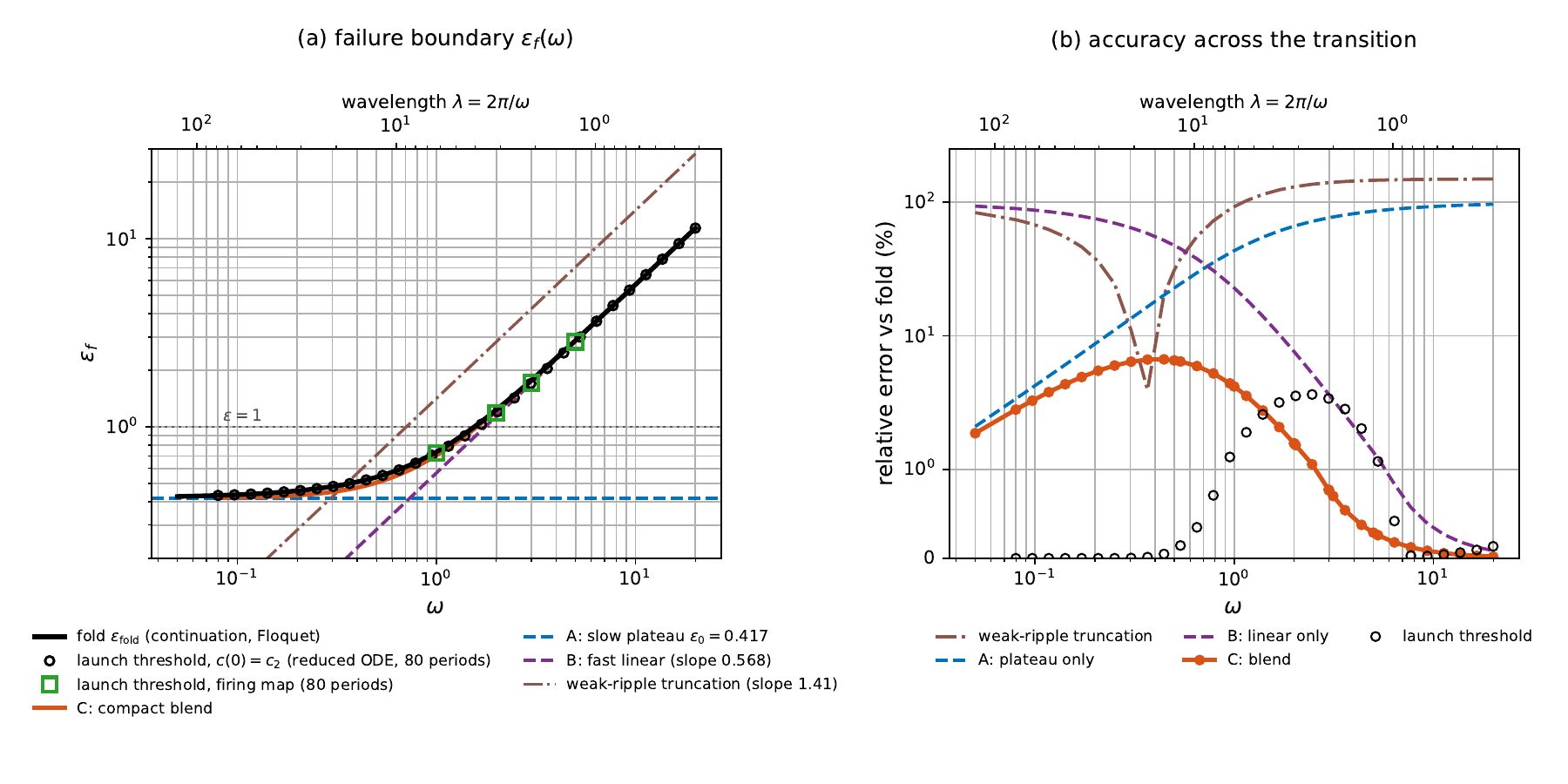}
  \caption{\label{fig:epsf} {\bf Failure boundary $\epsilon_f(\omega)$ under
  cosine inhomogeneity, and the accuracy of the asymptotic analyses.}
  {\bf (a)} Boundary on log--log axes. Black line: the fold $\epsilon_{\rm fold}$ at
  which the positive periodic speed profile ceases to exist (continuation of the
  periodic profile of Eq.~\ref{eq:cx} with a neutral Floquet multiplier); open
  circles: launch threshold for $c(0)=c_2$ started at the forcing maximum (bisection,
  near-zero stop, $80$ periods); green squares: launch threshold of the discrete
  first-spike firing map with the same launch protocol (Table~\ref{tab:netgrid},
  $d=0.0005$). Both launch thresholds are finite-distance ($80$-period) survival
  thresholds, not folds. The dotted line marks $\epsilon=1$, above which the modulated
  coupling changes sign (Sec.~\ref{sec:domain}); the fold crosses it at
  $\omega\approx1.55$. Analytical predictions: {\bf A}, slow-modulation plateau (blue dashed,
  $\epsilon_0=0.417$, Eq.~\ref{eq:r1}); {\bf B}, fast-modulation linear law
  (purple dashed, slope $a_c/B=0.568$, Eqs.~\ref{eq:r2}--\ref{eq:r2line});
  {\bf C}, compact blend (orange, Eq.~\ref{eq:blend}); weak-ripple truncation of
  the averaged equation (brown dash-dot, slope $1.41$, Eq.~\ref{eq:homog}), which
  overshoots by $2.5\times$ at this coupling.
  {\bf (b)} Relative error of each prediction, and of the launch threshold, against
  the fold (symlog axis: logarithmic above $1\%$, linear below). The plateau (A) is
  accurate only at small $\omega$ and the linear law (B) only at large $\omega$,
  whereas the blend (C) stays within $6.6\%$ over the sampled range. The
  launch threshold lies below the fold by up to $3.6\%$ at intermediate frequencies
  (a basin effect) and slightly above it at high frequencies (finite integration
  horizon). Default parameters (default parameter set, Sec.~\ref{sec:model}) $\tau_1=1,\tau_2=2,
  \sigma=1, V_T=1, g_{syn}=10$ give $c_1=0.1492$, $c_2=3.3508$, $B=5$.}
\end{figure*}

Spatial averaging and homogenization theory, used previously to estimate
average wave speed and the success/failure transition in inhomogeneous neural
media~\cite{Paul2001, kilpatrick2008, Keener2000-1}, rest on the joint
assumption of small amplitude and short wavelength. Expanding the period average
to leading nontrivial order, $\langle(C+a\sin\theta)^{-1}\rangle \simeq
C^{-1}(1+a^2/2C^2)$, reduces Eq.~\ref{eq:slow} to the weakly inhomogeneous drift
equation $\sigma C' = -(C-c_1)(C-c_2)/C - c_1c_2 a^2/(2C^3)$, whose saddle-node
occurs at $c_1c_2 a^2/2 = G^* \equiv \max_C[-(C-c_1)(C-c_2)C^2]$. This yields the
homogenization estimate of the failure boundary,
\begin{equation} \label{eq:homog}
\epsilon_f^{\,\mathrm{H}}(\omega) \simeq \frac{a_H}{B}\,\omega, \qquad
a_H = \sqrt{\frac{2G^*}{c_1 c_2}} .
\end{equation}
For the default parameters $a_H = 7.073$, a slope $a_H/B = 1.4147$, a factor
$2.5$ larger than the exact value $0.5683$ (Fig.~\ref{fig:epsf}, magenta). We
stress what is being compared: Eq.~\eqref{eq:homog} is the weak-ripple
(quadratic-fluctuation) truncation of the averaged equation for \emph{this} model,
not a result taken from the homogenization literature applied to its own models,
and the comparison is at the default coupling. The origin of the discrepancy is
explicit: the estimate \eqref{eq:homog} requires $a_H = 7.07 > c_2 = 3.35$, so the
weak-ripple expansion is violated at its own predicted failure point.
Equation~\ref{eq:af} makes this sharper: since $a_f\ge a_c\approx2.84$ everywhere on
the failure boundary at this coupling, the relevant small parameter $a/C$ is of order
unity along the \emph{entire} curve, not merely asymptotically: at the default
coupling $a_c/C_\ast = 2.84/3.07 = 0.93$, so the expansion parameter is within
$7\%$ of $1$ exactly where the estimate is being used. The truncation
does \emph{not} miss the nonlinearity; it correctly retains the leading convex
(variance) correction $a^2/2C^2$, but it \emph{truncates} the full average
$\langle1/c\rangle=1/\sqrt{C^2-a^2}=C^{-1}(1+a^2/2C^2+3a^4/8C^4+\cdots)$ after that
term. The truncation is excellent while $a\ll C$, but as $a$ approaches $C$ the
speed dips toward small values, where $1/c$ is large, and these excursions come to
dominate the average. The size of the error is therefore parameter-dependent: near
the homogeneous onset $g_{\min}=5.83$ the critical ripple is small ($a_c/C_\ast=0.047$
at $g=5.83$) and the truncated slope is accurate to $0.1\%$; at $g=10$ the ripple is
$a_c/C_\ast=0.93$ and the slope is off by $2.5\times$. The error thus spans three
orders of magnitude across the coupling range, and is not a fixed property of the
method. Being a short-wavelength theory, the truncation also does not address
the quasi-static plateau $\epsilon_0$ that governs failure at small $\omega$
(its extrapolation to $\omega\to0$ predicts $\epsilon_f\to0$, a relative error
approaching $100\%$ of the true plateau). Retaining the full $O(1)$-amplitude
average, Eqs.~\ref{eq:r2}--\ref{eq:r2line}, is what captures the fast limit
correctly; the slow limit requires the adiabatic analysis of Sec.~\ref{sec:r1}
instead.

As shown in Fig.~\ref{fig:epsf}, the fold settles onto the quasi-static plateau
(Eq.~\ref{eq:r1}) as $\omega\to0$ and onto the linear law (Eq.~\ref{eq:r2line})
as $\omega\to\infty$, with representative values
$\epsilon_{\rm fold} = 0.426, 0.540, 0.736, 1.230, 3.442, 5.702$ at
$\omega = 0.05, 0.5, 1, 2, 6, 10$; the corresponding launch thresholds for
$c(0)=c_2$ are $\epsilon_{\rm launch}=0.426, 0.539, 0.725, 1.187, 3.420, 5.704$.
In the slow regime the two coincide (the launch relaxes onto the unique stable
profile within the first period); at intermediate frequencies the launch fails a
few percent before the profile disappears. Propagation is thus most fragile for slow
inhomogeneity, where an amplitude as small as $\epsilon_f\approx0.42$ stops the
wave, and increasingly robust as the inhomogeneity becomes finer-grained, the
network effectively averaging over the rapid spatial modulation.

\subsubsection{\label{sec:hb} Harmonic balance through the fold}
The perturbation series of the preceding section is an expansion about the
stable periodic solution and, at any finite order, is a polynomial in $\epsilon$;
it does not display the fold at which that solution disappears. A nonperturbative
trigonometric representation, obtained by harmonic balance, removes this
restriction. Multiplying Eq.~\ref{eq:cx} by $c$ gives the polynomial form
\begin{equation}\label{eq:poly}
\sigma c\,c' + (c-c_1)(c-c_2) - \sigma B\epsilon\,c\cos(\omega x) = 0,
\end{equation}
into which we substitute the truncated series
$c = \bar c + \sum_{k=1}^{K}[a_k\cos(k\omega x)+b_k\sin(k\omega x)]$ and project onto
$\{1,\cos k\omega x,\sin k\omega x\}_{k\le K}$, obtaining $2K+1$ algebraic
equations for the coefficients. These are solved self-consistently and carry no
restriction on $\epsilon$; propagation failure is the fold (saddle-node) at
which the periodic solution ceases to exist.

For $K=1$ the system closes analytically. With $s\equiv2\bar c-(c_1+c_2)$, the
$\cos$ and $\sin$ projections give
\begin{equation}\label{eq:hbamp}
a_1 = \frac{\sigma B\epsilon\,\bar c\,s}{s^2+(\sigma\omega \bar c)^2},\qquad
b_1 = \frac{\sigma^2 B\epsilon\,\omega\,\bar c^2}{s^2+(\sigma\omega \bar c)^2},
\end{equation}
and the mean projection reduces to a single relation between $\epsilon$ and the
mean speed $\bar c$,
\begin{equation}\label{eq:hbbranch}
\epsilon^2(\bar c) = \frac{-2(\bar c-c_1)(\bar c-c_2)\,[\,s^2+(\sigma\omega \bar c)^2]}
{(\sigma B)^2\,\bar c\,[(c_1+c_2)-\bar c]}.
\end{equation}
This is the periodic-solution branch: $\bar c=c_2$ at $\epsilon=0$, with $\bar c$
decreasing as $\epsilon$ grows. The branch folds where $d\epsilon^2/d\bar c=0$, so
the $K=1$ estimate of the failure amplitude is explicit,
\begin{equation}\label{eq:hbfold}
\epsilon_f^2 = \max_{c_1<\bar c<c_2}\epsilon^2(\bar c),
\end{equation}
attained at a mean speed $\bar c^{\,\ast}$; it is a one-harmonic estimate at every $\omega$.

At its fold the $K=1$ profile is not an admissible speed profile (its minimum is
negative, $-0.40$ at $\omega=1$), so Eq.~\ref{eq:hbfold} is a threshold estimate.
Against the computed fold (Fig.~\ref{fig:epsf}) it overestimates by $22$--$39\%$
over $0.05\le\omega\le20$; two and three harmonics reduce the largest error to
$13\%$ and $6\%$. The mode equations beyond $K=1$, the comparison with the
perturbation series, and these error curves are given in Sec.~\ref{S-app:hb} of the
Supplemental Material~\cite{SM}.

\subsubsection{\label{sec:netval} Locality tested against the discrete firing map}
\begin{figure}[htb]
 \centering
  \includegraphics[width=\columnwidth]{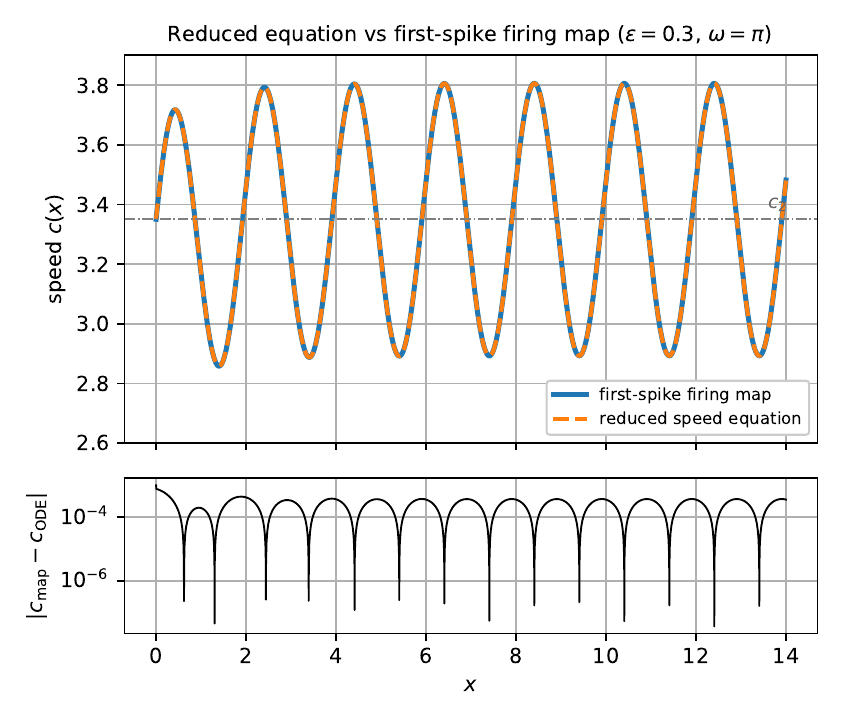}
  \caption{\label{fig:netval} {\bf Reduced equation versus the discrete firing map
  ($\epsilon=0.3$, $\omega=\pi$).} Both start from the homogeneous speed $c_2$ at
  the forcing maximum (for the firing map, homogeneous prehistory; grid
  $d=5\times10^{-4}$). {\bf (top)} Speed profile $c(x)$ about $c_2$ (dash-dot): the
  first-spike firing map (blue) and the reduced speed equation (orange dashed) are
  indistinguishable. {\bf (bottom)} Their pointwise difference stays below
  $10^{-3}$ throughout and below $4\times10^{-4}$ after $x=2$
  (\texttt{fig\_netval.py}). }
\end{figure}

The reduced equation Eq.~\ref{eq:simple} rests on the locality of the
exponential-kernel firing map, and the modulated problem is where that locality
can be tested sharply. In the homogeneous network the wave simply relaxes to
$c_2$, and a reduction that quietly retained some memory of the wave's history
would be difficult to distinguish from one that did not; periodic modulation
breaks the translation invariance that hides the difference, so that if the
firing map carried any dependence on where the front had been, the acceleration
could not be a function of $c$ and $x$ alone. We therefore simulate the
integrate-and-fire firing map directly (ordered first spikes on a grid of
neurons), using an exact two-accumulator recursion for the exponential kernel
that solves each neuron's threshold crossing without truncating the synaptic
history. This validates the discretized leading-edge dynamics; it does not
address repeated spiking, and the statements about periodic orbits and their
attraction below refer to the spatial speed flow, not to the temporal stability
of the network. With $\epsilon = 0$ the
simulation reproduces the homogeneous stable speed to a relative error of
$3\times10^{-7}$. Under cosine inhomogeneity the network speed profile $c(x)$
locks onto the periodic orbit of the reduced equation after a brief transient
(Fig.~\ref{fig:netval}). With the launch prepared as in the reduced equation
(homogeneous prehistory at speed $c_2$, modulation phase $\cos\omega x$ from
$x=0$, $80$ periods), the network's launch threshold converges with the grid
spacing to the reduced-equation launch threshold ($\epsilon_{\rm launch} = 0.725,
1.187, 1.709, 2.840$ at $\omega = 1, 2, 3, 5$; Table~\ref{tab:netgrid}, green
squares in Fig.~\ref{fig:epsf}). Both are finite-distance launch observables, not
folds. Started instead from a shocked (initially firing) region, as in
Sec.~\ref{sec:const}, the network gives $0.736, 1.214, 1.718, 2.765$; the difference
is the launch preparation. The reduced equation thus
describes the leading edge of the discrete network for consistently prepared
launches.

\begin{table}[htb]
\caption{\label{tab:netgrid} Launch threshold of the discrete first-spike firing map
versus grid spacing $d$, with the launch protocol matched to the reduced equation
(homogeneous prehistory at speed $c_2$, modulation phase $\cos\omega x$ from $x=0$,
$80$ periods, near-zero stopping criterion). The thresholds converge with $d$ toward
the reduced-ODE launch thresholds quoted in the text ($0.725$, $1.187$, $1.709$,
$2.840$ at $\omega=1,2,3,5$); the $d=0.0005$ column is plotted in
Fig.~\ref{fig:epsf}. Shocked-region runs (MATLAB \texttt{fullnet\_sim.m}) give
$0.736,1.214,1.718,2.765$ and differ through their launch preparation.}
\begin{ruledtabular}
\begin{tabular}{lccc}
$\omega$ & $d=0.002$ & $d=0.001$ & $d=0.0005$ \\ \hline
$1$ & $0.72527$ & $0.72528$ & $0.72529$ \\
$2$ & $1.18625$ & $1.18643$ & $1.18651$ \\
$3$ & $1.70767$ & $1.70831$ & $1.70863$ \\
$5$ & $2.83530$ & $2.83777$ & $2.83899$ \\
\end{tabular}
\end{ruledtabular}
\end{table}

\subsection{\label{sec:shapes} Comparison across modulation shapes}
\begin{figure}[htb]
\centering
\includegraphics[width=\columnwidth]{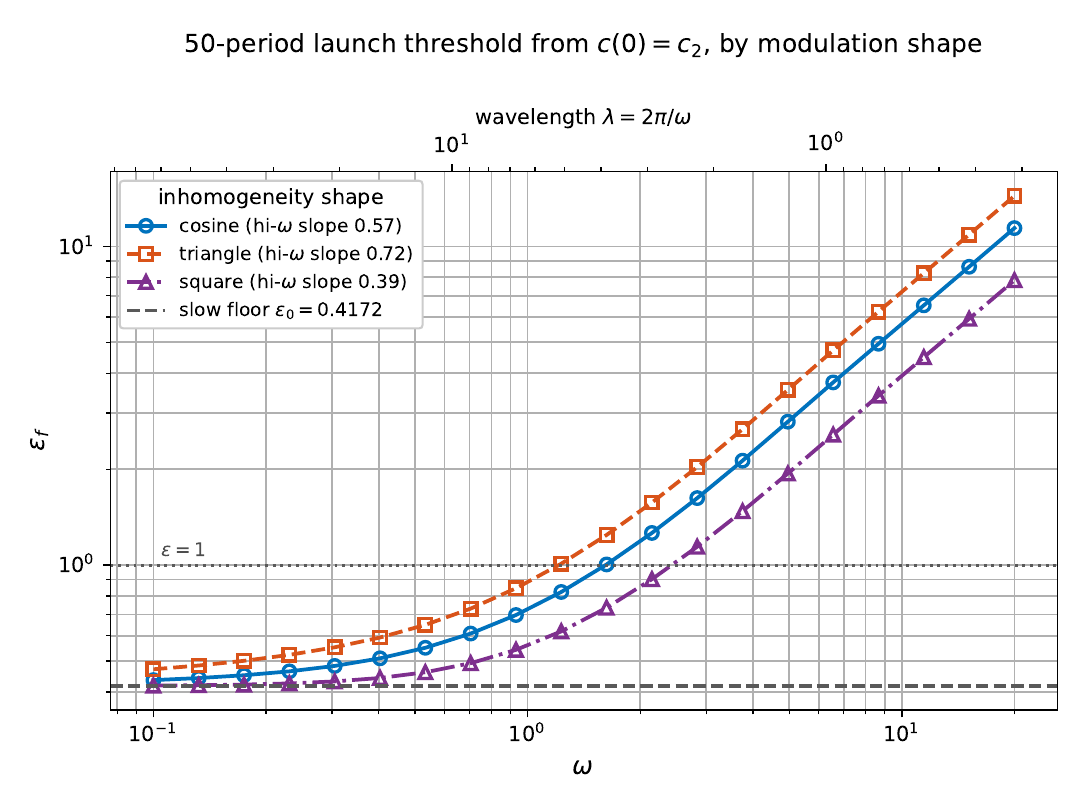}
\caption{\label{fig:shapes} {\bf Launch threshold ($50$ periods, from $c(0)=c_2$)
for three periodic inhomogeneity shapes} (square/alternating, triangle, cosine; amplitude
$\epsilon$, period $2\pi/\omega$, normalized to $\min k=-1$). All approach the common
slow-limit floor $\epsilon_0=(\sqrt{c_2}-\sqrt{c_1})^2/(\sigma B)=1-g_{\min}/g$
(dashed; $0.42$ at the default coupling $g=10$, Eq.~\ref{eq:eps0g}), set by the
trough saddle-node; their high-frequency slopes ($0.39$, $0.72$, $0.57$) differ,
set by averaging over the full modulation profile (slopes fitted over
$3.75\le\omega\le20$). Reduced-ODE bisection; these are finite-distance launch
thresholds, not folds. The dotted line marks $\epsilon=1$ (Sec.~\ref{sec:domain}).
Default parameters; the wavelength $\lambda=2\pi/\omega$ is shown on top.}
\end{figure}

We now make the general statements~(i)--(iii) concrete on the three standard
shapes, the cosine (smooth), the triangle (kinked), and the square
(discontinuous, solved exactly in Sec.~\ref{sec:const}), all of amplitude
$\epsilon$ and period $2\pi/\omega$. By statement~(i) the three boundaries
collapse onto the common trough-controlled floor $\epsilon_0=0.42$ as
$\omega\to0$ (Fig.~\ref{fig:shapes}); their high-frequency slopes, however,
differ, because each shape carries a different sojourn density $\rho(\xi)$ in the
Stieltjes transform $\mathcal{S}(C)$ of Eq.~\ref{eq:gen-fast}. That transform is
elementary in every case: the cosine's arcsine density gives
$\mathcal{S}=(C^2-a^2)^{-1/2}$; the square's \emph{uniform} density (its
$\int\!K$ is a triangle wave) gives the logarithm
$\mathcal{S}=\operatorname{arctanh}(a/C)/a$, whose saddle-node has the clean
characterization $C_*^2=a_c^2+c_1c_2$; and the triangle's parabolic $\int\!K$
gives an arctangent--arctanh combination (all derived in
Sec.~\ref{S-app:triangle} of the Supplemental Material~\cite{SM}). The resulting high-frequency slopes are $0.57$,
$0.72$, and $0.39$ for cosine, triangle, and square. The slopes of the
$50$-period launch thresholds agree with them to within $1\%$
(Fig.~\ref{fig:shapes}, Table~\ref{tab:shapeslope}); this compares launch
thresholds, not folds. For the cosine the two differ by up to $2\%$ at the low end of
the fit range and by less than $0.2\%$ for $\omega\gtrsim7$ (Fig.~\ref{fig:epsf}). The failure boundary is therefore pinned at low
frequency by the modulation's \emph{extreme} and shaped at high frequency by its
\emph{full profile}.

Statement~(iii) is borne out as well: only the first-order term $h_1$ depends on
$k$, while every higher order follows the same recursion (Sec.~\ref{S-app:pert} of the Supplemental Material~\cite{SM});
the shape still affects every higher coefficient through $h_1$, so convergence
rates must be measured shape by shape (Sec.~\ref{S-app:triangle} of the Supplemental Material~\cite{SM}). In sum, for a bounded zero-mean periodic modulation the
fold amplitude rises from the trough-controlled floor
(Eq.~\ref{eq:gen-slow}) to the profile-controlled linear regime
(Eq.~\ref{eq:gen-fast}); the cosine, triangle,
and square are the explicitly worked smooth, kinked, and discontinuous standard
forms.

\begin{table}[htb]
\caption{\label{tab:shapeslope} High-frequency slope $d\epsilon_f/d\omega$ of the
failure boundary for the three standard shapes: the closed-form averaging
prediction (the Stieltjes transform of the occupation density of $\int\!k$,
Eq.~\ref{eq:gen-fast}, with the speed-positivity constraint) versus the slope of the
$50$-period launch threshold from $c(0)=c_2$ (reduced-ODE bisection, least-squares
line over $3.75\le\omega\le20$). This is a finite-distance launch observable, not
the fold; the two agree to $\lesssim1\%$ here.}
\begin{ruledtabular}
\begin{tabular}{lccc}
shape & $\max\int\!k$ & averaging & launch \\ \hline
cosine (smooth)        & $1.00$  & $0.568$ & $0.571$ \\
triangle (kinked)      & $0.785$ & $0.717$ & $0.721$ \\
square (discontinuous) & $1.57$  & $0.390$ & $0.391$ \\
\end{tabular}
\end{ruledtabular}
\end{table}

These three shapes are representatives of a broader pattern. The high-frequency
slope sees the modulation only through $\mathcal{S}(C)$, which is elementary
precisely when $\xi=\int\!K$ stays within a single uniformizable function class,
and two families qualify. Piecewise-\emph{polynomial} $K$ keeps $\xi$ polynomial
and $\mathcal{S}(C)$ elementary at every degree; the triangle is simply the first
continuous member, the higher tents closed-form but increasingly unwieldy. And
bowing the square wave's plateaus into \emph{exponential} ramps gives $\xi$ a
sojourn density that is reciprocal in a shifted coordinate and $\mathcal{S}(C)$ a
single logarithm, with the square recovered as the ramps flatten ($b\to0$) and the
slope sweeping continuously through the cosine and triangle values as the bend
grows. For a \emph{continuous} exponential tent the nonzero mean adds a linear
pedestal to the exponential $\xi$, and we have not found an elementary form
(Sec.~\ref{S-app:expfamily} of the Supplemental Material~\cite{SM}).

The two asymptotic limits probe different features of the modulation. The fast slope is a functional of the entire
profile through the sojourn density $\rho(\xi)$ of $\int\!k$, so it separates the
shapes of Table~\ref{tab:shapeslope}; distinct primitives with the same value
distribution would, however, give the same slope. The slow-limit approach to the failure floor
(Sec.~\ref{S-app:coasting} of the Supplemental Material~\cite{SM}), by contrast, sees only the
\emph{local} geometry of the trough, through the power $p$ with which $k$ leaves its
minimum. Table~\ref{tab:twolimit} sets the two side by side; the cosine and the
square differ in both limits. In the slow limit failure is controlled by the
trough; in the fast limit, by the value distribution of the integrated profile.

\begin{table*}[htb]
\caption{\label{tab:twolimit} The fast and slow limits are independent probes of the
modulation shape. Fast limit: the high-frequency slope $\propto\mathcal{S}(C)$, the
Stieltjes transform of the sojourn density $\rho(\xi)$ of $\xi=\int\!k$
(Table~\ref{tab:shapeslope}). Slow limit: the approach to the failure floor
$\epsilon_f-\epsilon_0\propto\omega^{2p/(p+2)}$, set by the local trough power $p$
for a two-sided trough (Sec.~\ref{S-app:coasting} of the Supplemental Material~\cite{SM}). The bowed exponential ramp
has a one-sided trough, to which the two-sided argument does not apply; its
measured effective exponent at the periods computed is $\approx0.8$.}
\begin{ruledtabular}
\begin{tabular}{lcc}
modulation & fast: $\rho(\xi)\to\mathcal{S}(C)$ & slow: trough $p\to$ exponent \\ \hline
cosine     & arcsine $\to(C^2-a^2)^{-1/2}$          & smooth $p{=}2$: $\omega^{1}$ \\
triangle   & parabolic $\to\arctan{+}\operatorname{arctanh}$ & corner $p{=}1$: $\omega^{2/3}$ \\
exp.\ ramp & shifted reciprocal $\to$ logarithm      & one-sided corner: not derived \\
square     & uniform $\to\operatorname{arctanh}$     & flat $p{\to}\infty$: $\omega^{2}$ \\
\end{tabular}
\end{ruledtabular}
\end{table*}

 \section{CONCLUSION}
We extended our previous analysis of constant-speed traveling waves in a
homogeneous integrate-and-fire network~\cite{PRE2016} to spatially modulated
synaptic coupling, deriving the acceleration--speed relation
$a(x) = -(c-c_1)(c-c_2)/\sigma + Bc\,K(x)$ for a general modulation $K(x)$. For
a piecewise-constant alternating modulation we obtained the periodic speed
profile in closed form and located its failure as the fold at which the stable
and unstable periodic profiles merge. Building on that solvable case, we
analyzed an arbitrary bounded zero-mean periodic modulation: the fold
$\epsilon_{\rm fold}(\omega)$ is fixed in the two limits by two different
functionals of the modulation shape, a trough-controlled slow-limit plateau
$\epsilon_0=(\sqrt{c_2}-\sqrt{c_1})^2/(\sigma B)$ (with $\min k=-1$) and a
profile-controlled high-frequency slope set by the Stieltjes transform of the
sojourn density of $\int\!k$. We worked this out explicitly for the cosine,
triangle, and square (smooth, kinked, and discontinuous standard forms), for the
cosine constructing the small-amplitude series for the speed with its harmonic
support fixed order by order, and deriving the critical amplitude in both
asymptotic regimes. In the slowly varying limit the stable branch is lost through
a saddle-node at the trough at the bottleneck speed $c_\ast=\sqrt{c_1c_2}$,
strictly above the homogeneous threshold $c_1$ (this is the speed at which the
adiabatically tracked branch disappears, not the speed at which the wave finally
arrests, which is lower); in the rapidly varying limit the boundary is linear in
$\omega$. We kept two failure observables apart throughout: the fold, computed by
continuation with a neutral Floquet multiplier, and the threshold at which a
particular launch collapses; they differ by a few percent, in either direction,
and the constant $c_1$ is not a separatrix under forcing. Both predictions agree
with direct numerical integration and with first-spike firing-map simulations of
the network.

Comparing against the weak-ripple truncation of the averaged equation sharpened
the picture. Failure is set by the modulation's \emph{extreme} (the trough, via
a saddle-node) in the slow limit and by its \emph{full profile} in the fast
limit, but in neither by its spatial mean. The truncation overshoots the
high-frequency boundary by a parameter-dependent factor (about $2.5$ at the
default coupling, and negligible near the homogeneous onset $g_{\min}$, where the
critical ripple is small), and, being a short-wavelength theory, does not
address the slowly varying regime, where the threshold is a finite plateau. That
weak-modulation averaging struggles with bistable propagation failure, and that
more careful averaging or interface methods repair it, was known in other
models~\cite{Keener2000, Paul2001, kilpatrick2008, CoombesLaing2011}. The mechanism of
that difference is visible in the reductions themselves: the interface drift of
Ref.~\cite{CoombesLaing2011} carries the modulation convolved with the coupling
kernel, whereas here it enters pointwise. Rate and integrate-and-fire descriptions
of traveling waves are known to agree closely on wave speed~\cite{CremersHerz2002};
the present comparison suggests that the agreement does not extend to the failure
criterion, since in the rate description the front is lost as its speed vanishes,
whereas here, in the slow limit, the local bottleneck is lost at the strictly
positive speed $c_\ast$. Because the exponential-kernel firing map reduces exactly to a
local law for the speed~\cite{PRE2016}, periodic modulation of that law can be
solved order by order through a shape-independent recursion, and the loss of the
periodic wave is a fold of a scalar equation whose slow and fast limits are
explicit: the geometric-mean bottleneck, the profile functional that fixes the fast
slope, and the trough geometry that sets the finite-period corrections. The two descriptions are
complementary rather than redundant. A finite truncation of the series is a polynomial in $\epsilon$ and
displays no fold. In
the quasi-static limit of the cosine the relation is exact: the radius of
convergence of the quasi-static series is the failure amplitude $\epsilon_0$. Two scope statements belong with these
results. For modulation amplitudes $\epsilon\le1$ the coupling stays excitatory,
and this regime, which contains the plateau, the bottleneck law and the rise of
the boundary up to $\omega\approx1.6$, is the neural model proper; the linear
high-frequency law is reached only for $\epsilon>1$, where the modulated coupling
changes sign, and is a prediction of the complete model whose biological
counterpart is hypothesized rather than established (Sec.~\ref{sec:domain}). And
the kernel comparison of Sec.~\ref{S-sec:kernels} of the Supplemental Material~\cite{SM} covers four single-scale kernels;
multiscale kernels can carry more than two candidate speeds and were not analyzed.
With the model parameters known and a well-resolved speed profile, the reduced
law can in principle be inverted,
$k(\omega x)=[\sigma c'+c-(c_1+c_2)+c_1c_2/c]/(\sigma B\epsilon)$; this is a
conditional inverse problem, sensitive to noise through the derivative, not an
established experimental prediction.

Why should this matter beyond the model? Cortical tissue is not uniform: alongside
areas of near-homogeneous architecture such as neocortical layer IV, the visual and
somatosensory cortices carry approximately periodic columnar and barrel
structure~\cite{Rockland2010, AdamsHorton2009, Ermentrout2009}, and propagating activity has
been recorded crossing such tissue~\cite{Muller2018}. Within the model, two
qualitative expectations follow. First, the speed of a wave crossing a periodic
structure is modulated by the local coupling with a lag (the first-order response
carries the phase $\delta_1=\arctan(\sigma\omega/\gamma)$), so a speed profile
constrains the connectivity only through the inverse problem above. Second, in the
slowly varying regime the wave is blocked where the coupling is weakest, and the
local bottleneck is lost at the finite speed $\sqrt{c_1c_2}$ rather than at $c_1$.
Whether these survive repeated spiking, other kernels and calibrated parameters has
not been tested. For modelers, the results show that a description retaining only
the mean connectivity can misjudge the failure threshold substantially.

More broadly, propagation \emph{failure} is the under-explored complement to the
traveling waves now seen to organize cortical
computation~\cite{Muller2018}. In the slow regime of this model a wave is halted
by the \emph{extreme} of the heterogeneity it crosses rather than its average,
which is a caution for coarse-grained models that average over spatial structure.
Three extensions suggest themselves. First, the same reduction can be sought for the exact
mean-field and kinetic-theory descriptions of spiking networks underlying
next-generation neural-mass and neural-field
models~\cite{Montbrio2015, RanganKovacicCai2008, KovacicTaoRanganCai2009}, testing
whether the geometric-mean law survives macroscopic coarse-graining. Second, the local failure criterion of Sec.~\ref{S-sec:kernels} of the Supplemental
Material~\cite{SM} suggests a statistical next step: for slowly varying aperiodic or
\emph{disordered} heterogeneity, failure would be governed by the spatial infimum of
the local coupling, making the threshold an extreme-value problem set by the deepest fluctuation rather than by mean
connectivity; adding dynamical noise then turns the sharp boundary into a
first-passage problem, replacing it with a \emph{distribution} of block locations
and a failure probability, within the stochastic neural-field and Fokker--Planck
frameworks for fluctuation-driven
networks~\cite{Paul2001, kilpatrick2008, NewhallKovacic2010}. Third, because seizure fronts and
spreading depression propagate through heterogeneous cortex, an extreme-controlled
failure law may help locate where such pathological waves stall.

\begin{acknowledgments}
J.N. has previously published as Jie Zhang~\cite{PRE2016}. The early stages of this work were carried out at Georgia State University. We
thank G.~Cymbalyuk and R.~C.~Mure\c{s}an for helpful discussions, and R.~Leca
for his support and for many helpful discussions. R.O. acknowledges the support
and hospitality of the Transylvanian Institute of Neuroscience. This work was
supported in part by Decanex Inc., Toronto, Canada.
\end{acknowledgments}

\noindent\textbf{Author contributions.}
R.O. conceived the study, derived the acceleration--speed relation and its periodically
modulated extension, carried out the perturbative and asymptotic analysis, and wrote the
manuscript. J.N. worked on the derivations of the low-order coefficients and prepared the
first draft. R.E.-T. contributed to the biological framing and the literature on cortical
architecture, and critically revised the manuscript. R.B. and M.O. discussed the general
approach and provided valuable directions and checks. M.C. independently reproduced the
periodic-orbit fold computations, identified the distinction between loss of the periodic
wave and launch failure, and critically revised the manuscript. J.N. and R.E.-T.
contributed equally to this work. All authors read and approved the manuscript.

\clearpage
\onecolumngrid   
\setcounter{section}{0}\setcounter{equation}{0}\setcounter{figure}{0}\setcounter{table}{0}
\renewcommand{\thesection}{S\arabic{section}}
\renewcommand{\theequation}{S\arabic{equation}}
\renewcommand{\thefigure}{S\arabic{figure}}
\renewcommand{\thetable}{S\arabic{table}}
\begin{center}\textbf{\large Supplemental Material}\end{center}

\renewcommand{\thesection}{S\arabic{section}}
\renewcommand{\theequation}{S\arabic{equation}}
\renewcommand{\thefigure}{S\arabic{figure}}
\renewcommand{\thetable}{S\arabic{table}}
\setcounter{section}{0}\setcounter{equation}{0}\setcounter{figure}{0}\setcounter{table}{0}

\noindent This document supports the main text. It contains the reduction of the firing map to the
acceleration--speed relation (Sec.~\ref{S-app:accel-deriv}); additional figures for the alternating
(square-wave) modulation (Sec.~\ref{S-app:sqfigs}); the approach to the slow-limit failure floor
(Sec.~\ref{S-app:coasting}); the explicit low-order perturbation coefficients
(Sec.~\ref{S-app:pert}); harmonic balance beyond one harmonic (Sec.~\ref{S-app:hb}); the triangle-wave
modulation (Sec.~\ref{S-app:triangle}); modulation families with elementary fast-limit transforms
(Sec.~\ref{S-app:expfamily}); propagation/failure phase diagrams for the cosine modulation
(Sec.~\ref{S-app:phase}); and a comparison across coupling kernels (Sec.~\ref{S-sec:kernels}). Equation,
figure and table numbers carry an ``S'' prefix; references to unprefixed numbers are to the main text.

\section{Derivation of the acceleration--speed relation}\label{S-app:accel-deriv}
Here we carry out in full the reduction of the firing map Eq.~\ref{eq:main} to the
acceleration--speed relation Eq.~\ref{eq:simple}. The steps follow the homogeneous
analysis of Ref.~\cite{PRE2016}; the only change is that the inhomogeneity enters
through the factor $1+K(y)$ in the coupling. Let
\begin{equation}\label{S-eq:PQdef}
P(x) = e^{-x/\sigma},\quad \alpha_i(x) = e^{-t/\tau_i},\quad Q_i(x) = \int_{-\infty}^{x}e^{y/\sigma}e^{t^*(y)/\tau_i}(1+K(y))\,dy,\quad i = 1, 2.
\end{equation}
Thus, Eq.~\ref{eq:main} becomes
\begin{equation} \label{S-eq:0d}
\frac{2\sigma(1-\frac{\tau_1}{\tau_2})V_T}{g_{syn}} = (P\alpha_2Q_2 - P\alpha_1Q_1)(x).
\end{equation}
The first derivative of Eq.~\ref{eq:main} or Eq.~\ref{S-eq:0d} with respect to $x$ is
\begin{equation} \label{S-eq:1d}
0 = -P\alpha_2Q_2(\frac{1}{\sigma}+\frac{t'(x)}{\tau_2}) + P\alpha_1Q_1(\frac{1}{\sigma}+\frac{t'(x)}{\tau_1}),
\end{equation}
since,
\begin{align}\label{S-eq:Qprime}
Q_i'(x) &= \Big(\int_{-\infty}^{x}e^{y/\sigma}e^{t^*_y/\tau_i}(1+K(y))\,dy\Big)' = e^{x/\sigma}e^{t/\tau_i}(1+K(x)),\nonumber\\
&\Rightarrow\quad P(x)\alpha_i(x)Q_i'(x) = 1 + K(x).
\end{align}
By Eq.(\ref{S-eq:0d}) and Eq.(\ref{S-eq:1d}), we can solve for $P\alpha_1Q_1$ and $P\alpha_2Q_2$, which are the same as in the homogeneous network,
 \begin{align}
 P\alpha_1Q_1 =& \frac{V_T}{g_{syn}}2\sigma(1-\frac{\tau_1}{\tau_2})\frac{\frac{1}{\sigma}+\frac{1}{c\tau_2}}{\frac{1}{c}(\frac{1}{\tau_1}-\frac{1}{\tau_2})}, \\
 P\alpha_2Q_2 =& \frac{V_T}{g_{syn}}2\sigma(1-\frac{\tau_1}{\tau_2})\frac{\frac{1}{\sigma}+\frac{1}{c\tau_1}}{\frac{1}{c}(\frac{1}{\tau_1}-\frac{1}{\tau_2})}.
 \end{align}

Then we take the second derivative of Eq.~\ref{S-eq:0d} and obtain the following equation containing $t'$ and $t''$,
 \begin{align}\label{S-eq:2d}
0 = &P\alpha_2Q_2((\frac{1}{\sigma}+\frac{t'}{\tau_2})^2-\frac{t''}{\tau_2}) \\ \nonumber
& -P\alpha_1Q_1((\frac{1}{\sigma}+\frac{t'}{\tau_1})^2-\frac{t''}{\tau_1})  + (1+K(x))(\frac{t'}{\tau_1}-\frac{t'}{\tau_2}),
 \end{align}
where the boundary term arises from differentiating $P\alpha_iQ_i'=1+K(x)$ in
Eq.~(\ref{S-eq:Qprime}); the factor $1+K$ (not $K$ alone) is essential, since at $K=0$
it supplies the homogeneous relation. Using the results for $P\alpha_1Q_1$ and
$P\alpha_2Q_2$, and the speed/acceleration identities $c=1/t'$, $a=-c^3t''$
(the definition of $a(x)$ in Sec.~\ref{sec:model}), Eq.~(\ref{S-eq:2d}) rearranges into the acceleration--speed
relation Eq.~\ref{eq:simple} of the main text, with the homogeneous speeds $c_1,c_2$
as given there. A short check: writing $R_i=P\alpha_iQ_i$, the threshold and
first-derivative relations give, at the default parameters, $R_1=0.2c+0.1$ and
$R_2=0.2c+0.2$; substituting into $R_1'=-(1+1/c)R_1+(1+K)$ yields
$c'=3.5-c-0.5/c+5K$, which is Eq.~\ref{eq:cx-gen} with $c_1+c_2=3.5$, $c_1c_2=0.5$,
$B=5$. The reduction assumes ordered first firing ($t^*$ increasing in $x$), a
compatible prehistory, and positive speed.

\section{Additional figures for the alternating (square-wave) modulation}\label{S-app:sqfigs}
These figures complement Sec.~\ref{sec:const} of the main text.

\begin{figure}[htb]
 \centering
  \includegraphics[width=0.72\textwidth]{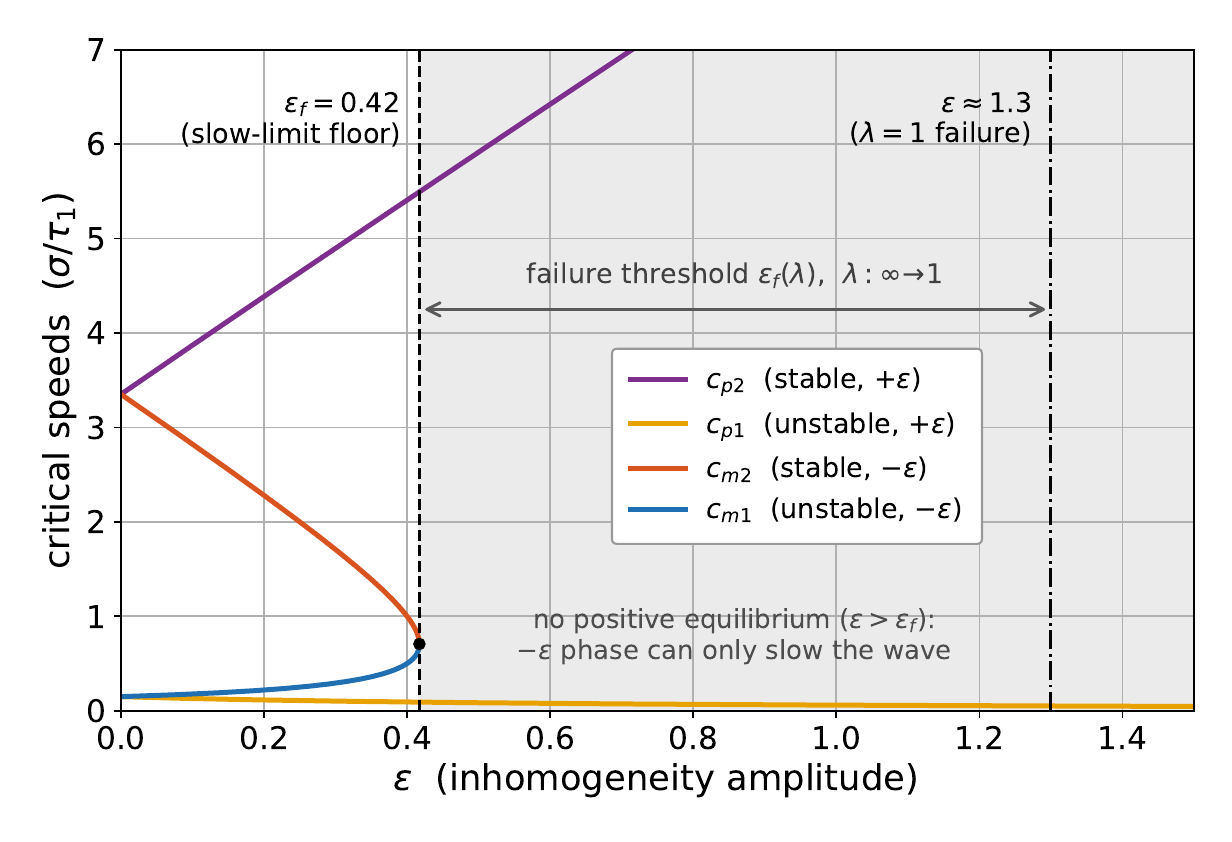}
  \caption{\label{S-fig:3} {\bf Per-phase equilibrium speeds vs.\ inhomogeneity
  amplitude $\epsilon$ (Eqs.~\ref{eq:cp12}--\ref{eq:cm12}).} Purple: $c_{p2}$
  (stable, $+\epsilon$); yellow: $c_{p1}$ (unstable, $+\epsilon$); red: $c_{m2}$
  (stable, $-\epsilon$); blue: $c_{m1}$ (unstable, $-\epsilon$). Shaded region
  ($\epsilon>\epsilon_f=0.42$): the negative phase has \emph{no positive
  equilibrium}, so the speed there can only decrease. (The roots $c_{m1},c_{m2}$
  are complex for $0.42<\epsilon<0.98$ and real but negative for $\epsilon>0.98$;
  this complex$\to$real switch is purely algebraic (a negative equilibrium speed
  is unphysical and unreachable) and has no dynamical consequence.) The dashed
  line at $\epsilon_f=0.42$ is the slow-limit
  ($\lambda\to\infty$) saddle-node floor, where $c_{m1},c_{m2}$ first merge;
  $\epsilon_f=(\sqrt{c_2}-\sqrt{c_1})^2/(\sigma B)$. The dash-dotted line at
  $\epsilon\approx1.3$ is the actual failure threshold for $\lambda=1$ (cf.\
  Fig.~\ref{fig:2}). Crucially, $\epsilon_f=0.42$ is \emph{not} where the
  $\lambda=1$ wave fails: a finite period postpones failure to $\epsilon\approx
  1.3$, and the failure $\epsilon$ decreases toward the $0.42$ floor as $\lambda$
  grows.
  }
\end{figure}

Figure~\ref{S-fig:3} shows how these equilibria move with $\epsilon$: $c_{p1}$ is
the lowest and $c_{p2}$ the highest. The disappearance of $c_{m1},c_{m2}$ at
$\epsilon_f=0.42$ does \emph{not}, by itself, cause failure. It is the slow-limit
threshold: for every $\epsilon>\epsilon_f$ the negative phase has no positive
equilibrium (the roots are complex up to $\epsilon=0.98$ and real but negative
beyond, never positive), so the speed there falls monotonically, and an
infinitely long phase ($\lambda\to\infty$) would fail at $\epsilon_f$. For a finite period the negative phase is too short to complete
the collapse, and the following positive phase restores the speed, provided it
has not dropped below $c_{p1}$, the level below which the positive phase can no
longer pull it back up (a necessary condition; the actual basin boundary is the
unstable periodic profile). This sustains propagation well past $\epsilon_f$: for
$\lambda=1$ the wave survives until $\epsilon\approx1.3$. Failure therefore
depends on both amplitude and period, and the failure threshold decreases
monotonically toward the $\epsilon_f=0.42$ floor as $\lambda$ increases.

\begin{figure}[htb]
 \centering
  \includegraphics[width=0.72\textwidth]{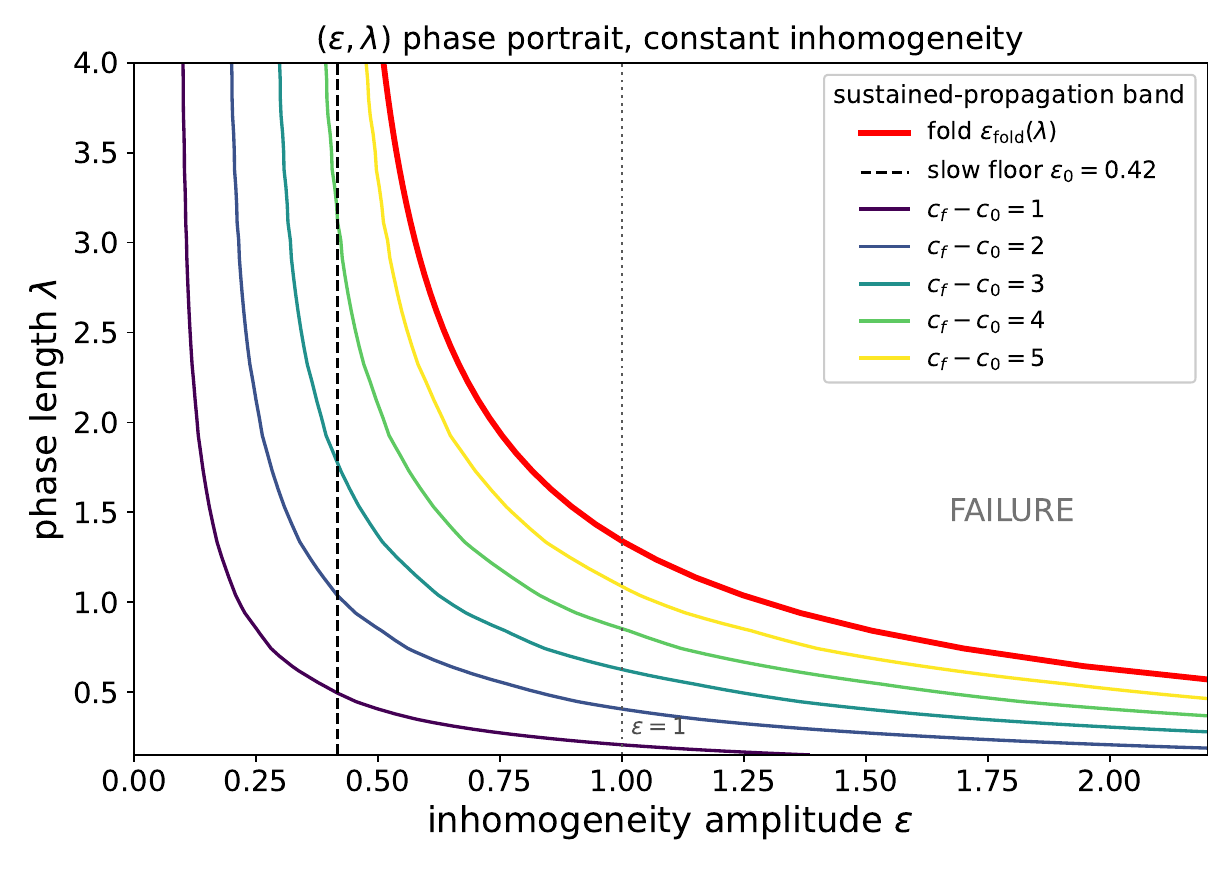}
  \caption{\label{S-fig:5} {\bf Propagation/failure phase portrait in the
  $(\epsilon,\lambda)$ plane (constant inhomogeneity).} Colored curves are contours
  of the periodic speed-band width $c_f-c_0$ (legend; settled band after $60$ periods
  of direct integration from $c_2$); the red curve is the fold $\epsilon_{\rm fold}(\lambda)$, the largest amplitude admitting a periodic wave,
  from the period-closure conditions (Eqs.~\ref{eq:lambda-p}--\ref{eq:lambda-m}); the
  dashed line is the slow-limit floor $\epsilon_0=0.42$ and the dotted line marks
  $\epsilon=1$ (Sec.~\ref{sec:domain}). Below the boundary the wave
  propagates with an oscillation band $[c_0,c_f]$ that widens as either $\epsilon$ or
  $\lambda$ grows; above it the wave fails. Smaller $\lambda$ tolerates much larger
  $\epsilon$ (the wave averages over the rapid switching), while the boundary descends
  to the floor $\epsilon_0$ as $\lambda\to\infty$.}
\end{figure}

Figure~\ref{S-fig:5} consolidates the speed-band picture across the whole
$(\epsilon,\lambda)$ plane. At small $\epsilon$ or $\lambda$ the band $c_f-c_0$ is
narrow and the speed stays near the homogeneous $c_2$; increasing either widens it,
and the wave fails once the periodic orbit can no longer close, along the red
boundary $\epsilon_f(\lambda)$. That boundary flares to large $\epsilon$ at small
$\lambda$ and descends to the floor $\epsilon_0=0.42$ as $\lambda\to\infty$,
consistent with the $\epsilon_f(\lambda)$ curve of Fig.~\ref{fig:epsf-lambda} and the
$(c_0,c_f)$ trace of Fig.~\ref{fig:4}.

\begin{figure}[htb]
 \centering
  \includegraphics[width=0.72\textwidth]{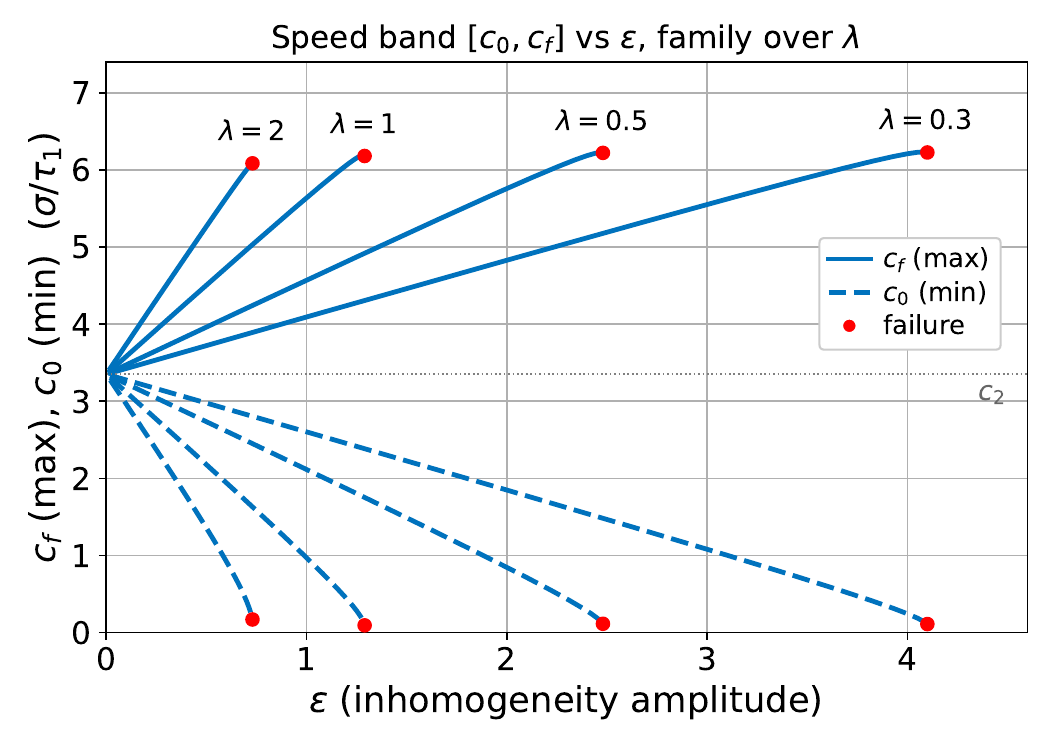}
  \caption{\label{S-fig:c0cf-fam} {\bf Speed band $[c_0,c_f]$ vs $\epsilon$, family
  over the phase length $\lambda$.} Solid: maximum speed $c_f$; dashed: minimum speed
  $c_0$. Each band opens from the homogeneous speed $c_2$ at $\epsilon=0$ and is
  continued in steps $\Delta\epsilon=0.01$; the red dot is the last amplitude at
  which the period-closure conditions converge, within $0.01$ below the fold
  $\epsilon_{\rm fold}(\lambda)$ of Fig.~\ref{fig:epsf-lambda}. The four bands are, left to
  right by failure amplitude, $\lambda = 2, 1, 0.5, 0.3$; smaller $\lambda$
  tolerates larger $\epsilon$ because the wave averages over the rapid sign changes
  of $K$, the short-wavelength regime in which homogenization is invoked (and where
  its failure boundary is the linear-in-$\omega$ law of Sec.~\ref{sec:r2}).}
\end{figure}

Figure~\ref{S-fig:c0cf-fam} consolidates this speed-band
picture across the whole $(\epsilon,\lambda)$ family. Each band opens from the
homogeneous speed $c_2$ at $\epsilon=0$; the upper branch $c_f$ and lower branch
$c_0$ separate as $\epsilon$ grows, and each band terminates (red marker) at the
failure boundary $\epsilon_f(\lambda)$. Smaller $\lambda$ (faster modulation)
sustains the wave to much larger amplitudes, while larger $\lambda$ fails just
above the floor, the same trend quantified in Fig.~\ref{fig:epsf-lambda}.

\section{Derivation of the approach to the slow-limit failure floor}\label{S-app:coasting}
This appendix spells out the short coasting argument summarized in
Sec.~\ref{sec:const} [Eqs.~(\ref{eq:coast})--(\ref{eq:floor-approach})], which fixes
how the failure amplitude $\epsilon_f(\lambda)$ relaxes onto its slow-limit floor
$\epsilon_0=(\sqrt{c_2}-\sqrt{c_1})^2/(\sigma B)$ as the period $\lambda$ grows.

\emph{Down-phase equilibria and their merging.} In the inhibitory ($K=-\epsilon$)
half-period the speed obeys $\sigma\,dc/dx = N_-(c)$ with
\begin{equation}\label{S-eq:Nminus}
N_-(c) = -\frac{(c-c_1)(c-c_2)}{c}-\sigma B\epsilon
       = -c+(c_1+c_2)-\frac{c_1c_2}{c}-\sigma B\epsilon .
\end{equation}
Its fixed points are the roots of $c^2-(c_1+c_2-\sigma B\epsilon)c+c_1c_2=0$,
\begin{equation}\label{S-eq:cmroots}
c_{m1,2}=\tfrac12\Big[(c_1+c_2-\sigma B\epsilon)\mp
\sqrt{(c_1+c_2-\sigma B\epsilon)^2-4c_1c_2}\,\Big],
\end{equation}
real and positive only while $c_1+c_2-\sigma B\epsilon\ge2\sqrt{c_1c_2}$. They collide
when the discriminant vanishes, i.e.\ when
$\sigma B\epsilon=c_1+c_2-2\sqrt{c_1c_2}=(\sqrt{c_2}-\sqrt{c_1})^2$, that is at exactly
$\epsilon=\epsilon_0$; the merged equilibrium then sits at $c_*=\sqrt{c_1c_2}$.

\emph{Normal form just above threshold.} For $\epsilon$ slightly above $\epsilon_0$ the
down-phase has no positive fixed point. Expanding $N_-$ about $c_*=\sqrt{c_1c_2}$ with
$\varphi=c-c_*$,
\begin{align}
N_-(c_*) &= (c_1+c_2)-2\sqrt{c_1c_2}-\sigma B\epsilon=-\sigma B(\epsilon-\epsilon_0)\equiv-D,\nonumber\\
N_-'(c_*) &= -1+\frac{c_1c_2}{c_*^2}=0,\qquad
N_-''(c_*)=-\frac{2c_1c_2}{c_*^3}=-\frac{2}{\sqrt{c_1c_2}},
\end{align}
so to leading order $\sigma\,\varphi'=N_-\approx-D-\varphi^2/\sqrt{c_1c_2}$, which is
Eq.~(\ref{eq:coast}): the vanished fixed point leaves a constant downward drift $-D$
plus a quadratic restoring term, the generic saddle-node normal form.

\emph{Coasting length.} Equation~(\ref{eq:coast}) is separable. With $q\equiv\sqrt{c_1c_2}$,
\begin{equation}
\sigma\frac{d\varphi}{\varphi^2/q+D}=-dx
\quad\Longrightarrow\quad
\sigma\sqrt{\tfrac{q}{D}}\,\arctan\!\frac{\varphi}{\sqrt{qD}}=-x+\text{const}.
\end{equation}
The speed enters the down-phase at $\varphi_0=c_0^{(+)}-c_*>0$ (the periodic maximum
reached in the preceding excitatory phase) and propagation fails when the speed
vanishes, $c\to0$, i.e.\ $\varphi\to-c_*=-\sqrt{c_1c_2}$. Integrating between these
endpoints,
\begin{equation}\label{S-eq:Lcoast-full}
L_{\rm coast}=\sigma\sqrt{\tfrac{q}{D}}\Big[\arctan\frac{\varphi_0}{\sqrt{qD}}
+\arctan\frac{c_*}{\sqrt{qD}}\Big].
\end{equation}
For $\epsilon$ near threshold the bottleneck is narrow, $\sqrt{qD}\ll\sqrt{c_1c_2}=c_*$,
so the failure endpoint sits deep in the tail of the arctangent,
$\arctan(c_*/\sqrt{qD})\to\tfrac{\pi}{2}$ (the rapid final plunge from the bottleneck
to $c=0$ adds only an $O(\sigma)$ length, subdominant to $L_{\rm coast}\sim D^{-1/2}$).
Thus
\begin{equation}\label{S-eq:Lcoast-asym}
L_{\rm coast}\simeq\sigma\sqrt{\tfrac{q}{D}}\Big[\tfrac{\pi}{2}+\arctan\frac{\varphi_0}{\sqrt{qD}}\Big]
=\frac{\pi\sigma\,(c_1c_2)^{1/4}}{\sqrt{D}}\times
\begin{cases}1 & \varphi_0\to\infty\ (\text{entry from }c\gg c_*),\\[2pt]
\tfrac12 & \varphi_0=0\ (\text{entry at the merge}).\end{cases}
\end{equation}
The prefactor is therefore $O(1)$, set by the entry speed $\varphi_0$ through the upper
limit of the arctangent; for the periodic orbit $\varphi_0$ is itself an $O(1)$ function
of $\epsilon$ near the fold, which is why the constant is left unspecified in the main text.

\emph{Approach to the floor.} The wave survives the down-phase iff it is shorter than
the coasting length, $\lambda\lesssim L_{\rm coast}$, with failure first occurring at
$\lambda=L_{\rm coast}$. Along the failure boundary $D=\sigma B(\epsilon_f-\epsilon_0)\to0$
as $\lambda\to\infty$, so $\sqrt{qD}\to0$ and both arctangents in
Eq.~(\ref{S-eq:Lcoast-asym}) saturate at $\pi/2$ (the down-swing is entered from a periodic
maximum well above $c_*$ and ends at $c=0$); hence $L_{\rm coast}\to\pi\sigma\sqrt{q/D}$.
Setting $\lambda=L_{\rm coast}$ and using $D=\sigma B(\epsilon_f-\epsilon_0)$ gives
\begin{equation}
\epsilon_f-\epsilon_0\;\simeq\;\pi^2\,\frac{\sigma\sqrt{c_1c_2}}{B\,\lambda^2},
\end{equation}
i.e.\ Eq.~(\ref{eq:floor-approach}) with $\kappa=\pi^2\approx9.9$. This normal-form value
agrees to within ${\sim}10\%$ with the prefactor extracted from the exact period-closure
boundary (Fig.~\ref{fig:epsf-lambda}), $\kappa\approx10.5$; the small excess comes from the
portion of the down-swing that lies outside the normal-form bottleneck. The cosine
blend $\epsilon_f\simeq\sqrt{\epsilon_0^2+(m_\infty\omega)^2}$ (Eq.~\ref{eq:blend}),
being even in $\omega$, expands at small $\omega=2\pi/\lambda$ as
$\epsilon_f-\epsilon_0\simeq(m_\infty^2/2\epsilon_0)\omega^2$; this quadratic form
matches the flat-trough (alternating) modulation of this section, but it does
\emph{not} describe a smoothly modulated coupling, whose floor is approached
\emph{linearly} in $\omega$, as we now show.

\begin{figure}[htb]
\centering
\includegraphics[width=0.9\textwidth]{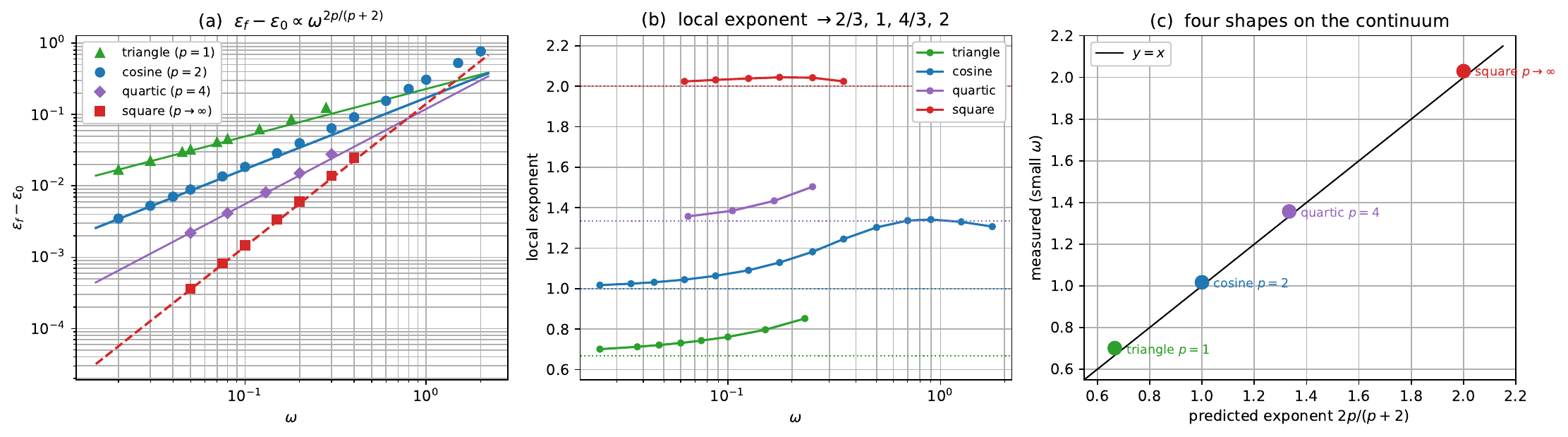}
\caption{\label{S-fig:floor-approach} {\bf The slow-limit floor $\epsilon_0$ is
approached at a rate set by the trough geometry,
$\epsilon_f-\epsilon_0\propto\omega^{2p/(p+2)}$.} Markers: failure thresholds from
direct integration of the speed flow
$\sigma c\,c'=-(c-c_1)(c-c_2)+\sigma B c\,\epsilon k(\omega x)$. {\bf (a)} \emph{Cosine}
(smooth, $p=2$): linear approach,
$\epsilon_f-\epsilon_0=\sqrt{\sigma\sqrt{c_1c_2}\,\epsilon_0/2B}\,\omega$ (Weber
ground state $\nu=1$, Eq.~\ref{S-eq:weber-linear}, parameter-free solid line).
\emph{Triangle} (corner, $p=1$): $\omega^{2/3}$, Airy (slope guide). \emph{Quartic}
(tuned two-harmonic trough $\propto4\cos\omega x+\cos2\omega x$, $p=4$): $\omega^{4/3}$.
\emph{Alternating} $\pm\epsilon$ (flat, $p\to\infty$): $\omega^2$,
$\epsilon_f-\epsilon_0=(\sigma\sqrt{c_1c_2}/B)\,\omega^2$ (coasting,
Eq.~\ref{eq:floor-approach}, parameter-free dashed line). {\bf (b)} The local exponent
$d\ln(\epsilon_f-\epsilon_0)/d\ln\omega$ converges to $2/3,1,4/3,2$ as $\omega\to0$.
{\bf (c)} The four shapes fall on the predicted continuum $2p/(p+2)$ ($y=x$ line);
points lie slightly above at the finite $\omega$ reached, descending onto the line as
$\omega\to0$. Defaults $\tau_1=1,\tau_2=2,\sigma=1,V_T=1,g_{syn}=10$.}
\end{figure}

\emph{Smooth troughs: a linear approach and the Weber reduction.} The $\lambda^{-2}$
law above relies on the drift $D$ being \emph{constant} through the down-phase, which
holds only for a piecewise-constant (flat-trough) profile. For a smoothly modulated
coupling the trough has nonzero curvature, and the approach to the floor changes
character. Near a smooth minimum of $K$ at $x_*$, expand
$K(x)\simeq K(x_*)+\tfrac12 K''(x_*)(x-x_*)^2$; for $K=\epsilon\cos\omega x$ the
trough is at $x_*=\pi/\omega$ with $K(x_*)=-\epsilon$ and $K''(x_*)=\epsilon\omega^2$.
The saddle-node normal form (\ref{eq:coast}) then carries a parabolic,
position-dependent drift,
\begin{equation}\label{S-eq:coast-parabolic}
\sigma\varphi' = -\frac{\varphi^2}{\sqrt{c_1c_2}} - D_0
 + \tfrac12\,\sigma B\epsilon\,\omega^2\,(x-x_*)^2,
\qquad D_0=\sigma B(\epsilon-\epsilon_0).
\end{equation}
The Riccati substitution $\varphi=\sigma\sqrt{c_1c_2}\,\psi'/\psi$ linearizes this
\emph{exactly} to the Weber (parabolic-cylinder) equation
\begin{equation}\label{S-eq:weber}
\psi'' = \big[a\,(x-x_*)^2 - b\big]\psi,\qquad
a=\frac{B\epsilon\,\omega^2}{2\sigma\sqrt{c_1c_2}},\quad
b=\frac{B(\epsilon-\epsilon_0)}{\sigma\sqrt{c_1c_2}}.
\end{equation}
Failure, the front decelerating to $c\to0$, corresponds to $\psi$ developing a
node, which first occurs when the single dimensionless group $\nu\equiv b/\sqrt{a}$
reaches the ground-state value $\nu=1$. Since
$\nu=(\epsilon-\epsilon_0)\,\omega^{-1}\sqrt{2B/(\sigma\sqrt{c_1c_2}\,\epsilon)}$,
setting $\nu=1$ (with $\epsilon\to\epsilon_0$ on the boundary) gives
\begin{equation}\label{S-eq:weber-linear}
\epsilon_f-\epsilon_0\ \simeq\ \sqrt{\frac{\sigma\sqrt{c_1c_2}\,\epsilon_0}{2B}}\;\omega
\ =\ \frac{(\sqrt{c_2}-\sqrt{c_1})\,(c_1c_2)^{1/4}}{\sqrt{2}\,B}\;\omega ,
\end{equation}
linear in $\omega$ (i.e.\ in $\lambda^{-1}$), not quadratic. Direct integration of the
cosine speed flow confirms both the exponent (the local slope
$d\ln(\epsilon_f-\epsilon_0)/d\ln\omega\to1.00$ as $\omega\to0$) and the
parameter-free prefactor: the fitted marginal group tends to $\nu_c\to1.00$ within
$0.5\%$ (Fig.~\ref{S-fig:floor-approach}). The two cases are therefore genuinely
distinct: a flat trough gives $\epsilon_f-\epsilon_0\propto\lambda^{-2}$, a smooth
trough $\propto\lambda^{-1}$, the exponent being decided by whether the trough
curvature $K''(x_*)$ vanishes. Because $a\propto K''(x_*)$, the finite-period
correction is governed entirely by the local curvature of the extreme, again a
property of the trough, not of any spatial average.

\emph{General trough geometry.} The Weber ($p=2$) and coasting ($p\to\infty$) cases
are two members of a one-parameter family. If the trough behaves as
$K(x)\simeq-\epsilon+\kappa\,|x-x_*|^{p}$ near a two-sided minimum, the same Riccati
linearization produces a linear equation with a $|x|^{p}$ potential, and the marginal
(ground-state) balance gives, formally,
\begin{equation}\label{S-eq:pexponent}
\epsilon_f-\epsilon_0\ \propto\ \omega^{\,2p/(p+2)} .
\end{equation}
A two-sided \emph{corner} trough ($p=1$, a triangle wave, whose branches meet the
extreme with nonzero slope) linearizes to the Airy equation and
approaches the floor as $\omega^{2/3}$; the smooth ($p=2$) and flat ($p\to\infty$)
cases recover the $\omega^{1}$ and $\omega^{2}$ laws above. The exponents measured
by direct integration are consistent with all three (Fig.~\ref{S-fig:floor-approach}). Two points are worth
noting. The exponent is fixed entirely by the local shape of the extreme (its
curvature $K''(x_*)$ or the lack of it), so, like $\epsilon_0$ itself, it is a
property of the trough and invisible to any spatial average. The bowed exponential profile of Sec.~\ref{S-app:expfamily} is different: it reaches its
infimum immediately before a jump, so its trough is one-sided, with a matching
condition at the switching point, and the two-sided argument above does not transfer
to it. Its measured effective exponent is $\approx0.8$ at the periods computed; a
separate derivation would be needed to fix its asymptotic value.

\emph{A continuum, not three points.} The exponent $2p/(p+2)$ is not confined to the
landmark shapes. Ordinary smooth modulations realize the \emph{intermediate} values
whenever the trough curvature is tuned to vanish: the two-harmonic profile
$K\propto4\cos\omega x+\cos2\omega x$ has $K''(x_*)=0$ but $K''''(x_*)>0$, a quartic
minimum ($p=4$), and approaches the floor as $\omega^{4/3}$, confirmed by direct
integration (Fig.~\ref{S-fig:floor-approach}c), a nontrivial exponent that is neither
$1$ nor $2$. Sharper-than-corner troughs ($p<1$; a cusp $|x-x_*|^{1/2}$ gives
$\omega^{2/5}$) fill the range below $2/3$. (These are fixed-$p$, $\omega\to0$
statements; the limit $p\to0$ would need a separate analysis.)

\section{Explicit low-order perturbation coefficients}\label{S-app:pert}
Substituting the expansion Eq.~\ref{eq:c_approx} into Eq.~\ref{eq:cx} uses the
Taylor expansion of $1/c$,
\begin{eqnarray}
\frac{1}{c(x)} = &\frac{1}{c_2} - \frac{h_1}{c_2^2}\epsilon +
(\frac{h_1^2}{c_2^3}-\frac{h_2}{c_2^2})\epsilon^2 - \nonumber\\
& (\frac{1}{c_2^4}h_1^3-\frac{2}{c_2^3}h_1h_2+ \frac{1}{c_2^2}h_3)\epsilon^3 + R_3(\epsilon),
\end{eqnarray}
with $R_3(\epsilon)\to0$ as $\epsilon\to0$. Collecting powers of $\epsilon$ in
Eq.~\ref{eq:cx} gives the order-by-order linear equations (with
$\lambda_0=-\gamma=-(c_2-c_1)/c_2$)
\begin{align}
\sigma h_1'(x) &= \lambda_0 h_1(x) + \sigma B \cos(\omega x), \\
\sigma h_2'(x) &= \lambda_0 h_2(x) - \frac{c_1}{c_2^2}h_1^2, \\
\sigma h_3'(x) &= \lambda_0 h_3(x) + \frac{c_1}{c_2^3}h_1^3-\frac{2c_1}{c_2^2}h_1h_2,
\end{align}
whose solutions are
\begin{align}
h_1(x) &= A_1\cos(\omega x +\phi_1),\\
h_2(x) &= A_2\cos(2\omega x+\phi_2)+C_{1},\\
h_3(x) &= A_3\cos(\omega x+\phi_3)+A_{4}\cos(3\omega x+\phi_{4}),
\end{align}
with (writing $\gamma=(c_2-c_1)/c_2$ and $\delta_k=\arctan(k\sigma\omega/\gamma)$)
\begin{align}
A_1 &= \frac{\sigma B}{\sqrt{\gamma^2+(\sigma\omega)^2}}, & \phi_1 &= -\delta_1, \\
A_2 &= -\frac{c_1 A_1^2}{2c_2^2\sqrt{\gamma^2+(2\sigma\omega)^2}}, & \phi_2 &= 2\phi_1-\delta_2, \\
C_1 &= -\frac{c_1 A_1^2}{2c_2^2\,\gamma}, & B &= \frac{g}{2V_t\tau_1}.
\end{align}
The third order applies the same first-order response to the $\omega$- and
$3\omega$-harmonics of its forcing $(c_1/c_2^3)h_1^3-(2c_1/c_2^2)h_1h_2$. With the
forcing amplitudes
\begin{equation}
m_1=\frac{3c_1A_1^3}{4c_2^3}-\frac{2c_1A_1C_1}{c_2^2},\quad
m_3=\frac{c_1A_1^3}{4c_2^3},\quad m_2=m_4=-\frac{c_1A_1A_2}{c_2^2},
\end{equation}
and the phasor sums $\mu_\omega=m_1e^{i\phi_1}+m_2e^{i(\phi_2-\phi_1)}$ and
$\mu_{3\omega}=m_3e^{3i\phi_1}+m_4e^{i(\phi_1+\phi_2)}$,
\begin{equation}
A_3=\frac{|\mu_\omega|}{\sqrt{\gamma^2+(\sigma\omega)^2}},\ \ \phi_3=\arg\mu_\omega-\delta_1,
\qquad
A_4=\frac{|\mu_{3\omega}|}{\sqrt{\gamma^2+(3\sigma\omega)^2}},\ \ \phi_4=\arg\mu_{3\omega}-\delta_3.
\end{equation}
For the default parameters this gives $A_3=3.40\times10^{-3}$, $\phi_3=-2.555$,
$A_4=3.70\times10^{-4}$, $\phi_4=0.973$ (with $\omega=\pi$).

\medskip
\noindent\emph{Harmonic support at every order.}
The low-order solutions show $h_1$ carrying the frequency $\omega$, $h_2$ the frequency
$2\omega$ plus a constant, and $h_3$ the frequencies $\omega$ and $3\omega$. The general
statement is that $h_n$ has Fourier support on the frequencies $k\omega$ with $|k|\le n$
and $k\equiv n\pmod 2$, and it follows from two facts. First, the order-$n$ equation has
the form $\sigma h_n'+\gamma h_n=F_n$, where $F_n$ is the coefficient of $\epsilon^n$ in
the expansion of $-c_1c_2/c$ about $c_2$ (with the forcing $\sigma B\cos\omega x$ at
$n=1$), i.e.\ a finite sum over all integer partitions $n=i_1+\dots+i_m$ ($m\ge2$) of
monomials $h_{i_1}\cdots h_{i_m}$ with constant coefficients (Eq.~\ref{eq:hngen}); all
partitions occur, not only products of two distinct factors (at order six, for instance,
$h_1h_2h_3$ contributes). Second, the periodic inverse of $\sigma\,d/dx+\gamma$ is
diagonal in the Fourier basis, multiplying the mode $e^{ik\omega x}$ by
$1/(\gamma+ik\sigma\omega)$, so it preserves Fourier support (this is where the
attenuation $1/\sqrt{\gamma^2+(k\sigma\omega)^2}$ and lag $\delta_k$ of each harmonic
come from). Now induct: $h_1$ has support $\{\pm1\}$. If each $h_i$ with $i<n$ has
support in $\{k:|k|\le i,\ k\equiv i\ (\mathrm{mod}\ 2)\}$, then a product
$h_{i_1}\cdots h_{i_m}$ has support in the sumset, whose elements satisfy
$|k|\le\sum i_j=n$ and $k\equiv\sum i_j=n\pmod 2$; the sum $F_n$ and the inverse
operator preserve this set, so $h_n$ has the stated support. Phases are carried along
automatically because the argument is made on complex exponentials rather than on
cosines. Consequently even orders contain only even harmonics (including a constant
mean shift) and odd orders only odd harmonics, each with its own amplitude and phase,
\begin{align}
h_{2n} &= \sum_{j=0}^n A_{2nj}\cos(2j\omega x+\phi_{2nj}), \\
h_{2n+1} &= \sum_{j=0}^n A_{(2n+1)j}\cos((2j+1)\omega x+\phi_{(2n+1)j}).
\end{align}
This is a statement about the finite-order structure of the series only; it says
nothing about its convergence in $\epsilon$, which is discussed in
Sec.~\ref{sec:approx}.

\begin{figure}[htb]
 \centering
  \includegraphics[width=0.96\textwidth]{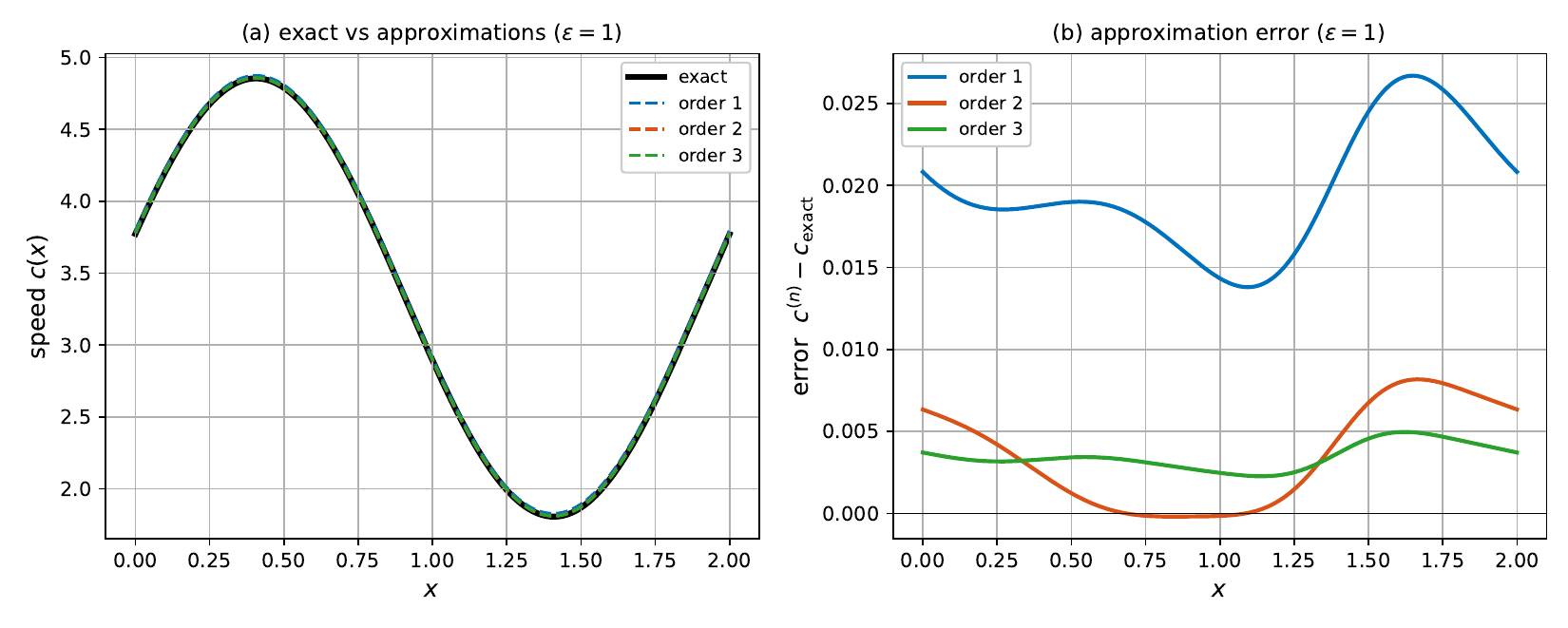}
  \caption{\label{S-fig:7} {\bf Speed approximations and their error ($\epsilon=1$,
  $\omega=\pi$).} {\bf (a)} The exact periodic speed $c(x)$ (black) over one
  period; the first-, second-, and third-order approximations
  $c^{(n)}=c_2+\epsilon h_1(+\epsilon^2 h_2)(+\epsilon^3 h_3)$ (dashed) lie
  essentially on top of it; the improvement with order is invisible here.
  {\bf (b)} The error $c^{(n)}-c_{\rm exact}$ makes it visible: it shrinks with
  each order, with maximum $2.7\times10^{-2}$, $8.2\times10^{-3}$,
  $5.0\times10^{-3}$ for orders one, two, three.}
\end{figure}

Even at $\epsilon=1$ the three approximations are visually indistinguishable from
the exact speed (Fig.~\ref{S-fig:7}(a)); the improvement with order shows up only in
the error (Fig.~\ref{S-fig:7}(b)), whose maximum over one period falls
monotonically: $2.7\times10^{-2}$, $8.2\times10^{-3}$, $5.0\times10^{-3}$ for
orders one, two, and three at the default parameters (default parameter set, Sec.~\ref{sec:model}) and
$\omega=\pi$.

\section{Harmonic balance: modes beyond $K=1$, relation to the perturbation series, and accuracy}\label{S-app:hb}
This section complements Sec.~\ref{sec:hb} of the main text, whose notation
(Eqs.~\ref{eq:poly}--\ref{eq:hbfold}) it uses.

Why $K=1$ closes, and what changes beyond it, is visible in the general
projection. Writing the truncated series in complex form,
$c=\sum_{|k|\le K}\hat c_k e^{ik\omega x}$ with $\hat c_0=\bar c$ and
$\hat c_k=(a_k-ib_k)/2$, Eq.~\ref{eq:poly} projects onto mode $n$ as
\begin{align}
&\Big(1+\tfrac{i n\omega\sigma}{2}\Big)\sum_m \hat c_m\hat c_{n-m}
 -(c_1+c_2)\,\hat c_n+c_1c_2\,\delta_{n0} \nonumber\\
&\qquad\qquad =\tfrac{\sigma B\epsilon}{2}\big(\hat c_{n-1}+\hat c_{n+1}\big),
\label{S-eq:hbmode}
\end{align}
with $\hat c_{-k}=\hat c_k^{\,*}$ and $\hat c_k=0$ for $|k|>K$. The linear
operator acting on harmonic $n$ is $s+in\sigma\omega\bar c$: the $K=1$ response
with $\omega\to n\omega$. At $n=1$ with $\hat c_2=0$, Eq.~\ref{S-eq:hbmode}
reduces to $\hat c_1(s+i\sigma\omega\bar c)=\sigma B\epsilon\,\bar c/2$, which is
Eq.~\ref{eq:hbamp}. At $K=2$ the $n=2$ balance closes the second harmonic in
terms of the first,
\begin{equation}\label{S-eq:hbc2}
\hat c_2=\frac{\tfrac{\sigma B\epsilon}{2}\hat c_1-(1+i\omega\sigma)\,\hat c_1^{\,2}}
{s+2i\sigma\omega\bar c},
\end{equation}
so $\hat c_2=O(\epsilon^2)$ and is driven by the square of the first harmonic
rather than forced directly. Substituting back makes the $n=1$ balance cubic in
$\hat c_1$ and couples it to the mean, so the single relation
Eq.~\ref{eq:hbbranch} between $\epsilon$ and $\bar c$ is lost: for $K\ge2$ the
coefficients satisfy a quadratic system that is still polynomial and still
eliminable, but $\epsilon(\bar c)$ is a root of a higher-degree polynomial rather
than a rational function. The closure at $K=1$ is exactly the statement
$\hat c_2=0$.

Two features of the $K$ dependence follow from Eq.~\ref{S-eq:hbmode}. The mean
projection reads
\begin{equation}\label{S-eq:hbmean}
(\bar c-c_1)(\bar c-c_2)+\tfrac12\sum_{k=1}^{K}\big(a_k^2+b_k^2\big)
=\tfrac{\sigma B\epsilon}{2}\,a_1,
\end{equation}
so each retained harmonic adds one positive term to the left, which at fixed
$\epsilon$ drives $(\bar c-c_1)(\bar c-c_2)$ more negative and moves the fold to
smaller $\epsilon$. Adding a harmonic also changes every solved coefficient, so this
is not a monotonicity theorem; in all cases computed below the estimates do
approach the reference from above. And the amplitude of harmonic $n$ carries the denominator
$s^2+(n\sigma\omega\bar c)^2$, which grows like $n^2$ at large $\omega$ but
collapses to $s^2$ for every $n$ as $\omega\to0$. Truncation in $K$ is thus rapidly
convergent for rapid modulation and uncontrolled in the slowly varying limit ---
the same regime in which averaging over the modulation fails, and for the same
reason: no scale separation between the harmonics of the response.

Two cautions apply to this truncation. Multiplying by $c$ removes the singularity
at $c=0$ from the projected equations, so a low-order trigonometric solution is
not guaranteed to stay positive: at the $K=1$ fold the reconstructed profile has a
negative minimum ($-0.40$ at $\omega=1$, $-0.47$ at $\omega=2$), so Eq.~\ref{eq:hbfold}
is a threshold estimate, not an admissible speed profile. And a small projected
residual does not imply a small residual of the original equation; that must be
checked on a fine grid. At $\omega=2$ the fold estimate moves from $1.50$ ($K=1$)
to $1.29$, $1.25$ and $1.230$ ($K=2,3,5$) against the continuation fold $1.2299$,
and the minimum speed is positive from $K=2$ on; the residual of the original
equation near the minimum improves more slowly than the threshold. We therefore
use the truncated profiles only as threshold estimates, not as accurate speed
profiles.

\begin{figure}[htb]
\centering
\includegraphics[width=0.96\textwidth]{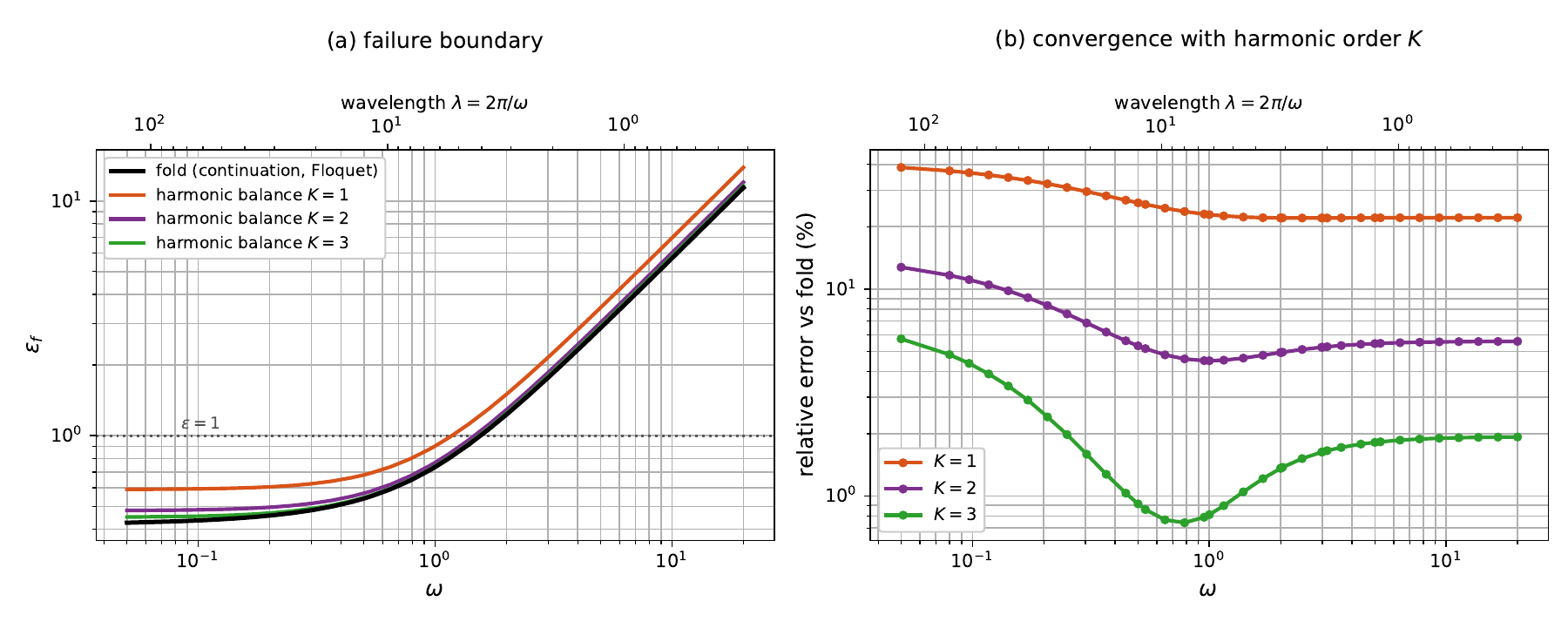}
\caption{\label{S-fig:hb} {\bf Failure boundary from harmonic balance, and its
improvement with $K$.} {\bf (a)} Reference fold $\epsilon_{\rm fold}(\omega)$
(black; continuation of the periodic profile with a neutral Floquet multiplier)
and the harmonic-balance folds for truncation $K=1,2,3$ (lines; $K=1$ is the
closed form Eq.~\ref{eq:hbfold}); in the computed range higher $K$ lies closer to
the reference. The horizontal line marks $\epsilon=1$, above which the modulated
coupling changes sign (Sec.~\ref{sec:domain}). {\bf (b)} Relative error of each
$K$ against the reference fold, which falls with harmonic order ($K=1$ overshoots
by $22$--$39\%$, $K=2$ by $4.5$--$13\%$, $K=3$ by $0.7$--$5.7\%$). The
$K=1$ profile is not positive at its fold (see text); the estimates are
threshold approximations at finite order.}
\end{figure}

Figure~\ref{S-fig:hb} compares the harmonic-balance folds with the fold obtained by
continuation of the periodic profile with a neutral Floquet multiplier (our
reference boundary). The $K=1$ estimate captures the boundary qualitatively across
the whole range (a finite plateau as $\omega\to0$ and linear growth at large
$\omega$) and overestimates it by $22$--$39\%$; $K=2$ and $K=3$ reduce the largest
error to $13\%$ and $6\%$. These are finite-order checks, not a convergence theorem; we use harmonic
balance as a compact analytic approximation and the continuation fold as the
reference.

It is worth contrasting this expansion with the harmonic-balance series
of Sec.~\ref{sec:hb}, as the two are complementary truncations
of the same periodic solution. The perturbation series truncates by \emph{power
of $\epsilon$}: it is anchored at the homogeneous mean $c_2$ and each order is a
linear response that adds one further harmonic, so a finite truncation is a
trigonometric polynomial whose mean is pinned near $c_2$ and which, being a
polynomial in $\epsilon$, does not display the fold at which propagation fails.
Harmonic balance instead truncates by \emph{number of harmonics} $K$: it keeps
the mean speed $c_0$ as a free unknown and solves for the amplitudes
self-consistently, to all orders in $\epsilon$ for the retained harmonics, so it
follows the periodic branch (and its terminating saddle-node) through failure and
to arbitrarily large $\epsilon$. They agree at first order: setting the
mean to $c_2$ and linearizing the first-harmonic balance in $\epsilon$ reproduces
$h_1$ exactly: the $K=1$ amplitudes (Sec.~\ref{sec:hb}) evaluated at $c_0=c_2$
have magnitude $\sigma B\epsilon/\sqrt{\gamma^2+(\sigma\omega)^2}=\epsilon A_1$ and
phase lag $\delta_1$, matching Eq.~\ref{eq:h1}. Beyond first order the mean shifts
(at $O(\epsilon^2)$) and a truncation at $K$ harmonics reproduces only those orders
whose harmonics it retains. Neither is contained
in the other at finite order: harmonic balance resums all powers of $\epsilon$
for $k\le K$ but discards higher harmonics, while the order-$n$ perturbation
series keeps every harmonic $k\le n$ but only to order $\epsilon^n$. The
perturbation series is therefore the tool of choice for the speed profile at
small amplitude, where its coefficients are explicit and its measured convergence
rapid; harmonic balance is what carries the analysis through the failure
boundary, where the mean speed has dropped well below $c_2$ and the amplitude is
of order unity.

\section{Triangle-wave inhomogeneity: the square-wave construction does not apply}\label{S-app:triangle}
The exact solvability of the alternating (square) case of
Sec.~\ref{sec:const} relies on $K$ being \emph{piecewise constant}, so that the
speed ODE is autonomous within each block. A triangle wave is piecewise
\emph{linear}, and the square-wave construction no longer applies. On a rising ramp
of a symmetric triangle of amplitude $\epsilon$ and period $2\pi/\omega$ we have
$K(x)=a+bx$ with $a=-\epsilon$, $b=2\epsilon\omega/\pi$. (We show below that the
standard reductions fail and that no elementary primitive is provided here; this
is not a proof that none exists.) Multiplying the reduced
equation Eq.~\ref{eq:cx} by $c$ gives
\begin{equation}\label{S-eq:tri-poly}
\sigma\,c\,\frac{dc}{dx} + c^2 - \big[c_1+c_2+\sigma B(a+bx)\big]\,c + c_1c_2 = 0,
\end{equation}
or, solving for the derivative,
\begin{equation}\label{S-eq:tri-abel}
\sigma\,\frac{dc}{dx} = -c + \big[c_1+c_2+\sigma B(a+bx)\big] - \frac{c_1c_2}{c}.
\end{equation}
Two special cases of Eq.~\ref{S-eq:tri-abel} are integrable, and the triangle is
neither:
\begin{itemize}
\item If $b=0$ (piecewise-constant $K$, the square wave), the right-hand side
carries no explicit $x$, so the equation is \emph{autonomous and separable},
$\int \sigma c\,dc/[-c^2+(c_1+c_2+\sigma B a)c-c_1c_2]=x$, yielding the closed
form Eq.~\ref{eq:x(c)}.
\item If the $c_1c_2/c$ term were absent, Eq.~\ref{S-eq:tri-abel} would be a
first-order \emph{linear} ODE with an $x$-dependent coefficient, again integrable.
\end{itemize}
The triangle retains \emph{both} the $1/c$ nonlinearity (inherited from the
exponential-kernel reduction) \emph{and} the explicit linear-in-$x$ forcing.
Equation~\ref{S-eq:tri-abel} is then an Abel equation of the second kind with a
non-constant coefficient,
\begin{equation}\label{S-eq:tri-abel3}
c\,\frac{dc}{dx} = \alpha(x)\,c - c^2 - c_1c_2,\qquad
\alpha(x)=c_1+c_2+\sigma B(a+bx)\ \ (\sigma=1),
\end{equation}
to which neither special case applies. We have not found an elementary solution
and make no claim that none exists; in practice the triangle is treated, like the
cosine, by the asymptotic and perturbative methods below.

What does survive is the asymptotic framework. In the slow limit
(Sec.~\ref{sec:r1}) the speed tracks the algebraic instantaneous fixed point
$c^2-[c_1+c_2+\sigma B K(x)]c+c_1c_2=0$ and fails at the trough $K=-\epsilon$, giving
the same $\epsilon_0=(\sqrt{c_2}-\sqrt{c_1})^2/(\sigma B)$ as every shape normalized to $\min k=-1$. In the
fast limit (Sec.~\ref{sec:r2}) one writes $c=C+\xi$ with
$\sigma\xi'=\sigma B\epsilon\,k$, so $\xi=(B\epsilon/\omega)\,\Phi(\omega x)$ where
$\Phi=\int k$ is a piecewise-\emph{parabolic} wave (the integral of the triangle).
The averaged drift $\sigma C'=-C+(c_1+c_2)-c_1c_2\,\mathcal{S}(C)$ depends on the
modulation only through $\mathcal{S}(C)=\langle(C+\xi)^{-1}\rangle$, the Stieltjes
transform of the occupation (sojourn) density of $\xi$ over one period. Writing
$a=\max\xi$, the three standard shapes give elementary closed forms,
\begin{align}
\text{cosine}\ (\xi=a\sin\theta):&\quad \mathcal{S}(C)=\frac{1}{\sqrt{C^2-a^2}},\label{S-eq:S-cos}\\
\text{square}\ (\xi\ \text{uniform on }[-a,a]):&\quad \mathcal{S}(C)=\frac{1}{a}\operatorname{arctanh}\frac{a}{C},\label{S-eq:S-sq}\\
\text{triangle}\ (\xi\ \text{parabolic}):&\quad \mathcal{S}(C)=\frac12\!\left[\frac{\arctan\sqrt{\tfrac{a}{C-a}}}{\sqrt{a(C-a)}}+\frac{\operatorname{arctanh}\sqrt{\tfrac{a}{C+a}}}{\sqrt{a(C+a)}}\right],\label{S-eq:S-tri}
\end{align}
reflecting the arcsine, uniform, and inverse-square-root sojourn densities of
$\xi$ respectively. The saddle-node of the drift (with $C+\xi>0$) fixes the
critical amplitude $a_c$ and the high-frequency slope. For the square it reduces
to the explicit pair
\begin{equation}\label{S-eq:sq-fold}
C_*^2=a_c^2+c_1c_2,\qquad (c_1+c_2)-C_*=\frac{c_1c_2}{a_c}\operatorname{arctanh}\frac{a_c}{C_*};
\end{equation}
the cosine gives $C_*(c_1+c_2-C_*)^3=(c_1c_2)^2$ (Eq.~\ref{eq:r2}), and the
triangle follows from Eq.~\ref{S-eq:S-tri}. The resulting slopes, $0.391$, $0.568$,
and $0.717$ for square, cosine, and triangle, reproduce the directly computed
boundary (Fig.~\ref{fig:shapes}, Table~\ref{tab:shapeslope}) to $\lesssim2\%$, the
square and cosine essentially exactly.

Exact solvability is, moreover, not needed in practice: the triangle is captured
by a rapidly converging approximation. Only the first-order term $h_1$ depends on
the modulation shape (it is the linear response to the Fourier series of
$k$), while every higher order follows the same shape-independent recursion
(Sec.~\ref{S-app:pert}). The small-amplitude series $c=c_2+\sum_n\epsilon^n h_n$
is equally explicit for the triangle, although the shape still enters every
higher coefficient through $h_1$. At $\omega=\pi$ the measured maximum error over
one period falls roughly geometrically with order (Table~\ref{S-tab:tri}), reaching
$\sim10^{-7}$ by third order at $\epsilon=0.1$. In the tested cases a handful of
terms reproduces the speed to numerical accuracy; these are measured errors, not a
convergence theorem.

\begin{table}[htb]
\caption{\label{S-tab:tri} Maximum error $E_n=\max_x|c^{(n)}-c|$ of the order-$n$
small-amplitude approximation for the \emph{triangle} inhomogeneity
($\omega=\pi$, default parameters). The series converges geometrically, as for the
cosine (Fig.~\ref{fig:conv}).}
\begin{ruledtabular}
\begin{tabular}{cccc}
$\epsilon$ & $E_1$ & $E_2$ & $E_3$ \\ \hline
$0.1$ & $1.2\times10^{-4}$ & $1.8\times10^{-6}$ & $1.5\times10^{-7}$ \\
$0.3$ & $1.1\times10^{-3}$ & $5.7\times10^{-5}$ & $1.2\times10^{-5}$ \\
$0.5$ & $3.3\times10^{-3}$ & $3.1\times10^{-4}$ & $9.5\times10^{-5}$ \\
\end{tabular}
\end{ruledtabular}
\end{table}
\section{Modulation families with elementary fast-limit transforms}\label{S-app:expfamily}
The elementary transforms of Sec.~\ref{S-app:triangle} are not isolated
coincidences. The averaged drift sees the modulation only through
$\mathcal{S}(C)=\langle(C+a\xi)^{-1}\rangle$ with $\xi=\int\!k$, and
$\int dx/(C+a\xi)$ is elementary whenever a single substitution rationalizes the
integrand, for example when $\xi$ is piecewise polynomial in $x$, exponential in a
common rate (the substitution $u=e^{\gamma x}$), or a trigonometric polynomial (the
Weierstrass substitution $t=\tan(\omega x/2)$). These are sufficient constructions,
not a classification. The cosine, square, and triangle are single-harmonic,
piecewise-constant, and piecewise-linear instances.

\begin{figure}[htb]
\centering
\includegraphics[width=0.85\textwidth]{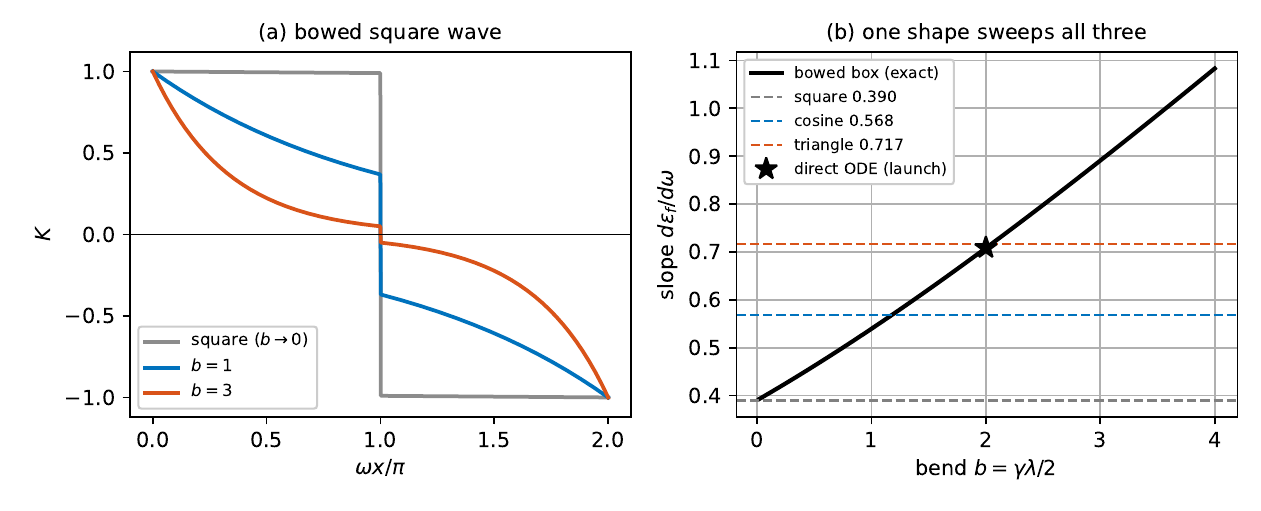}
\caption{\label{S-fig:expfamily} {\bf The exponential deformation of the square
wave.} (a) The bowed modulation Eq.~\ref{S-eq:Kbow} for several bends ($b\to0$ is the
square wave). (b) Its high-frequency slope (closed-form averaging via
Eq.~\ref{S-eq:S-bow}, solid) rises monotonically with the bend and sweeps continuously
through the square, cosine, and triangle values (dashed); the star is the slope of the
direct reduced-ODE launch threshold at $b=2$ ($50$ periods from $c(0)=c_2$, between
$\omega=10$ and $20$): $0.709$ against $0.708$ from averaging. Default parameters.}
\end{figure}

\emph{An exponential family deforming the square wave.} Replace the flat plateaus
of the square wave by exponential ramps: over one period $x\in[0,\lambda)$, with
decay length $\ell=\lambda/2b$ and bend $b=\gamma\lambda/2$,
\begin{equation}\label{S-eq:Kbow}
k(x)=\begin{cases}+\,e^{-x/\ell}, & 0\le x<\lambda/2,\\[2pt]
-\,e^{-(\lambda-x)/\ell}, & \lambda/2\le x<\lambda,\end{cases}
\end{equation}
zero-mean by antisymmetry, with $\min k=-1$ so the slow-limit floor is unchanged.
Each ramp's antiderivative is exponential. In the phase $\theta\in[0,\pi]$ of one
half-period the primitive is $u=\tfrac{\pi}{b}(1-e^{-b\theta/\pi})$ with mean $\mu$
(below), and the centered primitive $\phi=u-\mu$ has the density
$\rho_\phi(v)=1/[b(\pi/b-\mu-v)]$ on $-\mu\le v\le\tfrac{\pi}{b}(1-e^{-b})-\mu$,
reciprocal in the shifted coordinate $\pi/b-u$ (not in $\phi$ itself, which changes
sign). The Stieltjes transform is then a single logarithm,
\begin{equation}\label{S-eq:S-bow}
\mathcal{S}(C)=\frac{1}{b\,P}\ln\frac{PR-Q}{P-Q},\qquad
P=C+a\Big(\frac{\pi}{b}-\mu\Big),\quad Q=\frac{a\pi}{b},\quad R=e^{b},\quad
\mu=\frac{\pi}{b}-\frac{\pi}{b^{2}}\big(1-e^{-b}\big),
\end{equation}
with $a=B\epsilon/\omega$. As $b\to0$ the ramps flatten, the density
degenerates to the uniform one, and Eq.~\ref{S-eq:S-bow} reduces to the square-wave
transform Eq.~\ref{S-eq:S-sq} (with $\max\xi=\tfrac{\pi}{2}B\epsilon/\omega$): the
square is the $b=0$ member of this family. The high-frequency slope is
correspondingly tunable, rising monotonically with the bend from the square value
$0.39$ and passing through the cosine ($0.57$) and triangle ($0.72$) values
(Fig.~\ref{S-fig:expfamily}); a single bent shape thus realizes the entire range,
confirming that the slope is a functional of the whole profile rather than a trait
of a few special shapes.

\emph{Where closed form ends.} The logarithm Eq.~\ref{S-eq:S-bow} survives the
discontinuity of Eq.~\ref{S-eq:Kbow} but not its removal. A \emph{continuous}
exponential tent, with $k$ decaying for a length $\lambda/2$ into the trough and growing
back, is single-signed, so its period mean $m$ is nonzero, and the zero-mean
modulation $k-m$ has the antiderivative
\begin{equation}\label{S-eq:xi-tent}
\xi=\ell\big(1-e^{-x/\ell}\big)-m\,x,
\end{equation}
an exponential \emph{plus} a linear pedestal. The constant $-m$ in the kernel
integrates to $-mx$, and the substitution $u=e^{-x/\ell}$ that rationalizes the pure
exponential turns this term into $\ln u$,
\begin{equation}\label{S-eq:tent-fail}
\int\frac{dx}{C+a\xi}=\int\frac{-\ell\,du/u}{(C+a\ell)-a\ell\,u+am\ell\,\ln u},
\end{equation}
whose denominator contains both $u$ and $\ln u$, so the substitution no longer
rationalizes it. We have not found an elementary antiderivative and do not claim
that none exists; the discontinuity of the bowed square is what keeps $\xi$ free of
the pedestal and the substitution effective. The slow-limit floor, set by the trough alone, is unaffected, and the
slope remains computable numerically; it \emph{decreases} toward the floor as the
bend sharpens the trough.

\emph{The polynomial ladder.} The polynomial class, by contrast, never mixes:
piecewise-polynomial $k$ of any degree keeps $\xi$ piecewise-polynomial, and
$\int dx/(C+a\xi)$ is elementary for a polynomial denominator of every degree
(partial fractions). The triangle (degree-one $K$) is the simplest continuous
member; quadratic and higher tents remain closed-form, but their forms involve the
roots of the corresponding cubic and higher polynomials and offer no practical
advantage over the rapidly converging series of Sec.~\ref{S-app:pert}.

\section{Propagation/failure phase diagrams for the cosine modulation}\label{S-app:phase}
\label{S-sec:phase}
\begin{figure}[htb]
 \centering
  \includegraphics[width=0.96\textwidth]{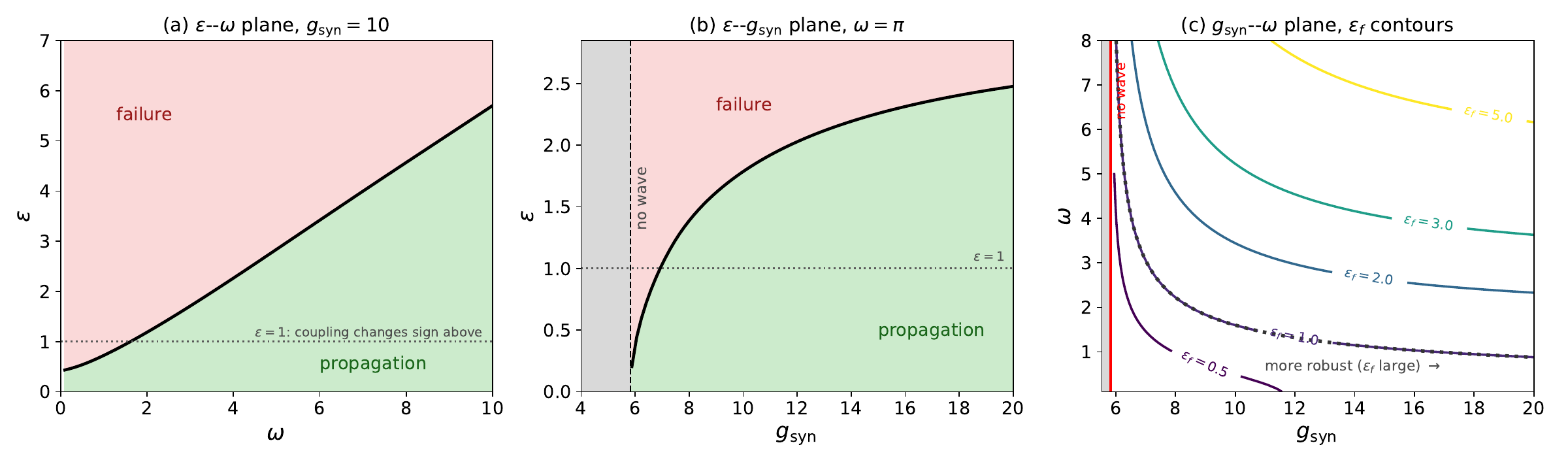}
  \caption{\label{S-fig:phase} {\bf Propagation/failure phase diagrams under cosine
  inhomogeneity.} In (a),(b) the black curve is the launch threshold $\epsilon_f$ for
  $c(0)=c_2$ from bisection on the reduced ODE (adaptive integration, $80$ periods,
  stop at $c=10^{-3}$); it is a finite-distance launch threshold, and the
  periodic-orbit fold differs from it by less than $4\%$ (Fig.~\ref{fig:epsf}); the green (lower) region sustains propagation,
  the red (upper) region fails. The dotted line $\epsilon=1$ in (a),(b), and the
  heavy dotted $\epsilon_f=1$ contour in (c), separate the excitatory-only regime
  (below) from the sign-changing extension of the model (above,
  Sec.~\ref{sec:domain}). {\bf (a)} $(\omega,\epsilon)$ plane at $g_{syn}=10$. {\bf (b)}
  $(g_{syn},\epsilon)$ plane at $\omega=\pi$; the gray strip $g_{syn}<g_{\min}=5.83$
  supports no traveling wave even in the homogeneous limit. {\bf (c)} The third pairing:
  the $(g_{syn},\omega)$ plane with contours of the failure amplitude $\epsilon_f$ itself
  (compact blend, Eq.~\ref{eq:blend}). $\epsilon_f$ rises with both coupling and
  frequency (a constant-$\epsilon_f$ contour trades higher $g_{syn}$ for lower
  $\omega$), and the gray strip is again $g_{syn}<g_{\min}$ (no wave). Default
  $\tau_1=1,\tau_2=2,\sigma=1,V_T=1$. }
\end{figure}

The complete picture is summarized by the propagation/failure phase diagrams of
Fig.~\ref{S-fig:phase}. In the $(\omega,\epsilon)$ plane at fixed $g_{syn}$
[panel (a)], the boundary rises from the quasi-static plateau
$\epsilon_0\approx0.42$ at small $\omega$ to the linear law at large $\omega$,
separating a lower propagation region from an upper failure region: fine-grained
inhomogeneity (large $\omega$) is tolerated up to large amplitudes, whereas slow
modulation halts the wave already at $\epsilon\approx\epsilon_0$. In the
$(g_{syn},\epsilon)$ plane at fixed $\omega$ [panel (b)], no traveling wave
exists below $g_{syn}=g_{\min}=5.83$, where the two homogeneous speeds $c_1$ and
$c_2$ coalesce (gray region); above $g_{\min}$ the failure amplitude grows
monotonically with $g_{syn}$, because stronger excitability ($B\propto g_{syn}$)
makes propagation more robust to inhomogeneity. The boundary emerges
continuously from zero at $g_{\min}$, where the marginal wave is destroyed by any
perturbation.

\section{\label{S-sec:kernels} Robustness across coupling kernels (exploratory)}
\begin{figure}[htb]
\centering
\includegraphics[width=0.82\textwidth]{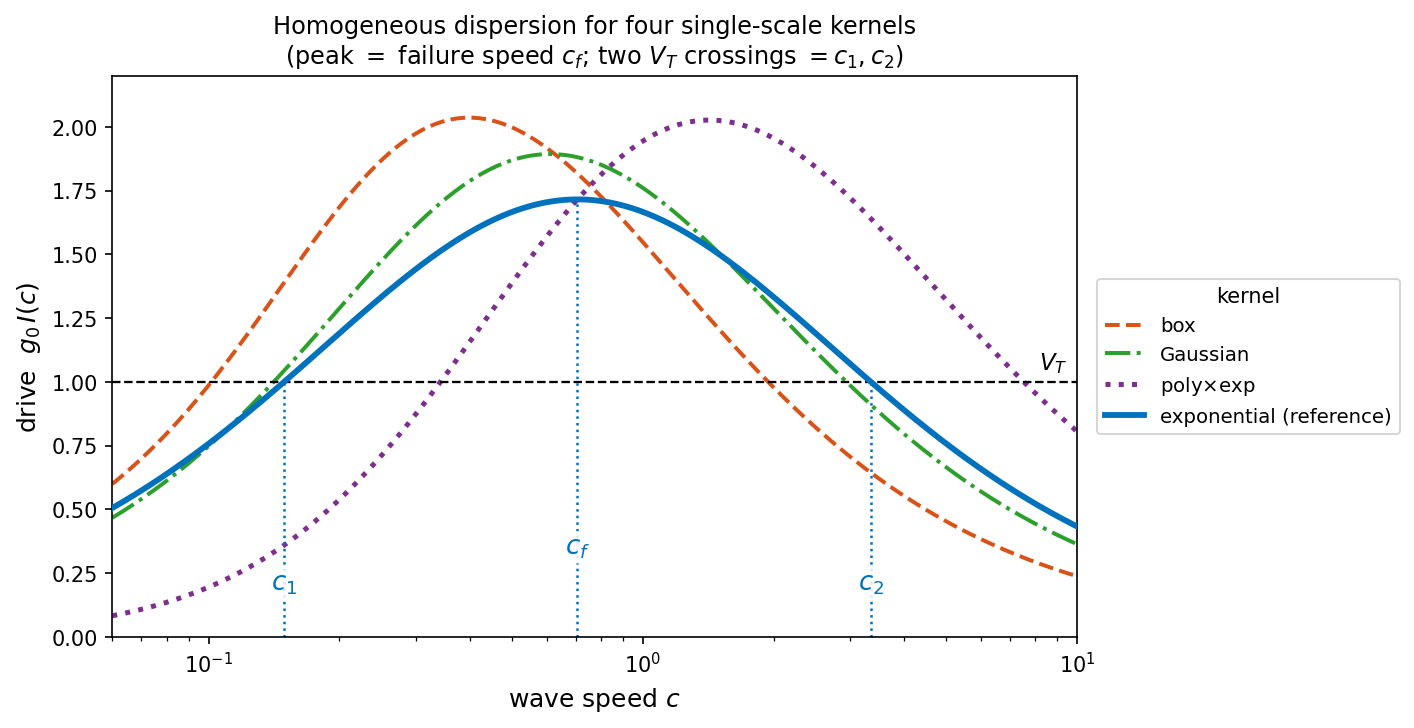}
\caption{\label{S-fig:kernels} {\bf The slow-limit failure mechanism is shared by four single-scale kernels.} Homogeneous
dispersion $g_0\int_0^\infty J(y)\,A(y/c)\,dy$ for four unit-mass kernels (exponential;
uniform/finite-support of Ref.~\cite{ErazoToscano2023}; Gaussian;
polynomial$\times$exponential; log speed axis). Each curve is a single-peaked bump:
two crossings of $V_T$ (dashed) give candidate speeds $c_1<c_2$, and, for a wave
tracking the upper branch, the slow-limit failure is the saddle-node at the peak, the bottleneck speed $c_\ast$ (labelled
$c_f$ in the figure; $c_1$, $c_2$, $c_\ast$ marked for the exponential reference).
The peaks sit at different speeds, ordered by each kernel's mean reach,
$c_\ast\approx\bar d/t_\star$ (an empirical approximation, see text), but the
structure is the same for these four kernels; only the exponential gives
$c_\ast=\sqrt{c_1c_2}$ exactly. Kernels with two well-separated scales can have
more than two crossings and are outside this comparison.}
\end{figure}

The closed forms above are specific to the exponential kernel, whose memorylessness
makes the leading-edge acceleration a local function of the speed. The failure
\emph{mechanism}, however, does not require this. In the slowly varying regime the wave
adiabatically tracks the local stable speed and dies where the local stable and unstable
speeds collide, a saddle-node fixed by the homogeneous dispersion
$g\int_0^\infty J(y)\,A(y/c)\,dy=V_T$ alone (the displacement $y$ between
pre- and postsynaptic neurons enters $A$ as the delay $y/c$), which exists for
any kernel. We therefore
repeated the slow-limit analysis for three further kernels of unit mass and comparable
range: the uniform finite-support kernel of Ref.~\cite{ErazoToscano2023}, a Gaussian, and
a polynomial$\times$exponential (Fig.~\ref{S-fig:kernels}).

Each of the four kernels, each with a single intrinsic spatial scale, has two
crossings of $V_T$ in Fig.~\ref{S-fig:kernels}, i.e.\ two candidate homogeneous
speeds $c_1<c_2$; if the wave tracks the upper branch in the adiabatic limit, that
branch is lost at the trough through a saddle-node at the dispersion peak, with an $O(1)$ failure
amplitude in all four cases ($\epsilon_f=0.42$--$0.51$). The mechanism, a
trough-controlled saddle-node of the local dispersion, is thus shared across the
kernels tested. It is not universal: a kernel with two well-separated spatial
scales (for instance an equal mixture of exponentials with $\sigma=1$ and
$\sigma=100$) has a dispersion curve with four crossings of $V_T$ at $g=15$, hence
more than two candidate speeds, and the branch followed by an incoming wave must
then be identified. Nor do dispersion crossings alone settle dynamic stability for
the non-exponential kernels, which we have not simulated. What is special to the
exponential is the exact \emph{law}: $c_\ast=\sqrt{c_1c_2}$ holds
because the inhomogeneity shifts only the linear coefficient of the quadratic dispersion,
leaving the product $c_1c_2$ fixed; for the other kernels the failure speed sits at the
dispersion peak and falls a few to ${\sim}10\%$ below the geometric mean (e.g.\ box
$0.40$ vs.\ $0.44$, Gaussian $0.61$ vs.\ $0.64$), while remaining well above $c_1$. The
phenomenology is general; the geometric-mean law is exponential-specific.

The \emph{ordering} of the peaks in Fig.~\ref{S-fig:kernels} (box leftmost, Gaussian and
exponential nearly on top of one another, polynomial$\times$exponential well to the
right) has a one-line reading. The synaptic filter $A(t)$ peaks at
$t_\star=\frac{\tau_1\tau_2}{\tau_2-\tau_1}\ln\frac{\tau_2}{\tau_1}=2\ln2\approx1.39$, so a
wave at speed $c$ recruits its afferents through a band of distances centred on
$d\approx c\,t_\star$. The homogeneous drive $\int_0^\infty\!J(d)\,A(d/c)\,dd$ is largest
when this recruitment band sits over the bulk of the kernel's mass, which places the peak
at $c_\ast\approx \bar d/t_\star$, where $\bar d=\int_0^\infty\!d\,J(d)\,dd\big/\!\int_0^\infty\!J(d)\,dd$
is the kernel's mean reach. The four kernels are simply ordered by reach: the box (hard cutoff,
$\bar d=\tfrac12$) keeps its mass closest to the source; the Gaussian
($\bar d=\sqrt{2/\pi}\approx0.80$) and exponential ($\bar d=1$) are intermediate and nearly
coincide; and the polynomial$\times$exponential $\tfrac12|d|e^{-|d|}$ \emph{vanishes} at the
origin and carries its mass out to $\bar d=2$. The estimate $c_\ast\approx\bar d/t_\star$
($0.36,\,0.58,\,0.72,\,1.44$) tracks the measured peaks ($0.40,\,0.61,\,0.71,\,1.42$) to
${\sim}10\%$ for these four kernels; it is an empirical approximation. This also confirms the natural
instinct that the polynomial factor is what does the work: $|d|^{\,n}e^{-|d|}$ has mean reach
$\bar d=n+1$, so raising the degree widens the hole at the origin and slides the peak
rightward as $c_\ast\propto(n+1)/t_\star$ ($n=0$ recovers the exponential). Peak \emph{height}
then sets the slow-limit threshold through $\epsilon_f=1-V_T/(g\,I_{\max})$: among these
four kernels the exponential has the lowest peak and the smallest $\epsilon_f$ ($0.42$).
This is a comparison of four kernels, not a general bound.

It is worth stating the slow-limit result in its period-free form, since nothing
in it relies on periodicity, provided the wave adiabatically tracks the stable
local branch. For a slowly varying modulation $K(x)$, periodic or aperiodic, the
tracked stable branch ceases to exist exactly where the local coupling first
reaches the marginal value, so the adiabatically tracked wave is lost if and only if
\begin{equation}\label{S-eq:gencrit}
\min_x g(x)<g_{\min}=\frac{V_T}{\displaystyle\max_{c>0}\int_0^\infty\! J(y)\,A(y/c)\,dy},
\qquad c_\ast=\arg\max_{c>0}\int_0^\infty\! J(y)\,A(y/c)\,dy,
\end{equation}
with $g(x)=g\,(1+K(x))$ the local coupling (the essential infimum for
discontinuous shapes): in this limit failure is governed by the single weakest
point of $g$, and Fig.~\ref{S-fig:kernels} is just Eq.~\ref{S-eq:gencrit} evaluated
for four kernels, the integral collapsing to a closed form only for the
exponential. Equation~\ref{S-eq:gencrit} is the strict $\lambda\to\infty$
envelope; a finite gradient lets the wave coast a little past threshold
(Eq.~\ref{eq:floor-approach}), so for the exponential kernel it is a one-sided
statement: $\min_x g(x)>g_{\min}$ guarantees that the tracked stable branch exists
for any modulation shape and rate (whether a given launch reaches it is the
separate basin question of Sec.~\ref{sec:hb}), whereas $\min_x g(x)<g_{\min}$
forces failure only as the gradient becomes slow. Finite-width defects and
disordered media add conditions of their own and are left to future work.

\begin{figure}[htb]
\centering
\includegraphics[width=0.99\textwidth]{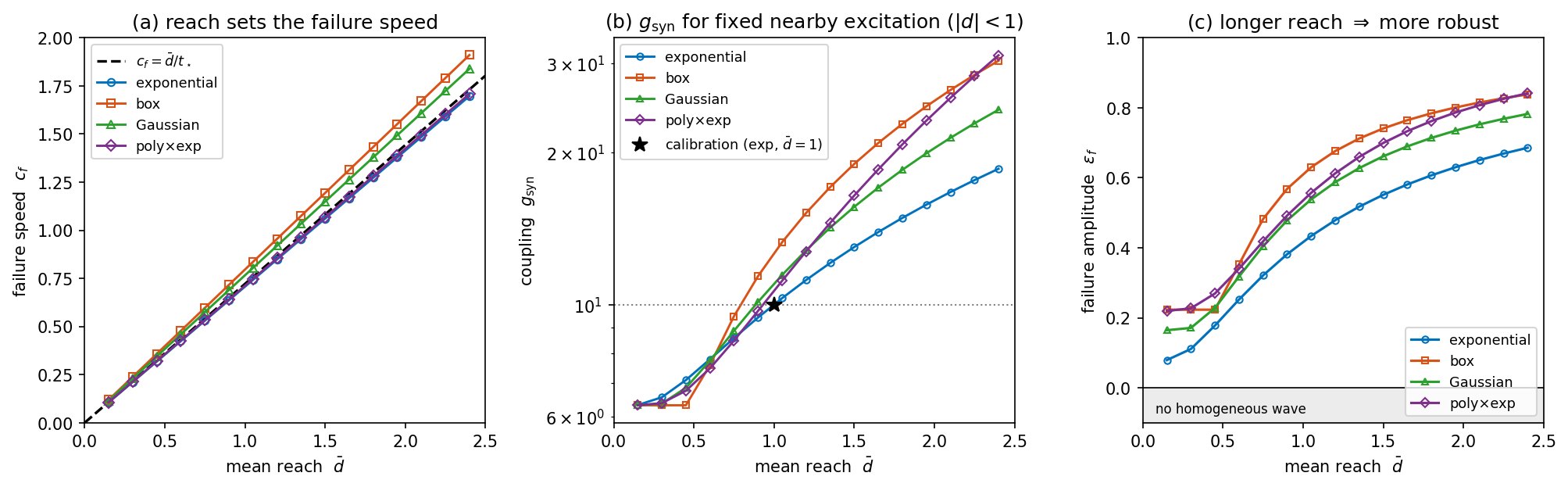}
\caption{\label{S-fig:reach} {\bf Reach sets the speed; at matched local drive the threshold rises with reach.}
Each unit-mass kernel of Fig.~\ref{S-fig:kernels} is swept over its spatial scale at
fixed nearby excitation: the synaptic input integrated over one relaxation length
($|d|<1$, close to $\ell\approx1.05$) is held equal across families through a per-family coupling
$g_\mathrm{syn}$. {\bf (a)} The failure speed collapses onto $c_\ast=\bar d/t_\star$ (labelled $c_f$ in the figure)
(dashed) for all shapes, with $\bar d$ the kernel's mean reach and $t_\star=2\ln2$
the synaptic-filter peak; every curve extrapolates through the origin
($c_\ast\to0$ as $\bar d\to0$), whereas (b) and (c) approach finite floors. {\bf (b)} Holding the local drive fixed requires more
total coupling as the kernel spreads out (log axis; $\star$ marks the exponential
calibration at $\bar d=1$). {\bf (c)} At matched local drive the failure
amplitude $\epsilon_f$ rises monotonically with reach, with shape setting the
offset; because total coupling grows along these curves, this is a normalization
comparison, not an isolated effect of reach.}
\end{figure}

The four kernels differ in two independent ways, their overall spatial
\emph{reach} and their \emph{shape}, and the slow-limit theory separates the
two cleanly. Rescaling any unit-mass kernel, $J_L(d)=L^{-1}J(d/L)$, leaves the
dispersion integral self-similar, $I_L(c)=I_1(c/L)$: the entire dispersion curve
simply stretches along the speed axis, and at fixed total coupling the slow-limit
threshold is unchanged. The speeds therefore scale linearly with
reach, $c_1,c_2,c_\ast\propto L$, while every dimensionless quantity is
reach-invariant. Taking the first moment $\bar d=\int_0^\infty\! d\,J\,dd/\!\int_0^\infty\! J\,dd$
as the reach measure collapses the failure speed across all four shapes onto a
single line, $c_\ast\simeq\bar d/t_\star$ with $t_\star=\frac{\tau_1\tau_2}{\tau_2-\tau_1}\ln\frac{\tau_2}{\tau_1}=2\ln2$
the synaptic-filter peak (Fig.~\ref{S-fig:reach}a; fitted slopes $0.71$--$0.80$
against $1/t_\star=0.72$, the residual ${\sim}10\%$ being the shape's signature).

To isolate how reach affects \emph{robustness} one must hold the local drive
fixed, so we give each family its own coupling
$g_\mathrm{syn}(L)=E_{\rm loc}/\!\int_{|d|<\ell}\! J\,dd$, equalizing the
synaptic input gathered within a radius $1$ (close to one relaxation length
$\ell=\sigma c_2/(c_2-c_1)\approx1.05$) across every kernel and every reach (calibrated so the exponential at $\bar d=1$
sits at $g_\mathrm{syn}=10$). Spreading a kernel out then costs more total
coupling, since mass leaks past the neighborhood (Fig.~\ref{S-fig:reach}b). The
reward is robustness: because $I_{\max}$ is reach-invariant while the local
fraction $\int_{|d|<\ell}\! J$ shrinks, the product $g_\mathrm{syn}I_{\max}$ grows
and the failure amplitude $\epsilon_f=1-V_T/(g_\mathrm{syn}I_{\max})$
\emph{increases monotonically with reach} (Fig.~\ref{S-fig:reach}c), rising from
$\epsilon_f\approx0.2$ for short-range coupling toward unity for the longest
reaches, with shape setting the offset. Because this normalization increases the
total coupling as the reach grows, the comparison does not isolate a mechanism
(such as coasting across a weak patch) or show that reach matters more than shape;
it answers the narrower question of how the threshold changes when local drive,
rather than total coupling, is held fixed.

\end{document}